\documentclass[journal]{IEEEtran}
\usepackage{cite}
\usepackage[T1]{fontenc}
\usepackage{amsmath,amssymb,amsfonts}
\usepackage{algorithmic}
\usepackage{graphicx}
\usepackage{textcomp}
\usepackage{subfigure}
\usepackage{varwidth}
\usepackage{url}
\usepackage{multirow}
\usepackage{color}
\usepackage{color}
	\definecolor{ForestGreen}{rgb}{0.0, 0.27, 0.13}
\definecolor{Dandelion}{rgb}{0.94, 0.88, 0.19}

\begin{document}

\title{Optimizing Co-flows Scheduling and Routing in Data Centre Networks for Big Data Applications}

\author{Sanaa Hamid Mohamed, Ali Hammadi, Taisir E.H. El-Gorashi, Jaafar Mohamed Hashim Elmirghani}
\maketitle
\begin{abstract}
This paper optimizes the scheduling and routing of the co-flows of MapReduce shuffling phase in state-of-the-art and proposed Passive Optical Networking (PON)-based Data Centre Network (DCN) architectures. A time-slotted Mixed Integer Linear Programming (MILP) model is developed and used for the optimization with the objective of minimizing either the total energy consumption or the completion time. The DCN architectures include four state-of-the-art electronic switching architectures which are spine-leaf, Fat-tree, BCube, and DCell data centres. The proposed PON-based DCN architectures include two designs that utilize ports in Optical Line Terminal (OLT) line cards for inter and possibly intra data centre networking in addition to passive interconnects for the intra data centre networking between different PON groups (i.e. racks) within a PON cell (i.e. number of PON groups connected to a single OLT port). The first design is a switch-centric design that uses two Arrayed Waveguide Grating Routers (AWGRs) and the second is a server-centric design. The study also considers different traffic patterns defined according to the distribution of map and reduce tasks in the servers and data skewness.
\end{abstract}


\maketitle

\section{Introduction}
\label{sec:introduction}
\IEEEPARstart{T}he limitations of current Data Centres Networks (DCNs) in terms of capacity, cost, and energy efficiency have triggered the need for new architectures capable of efficiently meeting the growing demands of cloud and fog computing distributed applications~\cite{REF1}.  The proven high-performance and cost-effectiveness of Passive Optical Networks (PONs) in access networks have motivated the use of these technologies in designing energy efficient, low cost, scalable, and elastic future cloud and fog DCNs. The integration of optical line terminals in access networks and data centres or additional processing devices for extended fog computing and caching capacities was suggested in~\cite{REF2,REF3,REF4,REF5}.

The benefits of using PONs in data centre networks include low equipment cost, low power consumption, data rate agnostic operation and high scalability of PONs compared to solely using Electronic Packet Switching (EPS).  Different PON technologies were considered for data centre networks, mainly while maintaining electronic Top-of-the-Rack (ToR) switches, including Orthogonal Frequency Division Multiplexing (OFDM), Wavelength Division Multiplexing (WDM) PON, and Arrayed Waveguide Grating Routers (AWGRs)~\cite{REF6,REF7,REF8,REF9,REF10}. In~\cite{REF11}, five novel PON-based designs were introduced to provide scalable, low cost, energy-efficient and high capacity intra and inter rack interconnections for future DCNs. These designs can replace typical access, aggregation and/or core switches in current DCNs with Optical Line Terminal (OLTs) and different passive intra-rack (i.e. between servers in the same rack) and inter-rack (i.e. between servers in different racks) interconnections. The third design, which is further discussed in~\cite{REF12} and~\cite{REF13}, utilizes AWGRs to provide high-performance interconnections between racks within each cell, and an array of photo detectors and tuneable lasers at each server for wavelength detection and transmission. An optimization study for the wavelengths assignment for inter-rack communication was presented in~\cite{REF12}. The energy efficiency of the design was assessed and compared to Fat-Tree and BCube data centres  and energy savings of 45$\%$ and 80$\%$ were achieved, respectively. The work in~\cite{REF14} introduced the fifth design which is a cost-effective server-centric PON DCN that does not require tuneable lasers. Instead, it utilizes Network Interface Cards (NICs) with non-tuneable optical transceivers in the servers for inter-rack communication. Experimental evaluations for the fifth design were provided in~\cite{REF15,REF16}. Benchmark studies were conducted against electronic, hybrid and optical DCNs through evaluating the completion time of sort operations performed on big data workloads in~\cite{REF1}, while resilience benchmark evaluations were reported in~\cite{REF17}.

To improve the performance and/or energy efficiency of MapReduce and related big data applications in different data centres, several studies were conducted. Hadoop clusters were modelled  in~\cite{REF18} and the influence of the network topology on the performance was studied. The completion time of MapReduce jobs in different topologies was estimated in~\cite{REF19} and compared to an optimal topology for MapReduce with dedicated links between servers. Also, intermediate data skew was examined and worse performance for all topologies was reported. In~\cite{REF20}, we examined the performance and energy efficiency of MapReduce in different electronic, hybrid and all-optical switching data centres and different rate-per-server values. The results indicated that optical switching technologies achieved an average power consumption reduction of 54$\%$ compared to electronic switching data centres with comparable performance. Virtual Machine (VM) assignment and traffic engineering were jointly considered in~\cite{REF21} to improve the energy efficiency of MapReduce-like systems. Energy savings of 60$\%$ in Fat-tree, and 30$\%$ in BCube data centres were achieved. The scheduling of big data traffic was addressed at the co-flow level, which is more applications-aware, in~\cite{REF22,REF23,REF24}. The authors in~\cite{REF22}  proposed \textit{Varys} to schedule inter-coflow traffic in data centres. The results indicated 3.66$\times$, 5.53$\times$, and 5.65$\times$ improvements compared to fair sharing, per-flow scheduling, and FIFO, respectively. For energy-efficient scheduling of MapReduce workloads, the authors in~\cite{REF25} optimized the parameters of the Low Power Idle (LPI) link sleep mode of the Energy Efficient Ethernet (EEE) standard~\cite{REF26} and savings between 5 and 8 times compared to legacy Ethernet were achieved. 

In this paper, we first introduce two of the proposed PON-based DCN designs which are an AWGR-centric design and a server-centric design ~\cite{REF11}. Then, a time-slotted Mixed Integer Linear Programming (MILP) model is developed and utilized to optimize the scheduling and routing of the co-flows of MapReduce shuffling phase with two objectives in different DCNs including the two PON-based designs. The completion time and energy consumption results based on the objective and the traffic patterns, defined according to the distribution of tasks, total data size and data skewness were then used to compare the performance and energy efficiency of these data centres. The remainder of this paper is organized as follows: Section~\ref{SECII} briefly describes the technologies used and requirements in the two PON-based designs. Section~\ref{SECIII} provides a MILP model to optimize the connections and wavelength routing and assignment in the AWGR-centric  design and the results. Section~\ref{SECIV} describes the system model used for optimizing the routing and scheduling in data centre networks given the characteristics of the considered MapReduce workloads. Section~\ref{SECV} presents the time-slotted MILP model, while Section~\ref{SECVI} provides the results and discussions. Finally, Section~\ref{SECVII} provides the conclusions.

\section{PON Technologies and Design Requirements in Proposed Data Centres}\label{SECII}

In access networks, PONs provide high speed broadband voice, data and video streaming services through efficient and flexible protocols to end user in premises. A single strand of fiber connected to an OLT port is passively split via splitters or Arrayed Waveguide Gratings (AWGs) to provision services for 128-256 end locations each equipped with an Optical Network Unit (ONU) or an Optical Network Terminal (ONT) at distances of up to 60 km~\cite{REF27}. Typically, this design can be a tree-based or a point to point design. The OLT switches at the central office are responsible for channel access arbitration and upload and download bandwidth allocation. In access networks, ONU to ONU communication is not a concern as the traffic is mostly transmitted from the OLT to users (i.e. download) or from the users to the OLT (i.e. upload). Hence, the tree topology and the lower upload bandwidth compared to the download bandwidth are suitable to meet residential applications requirements~\cite{REF11,REF28}.

For data centre networking, a Time Division Multiplexing (TDM) and a hybrid TDM-Wavelength Division Multiplexing (WDM) PONs were proposed in~\cite{REF11} by connecting a number of racks containing servers passively to OLT ports with flexible bandwidth allocation ratios. The racks are organized in cells where each cell contains a number of racks and is connected to an OLT port. These designs replace electronic access and aggregation switches by passive connections and core switches by the OLT. In these designs, the servers utilize OLT ports to communicate with other servers in the same cell or in other cells. Thus, this design limits the per server share of an OLT port bandwidth. To improve intra rack communication (i.e. within the rack), three passive technologies were proposed in~\cite{REF11,REF28}. Those are a star coupler, a Fiber Bragg Grating, and a Polymer optical backplane that was proposed in~\cite{REF29}. This backplane can provide non-blocking full mesh connectivity with a total of 1 Tbps capacity with multimode polymer waveguides each operating at 10 Gbps rate. This reduces the usage of OLT ports for intra rack traffic. The servers within the rack are interconnected with the remaining servers in the cell and with the OLT passively through directional splitters. For the inter rack communication within a cell, three designs were proposed in~\cite{REF11,REF28}. Two of these designs are further discussed in this work which are an AWGR-centric design~\cite{REF12,REF13,REF30} and a server-centric design~\cite{REF14}.

\subsection[The AWGR-centric Design (PON3)]{The AWGR-centric Design (PON3)}

For intra cell connections, the use of two AWGRs was proposed for the AWGR-centric design to achieve non-blocking connection between the racks and with the OLT~\cite{REF11,REF12,REF30,REF13}. An AWGR provides passive $ M\times M$ links between input and output ports. Each input port routes different wavelengths to different output ports and each output port  receives a different wavelength from an input port. This can be realized with cyclic and acyclic designs for the wavelengths routing~\cite{REF31}. The number of AWGR ports required (i.e. $M$) is a function of the number of racks within the cell and number of OLT ports used. In this design, each rack is connected to a single input port and a single output port in any of the two AWGRs. The OLT port is connected to both AWGRs through a single input and a single output port in each AWGR. The remaining ports connect the two AWGRs to achieve full connectivity between the racks and with the OLT.

\begin{figure}[ht]
\centering
	\scalebox{1.2}{\includegraphics{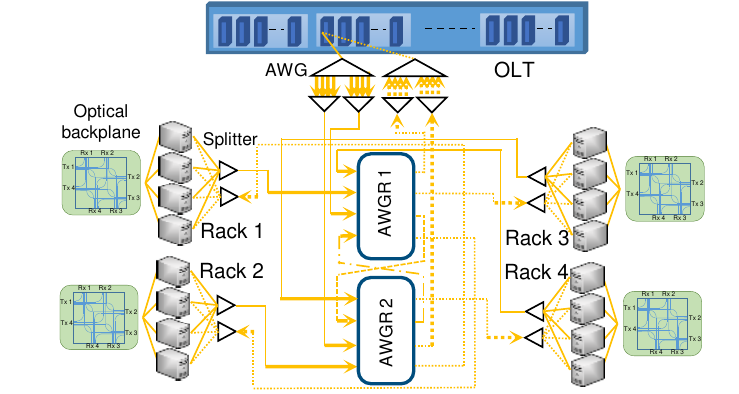}}
\caption[An example of the AWGR-centric design.]{An example of the AWGR-centric design.}
\label{FIGURE1}
\end{figure}


If intra rack communication is to be realized only through one of the passive solutions proposed in this Section, then a total of $G$-1 wavelengths are required for the inter rack connections, where $G$ is the total number of racks and OLT ports communicating. An example of this design with four racks and connection to a single OLT port (i.e. $G$=5) is depicted in Figure~\ref{FIGURE1}. This design requires two 4$\times$4 AWGRs and four wavelengths and allows up to 20 simultaneous connections between different racks and with the OLT. Section~\ref{SECIII} provides a MILP model developed to optimize the connections of the racks and OLT ports to the AWGRs ports and the wavelength assignment for the AWGR-centric design. 

The servers and the OLT port are  equipped with tuneable transceivers. In addition to offloading intra rack traffic, this design also reduces the usage of the OLT port for inter cell traffic.  If 10 Gbps tuneable transceivers are used in the example in Figure~\ref{FIGURE1}, a bisection bandwidth of 200 Gbps can be achieved. 

\subsection[The Server-centric Design (PON5)]{The Server-centric Design (PON5)}

\begin{figure}[ht]
	\centering
	\scalebox{1.2}{\includegraphics{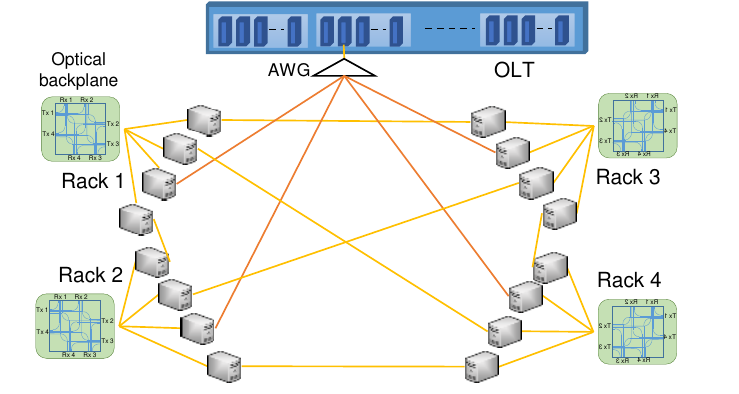}}
\caption[An example of the server-centric design.]{An example of the server-centric design (PON5).}
\label{pon5}
\end{figure}

The server-centric design is depicted in Figure~\ref{pon5}. This design utilizes Network Interface Cards (NICs) in servers to forward intra cell traffic between different racks and connects each rack with the OLT port through a single server in that rack. If a single wavelength is to be used, a star coupler can connect the OLT port with the racks and all servers share the bandwidth of that port through TDM only. If WDM is used,  an AWG is used instead of the star coupler. This design provides multiple paths between servers in different racks at reduced cost compared to the AWGR-centric design. 

\section{Optimizing the AWGR-centric Design}\label{SECIII}
This Section provides the MILP model developed to optimize the connections and wavelength routing and assignment in the AWGR-centric design. It also provides the optimization results when considering four racks and an OLT port. The MILP model takes an initial topology where all input ports of racks, OLT ports and AWGRs are connected to all output ports of AWGRs, and all output ports of racks, OLT ports and AWGRs are connected to all input ports of AWGRs. It then maximizes the number of achieved connections between the racks and the OLT port by assigning wavelengths to these connections while selecting unique port connections and maintaining correct wavelength routing and AWGRs use.

\subsection{MILP Model for optimizing the connections and wavelength routing and assignment}
The sets, parameters, and variables used in this model are provided below:

\begin{table}[ht!]
\textbf{Sets and parameters:}\\
\resizebox{\columnwidth}{!}{%
\begin{tabular}{ll}
$G$ & Number of communicating vertices including OLT   \\
& ports and PON groups (i.e. set of racks)\\
$\mathbb{W}$ & Set of wavelengths used (count to $G-1$) \\
$\mathbb{K}$ & Set of AWGRs \\
$M$ & The size of an AWGR (i.e. $M \times M$) which is equal  \\
& to the number of wavelengths  needed (i.e. $G-1$) \\
$\mathbb{T}$ & Set of OLT ports (initially one port is needed per  \\
& PON cell) \\
$\mathbb{R}$ & Set of PON groups \\
\end{tabular}
}
\end{table}

\begin{table}[ht!]
\resizebox{\columnwidth}{!}{%
\begin{tabular}{ll}
$\mathbb{P}$ & = $\mathbb{T} \cup \mathbb{R}$, Set of all communicating vertices \\
$\mathbb{I}_k$ & Set of input ports of AWGR $k; k \in \mathbb{K}$ \\
$\mathbb{O}_k$ & Set of output ports of AWGR $k; k \in \mathbb{K}$ \\
$\mathbb{N}$ & Set of all vertices (i.e. OLT ports, PON groups and   \\
&  AWGRs ports) in a cell\\ 
$\mathbb{N}_m$ &  Set of potential neighbors of vertex $m$; $m \in \mathbb{N}$\\ 
\end{tabular}
}
\end{table}

\begin{table}[ht!]
\textbf{Variables:}\\
\resizebox{\columnwidth}{!}{%
\begin{tabular}{ll}
$\beta_{mn}$ & Binary variable which is equal to one if vertex  \\
 &  $m$ is chosen to be connected with vertex $n$ \\
&  and is equal to zero otherwise; $m \in \mathbb{N},$\\
&   $n \in \mathbb{N}_m$ \\
$\chi_{jmn}^{sd}$ & Binary variable which is equal to one if wave- \\
& length $j$ is used in link ($m,n$)  if it is chosen to\\
&    connect vertex $s$ and vertex $d$ and is equal to \\
 &    zero otherwise; $j \in \mathbb{W}, m \in \mathbb{N}, n \in \mathbb{N}_m,$ \\
& $s,d \in \mathbb{P}, s \neq d$\\
$\mu_j^{sd}$ & Binary variable which is equal to one if wave-\\
& length $j$ is chosen to connect vertex $s$ and  \\
&vertex $d$ and is equal to zero otherwise; $j  \in \mathbb{W},$\\
&   $s,d \in \mathbb{P}, s \neq d$\\
\end{tabular}
}
\end{table}

The objective is to maximize the connections between vertices $s,d \in \mathbb{P}, s \neq d$ which can be expressed as:

\begin{gather} \label{obj_pc}
\max \sum_{\substack{j \in \mathbb{W}, s,d \in \mathbb{P}\\ s \neq d}} \mu_j^{sd}.
\end{gather}

Subject to the following constraints:

\begin{enumerate} 

\item Flow conservation: The allocation of the links and wavelengths to connections follows the flow conservation law~\cite{REF28}.
\begin{gather} 
 \displaystyle\sum_{n \in \mathbb{N}_m} \chi_{jmn}^{sd} -  \displaystyle\sum_{n \in \mathbb{N}_m} \chi_{jnm}^{sd} = \left\{\begin{matrix}
\mu_j^{sd} & m=s\\ 
-\mu_j^{sd} & m=d \\ 
 0 & otherwise
\end{matrix}\right. , \nonumber \\
\forall s,d \in \mathbb{P}, s \neq d, m \in \mathbb{N}, j \in \mathbb{W}  \label{Cons2II}
\end{gather}

\item Wavelength allocation: Constraint~\eqref{Cons3III} ensures that a single wavelength is selected per communicating pair. Constraint~\eqref{Cons4III} ensures that each destination receives from different sources through different wavelengths. Constraint~\eqref{Cons5III} ensures that each source transmits to different destinations through different wavelengths~\cite{REF28}.
\begin{gather}
 \displaystyle\sum_{j \in \mathbb{W}}\mu_j^{sd} \leq 1, \forall s,d \in \mathbb{P}, s \neq d. \label{Cons3III} \\
 \displaystyle\sum_{s \in \mathbb{P}, s \neq d}\mu_j^{sd} \leq 1, \forall d \in \mathbb{P}, j \in \mathbb{W}. \label{Cons4III} \\
 \displaystyle\sum_{d \in \mathbb{P}, s \neq d}\mu_j^{sd} \leq 1, \forall s \in \mathbb{P}, j \in \mathbb{W}. \label{Cons5III}
\end{gather}

\item Routing constraints: Constraint~\eqref{Cons6III} ensures that the flow for a connection between any pair is not relayed by any other vertex in $\mathbb{P}$.  Constraints~\eqref{Cons7III} and~\eqref{Cons8III} are for routing within the AWGRs, the first ensures that flows are only directed from input to output ports and the second ensures that each input port in an AWGR sends a different and single wavelength to each output port~\cite{REF28}.

\begin{gather}
 \displaystyle\sum_{\substack{s, d \in \mathbb{P}, s \neq d \\ n \in \mathbb{N}_i, j \in \mathbb{W}}}\chi_{jin}^{sd} -  \displaystyle\sum_{\substack{d \in \mathbb{P}, d \neq i \\ j \in \mathbb{W}}} \mu_j^{id} \leq 0, \forall i \in \mathbb{P}.\label{Cons6III} \\ \nonumber
 \displaystyle\sum_{n \in \mathbb{I}_k}\chi_{jmn}^{sd} \leq 0, \\
\forall s,d \in \mathbb{P}, s \neq d, k \in \mathbb{K}, m \in \mathbb{O}_k, j \in \mathbb{W}. \label{Cons7III} \\ \nonumber
 \displaystyle\sum_{\substack{s, d \in \mathbb{P}, s \neq d \\ j \in \mathbb{W}}}\chi_{jmn}^{sd} \leq 1, \\
\forall k \in \mathbb{K}, n \in \mathbb{O}_k, m \in \mathbb{I}_k. \label{Cons8III}
\end{gather}

\item Constraint to ensure that flows are routed only between connected communicating vertices selected according to constraints~\eqref{Cons10III}-\eqref{Cons18III}. Constraint~\eqref{Cons9III} ensures that the sum of traffic in link $(m,n)$ (i.e. $\sum_{s,d \in \mathbb{P}, s \neq d}\chi_{jmn}^{sd}$), which can maximally equal to one according to constraint~\eqref{Cons8III}, is equal to zero if $\beta_{mn}$ is equal zero and allows the sum to equal to one if  $\beta_{mn}$ is equal to one.
\begin{gather}
\displaystyle\sum_{s \in \mathbb{P}, d \in \mathbb{P}, s \neq d} \chi_{jmn}^{sd} \leq \beta_{mn}, \forall m \in \mathbb{N}, n \in \mathbb{N}_m, j \in \mathbb{W}. \label{Cons9III} 
\end{gather}

\item Constraints to determine the connections of the input and output ports of each AWGR with the OLT ports, PON groups and input and output ports of the other AWGR. Constraint~\eqref{Cons10III} ensures that each PON group is connected to a single AWGR input port, while Constraint~\eqref{Cons11III} ensures that each PON group is connected to a single AWGR output port. Constraint~\eqref{Cons12III}  assigns a single input port in each AWGR to the connection with the OLT port, while Constraint~\eqref{Cons13III} assigns a single output port in each AWGR to the connection with the OLT port. Constraint~\eqref{Cons14III}  ensures that each \textit{input} port in an AWGR has a unique connection with either a PON group, OLT port, or an \textit{output} port in the other AWGR. Constraint~\eqref{Cons15III} ensures that each \textit{output} port in an AWGR have a unique connection with either a PON group, OLT port, or an \textit{input} port in the other AWGR. Constraint~\eqref{Cons16III} ensures that all input and output ports of an AWGRs are internally connected. Constraint~\eqref{Cons17III}  ensures that the remaining output ports of each AWGR are connected to the remaining input ports of the other AWGR. Constraint~\eqref{Cons18III} ensures mutual neighboring between connected vertices.

\begin{gather}
\displaystyle\sum_{k \in \mathbb{K}, n \in \mathbb{I}_k} \beta_{mn} \leq 1, \forall m \in \mathbb{R}. \label{Cons10III} \\
\displaystyle\sum_{k \in \mathbb{K}, n \in \mathbb{O}_k} \beta_{mn} \leq 1, \forall m \in \mathbb{R}. \label{Cons11III} 
\end{gather}

\begin{table*}[t]
\centering
\caption{MILP results for the wavelengths assignment to OLT ports and PON groups communication in the AWGR-based PON DCN}
\label{TABLE1}
\resizebox{0.73\textwidth}{!}{%
\begin{tabular}{|c|c|c|c|c|c|}
\hline
 & OLT port 1 & PON group 1 & PON group 2 & PON group 3 & PON group 4 \\ 
 &$AWGR_1, \mathbb{O}_1 =1 $  &$AWGR_2, \mathbb{O}_2 =4 $ & $AWGR_1, \mathbb{O}_1 =4 $ & $AWGR_1, \mathbb{O}_1 =2 $ &$AWGR_2, \mathbb{O}_2 =1 $ \\ 
 &$AWGR_2, \mathbb{O}_2 = 3$  &  &  & &  \\ \hline
OLT port 1 & - & {\color{red}{$\lambda_3$}} & {\color{cyan}{$\lambda_2$}}  & {\color{ForestGreen}{$\lambda_1$}} & {\color{Dandelion}{$\lambda_4$}} \\ 
$AWGR_1, \mathbb{I}_1 =3 $  &  & 1 hop & 1 hop & 1 hop & 1 hop \\ 
$AWGR_2, \mathbb{I}_2 =3$ &  &  &  &  &  \\ \hline
PON group 1 & {\color{cyan}{$\lambda_2$}} & - & {\color{red}{$\lambda_3$}} & {\color{Dandelion}{$\lambda_4$}} & {\color{ForestGreen}{$\lambda_1$}} \\ 
$AWGR_1, \mathbb{I}_1 = 2$ & 1 hop &  & 1 hop  & 1 hop & 2 hops\\ \hline
PON group 2 & {\color{ForestGreen}{$\lambda_1$}} & {\color{Dandelion}{$\lambda_4$}} & - & {\color{cyan}{$\lambda_2$}} & {\color{red}{$\lambda_3$}} \\ 
$AWGR_2, \mathbb{I}_2 = 4$ & 1 hop & 1 hop &  & 2 hops & 1 hop \\ \hline
PON group 3 & {\color{red}{$\lambda_3$}} & {\color{ForestGreen}{$\lambda_1$}} & {\color{Dandelion}{$\lambda_4$}} & - & {\color{cyan}{$\lambda_2$}} \\ 
$AWGR_2, \mathbb{I}_2 = 1$ & 1 hop & 1 hop & 2 hops &  & 1 hop \\ \hline
PON group 4 & {\color{Dandelion}{$\lambda_4$}} & {\color{cyan}{$\lambda_2$}} & {\color{ForestGreen}{$\lambda_1$}} & {\color{red}{$\lambda_3$}} & - \\ 
$AWGR_1, \mathbb{I}_1 = 1$ & 1 hop & 2 hops & 1 hop & 1 hop & \\ \hline
\end{tabular}%
}
\end{table*}

\begin{figure}[!ht]
	\centering
	\scalebox{.6}{\includegraphics{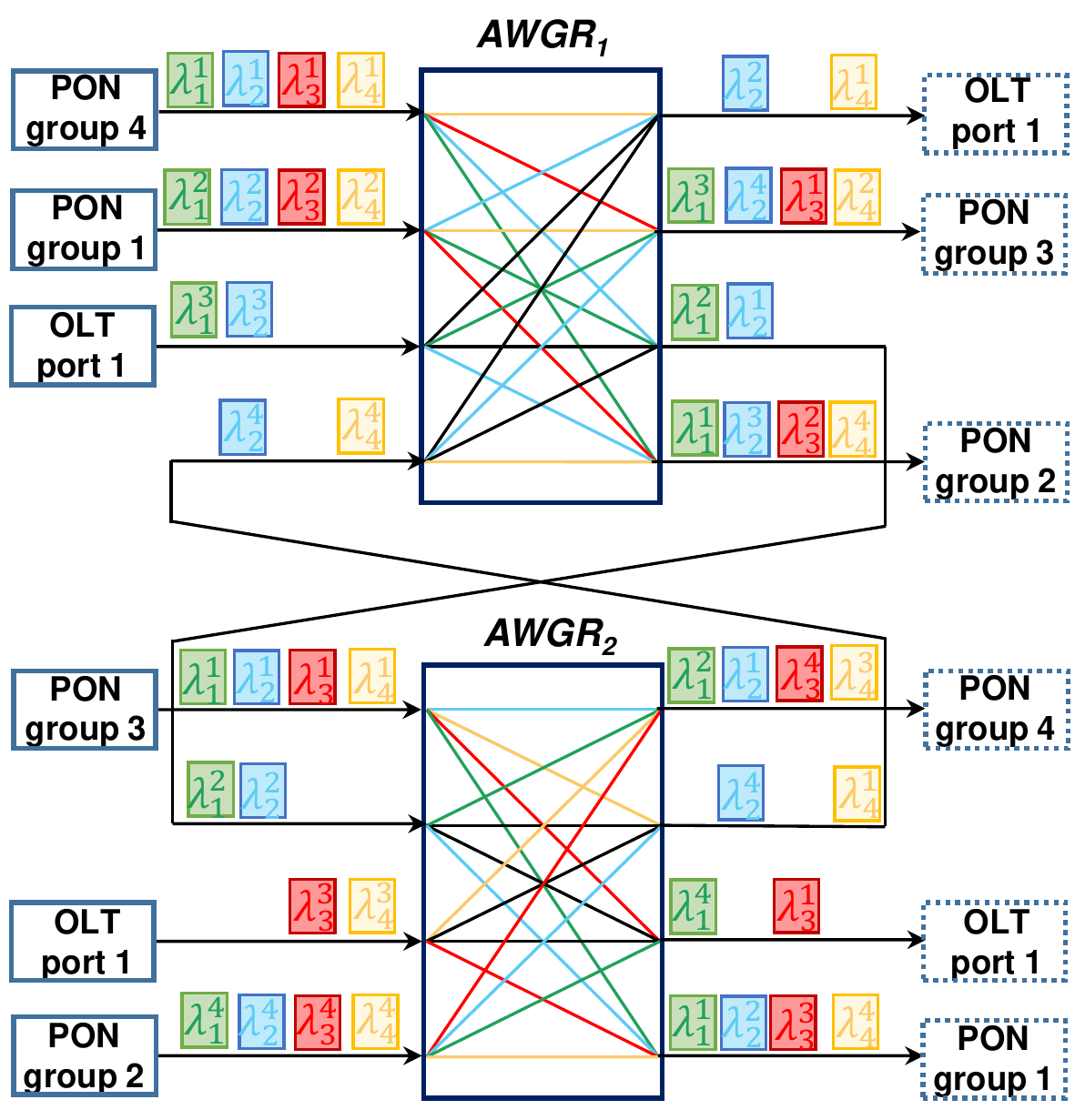}}
	\caption{MILP results for the wavelength assignments and the connections showing the wavelength continuity. Rectangles represent input ports of PON groups (i.e. racks) and dashed-line rectangles represent output ports of PON groups.}\label{FIGURE3}
\end{figure}

\begin{gather}
\displaystyle\sum_{n \in \mathbb{I}_k} \beta_{mn} \leq 1, \forall k \in \mathbb{K}, m \in \mathbb{T}. \label{Cons12III} \\
\displaystyle\sum_{n \in \mathbb{O}_k} \beta_{mn} \leq 1, \forall k \in \mathbb{K}, m \in \mathbb{T}. \label{Cons13III} \\
\displaystyle\sum_{m \in \mathbb{P} \cup \mathbb{O}_q} \beta_{mn} \leq 1, \forall k \in \mathbb{K}, q \in \mathbb{K}, k \neq q, n \in \mathbb{I}_k. \label{Cons14III} \\
\displaystyle\sum_{m \in \mathbb{P} \cup \mathbb{I}_q} \beta_{mn} \leq 1, \forall k \in \mathbb{K}, q \in \mathbb{K}, k \neq q, n \in \mathbb{O}_k. \label{Cons15III} \\
\beta_{mn} = 1, \forall k \in \mathbb{K}, m \in \mathbb{I}_k, n \in \mathbb{O}_k. \label{Cons16III} \\
\displaystyle\sum_{m \in \mathbb{O}_k, n \in \mathbb{I}_q} \beta_{mn} \leq \frac{M}{2}-1, \forall k \in \mathbb{K}, q \in \mathbb{K}, k \neq q. \label{Cons17III} \\
\beta_{mn} = \beta_{nm}, \forall m \in \mathbb{N}, n \in \mathbb{N}_m. \label{Cons18III} 
\end{gather}

\end{enumerate}

\subsection{Connections and Wavelength Routing and Assignment Results}

In an architecture where four racks, a single OLT port and two 4$\times$4 AWGRs are used, the MILP-based optimization results for the connections and the wavelength assignment for communication between the racks and the OLT port are provided in Figure~\ref{FIGURE3} and summarized in Table~\ref{TABLE1}. This Table shows the wavelengths that individual servers in racks and the OLT port should tune to in order to communicate with other servers in different racks or the OLT port. Such a setup can achieve up to 200 Gbps bisection capacity and allows arbitrary WDM/TDM routing. This bisection bandwidth for the single cell increases to 2 Tbps if 100 Gbps links are used. In the following Section, we developed and utilized a time-slotted MILP model to optimize the scheduling and routing in data centre networks with the objective of minimizing the total energy consumption or the completion time of MapReduce jobs. This enables us to compare the performance and energy efficiency of the proposed PON-based AWGR-centric architecture and the server-centric architecture with state-of-the-art data centre architectures.

\section{System Model and Parameters}\label{SECIV}

To quantitatively assess the impact of the data centre topology and workloads characteristics on the completion time and the energy efficiency of MapReduce shuffling operations, a MILP model that minimizes either the completion time or the total energy consumption while optimizing the scheduling and routing of co-flows was developed. The problem of optimizing the routing of co-flows with known sources and destinations through a capacitated network that performs shuffling operations can be categorized as a Multi-Commodity Flow problem which is NP-complete, but can be solved with solvers such as CPLEX. The MILP model developed contains additional constraints to model the routing requirements of each data centre topology. Also, the model can be considered time-slotted as a discrete time dimension is introduced in the variables to account for the scheduling of flows or the remainder of flows in the following scheduling time slots at the granularity of a second or less. Moreover, as the workload characteristics are highly coupled with the generated traffic in the data centre, the impact of the intermediate data skewness on the performance and energy efficiency of the routing and scheduling of the shuffling co-flows is also examined. The following Subsections provide the data centre models and the workloads modelling, in addition to the parameters and assumptions considered.

In this work, we optimized the routing and scheduling for pre-allocated map and reduce tasks. Although optimizing the tasks placement to improve the data locality at different stages of MapReduce can improve the performance and energy efficiency in lightly loaded data centres, with larger data sizes and larger data centre scales, it becomes harder to maintain locality for all tasks. Hence, we evaluate the impact of the data centre topology under random tasks allocation which also complies with the random allocation in native unmodified frameworks such as Hadoop~\cite{REF32}. The comparison between the data centres was not performed under similar bisection bandwidth or network diameter (i.e. number of hops between servers) as the work in~\cite{REF33}. Instead, we compared the performance and energy consumption required to shuffle the same amount of data when placing the map and reduce tasks in a fixed number of servers (i.e. 16 servers interconnected using different architectures). Furthermore, the data centre topologies are compared based on available technologies such as unifying the maximum data rate per transponder per wavelength to 10 Gbps while using the most suitable commodity hardware required for each architecture as will be detailed in the following Section. 

\subsection{Data Centre Models}\label{SECIV:1}
\begin{figure*}[!t]
\centering
\subfigure[]
{
\includegraphics[width=0.38\linewidth]{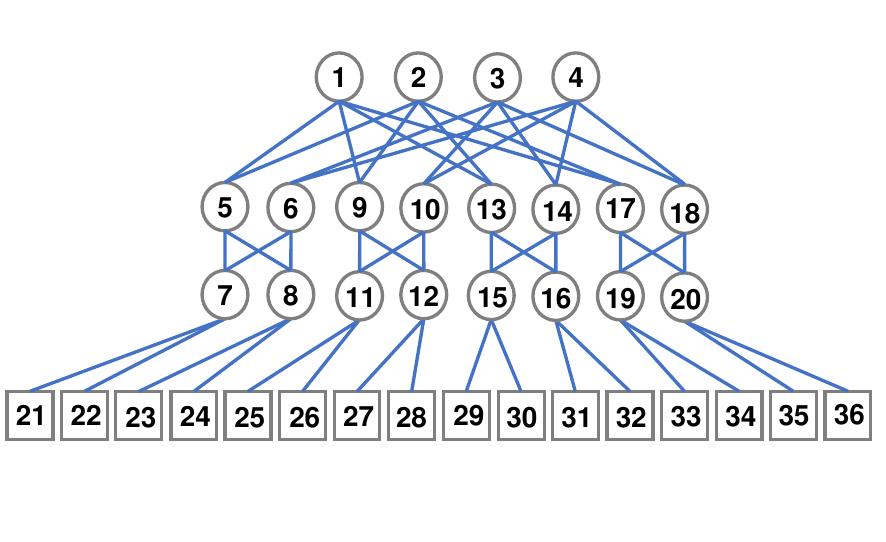}
\label{FIGURE4a}
}
\subfigure[]
{
\includegraphics[width=0.38\linewidth]{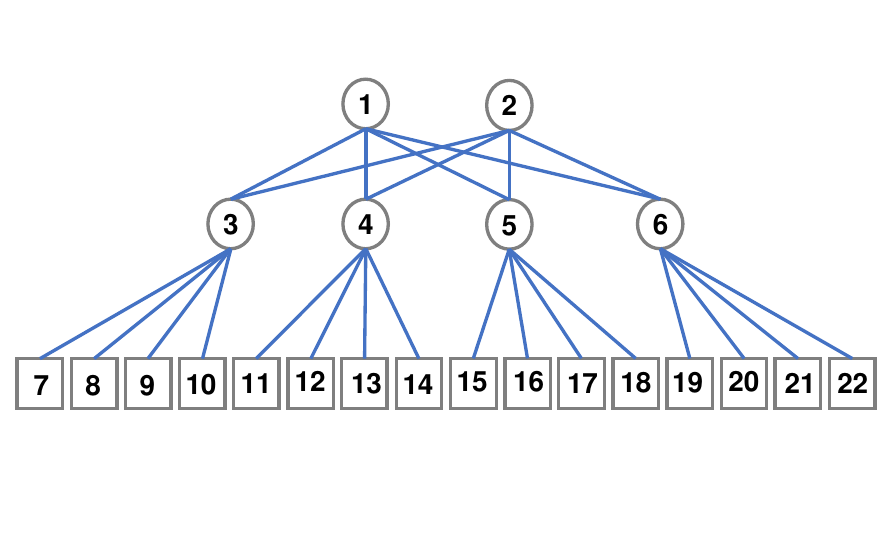}
\label{FIGURE4b}
}
\subfigure[]
{
\includegraphics[width=0.38\linewidth]{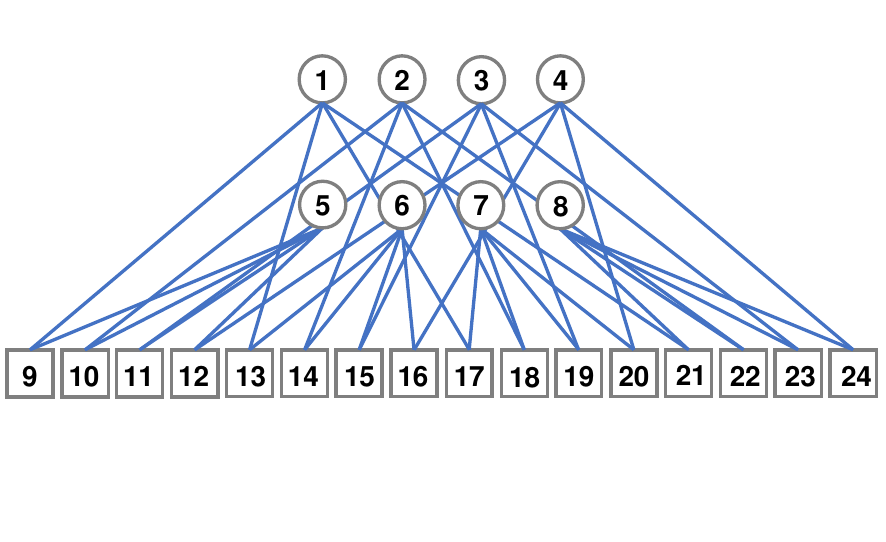}
\label{FIGURE4c}
}
\subfigure[]
{
\includegraphics[width=0.38\linewidth]{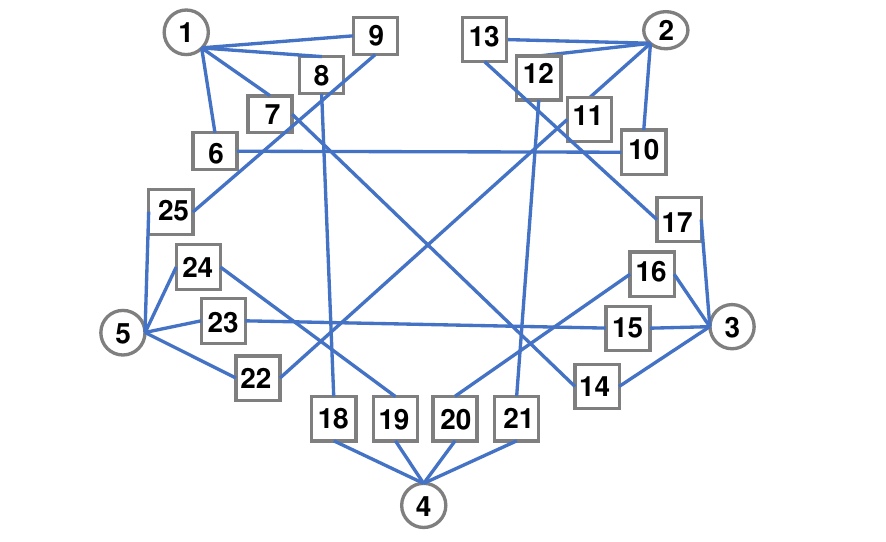}
\label{FIGURE4d}
}
\caption{Electronic DCNs graph representation. Squares and circles represent servers and switches, respectively. Blue edges represent EPS bidirectional links. (a) Fat-tree, (b) Spine-leaf, (c) BCube, and (d) DCell.}\label{FIGURE4}  
\end{figure*}

\begin{figure*}[!t]
\centering
\subfigure[]
{
\includegraphics[width=0.38\linewidth]{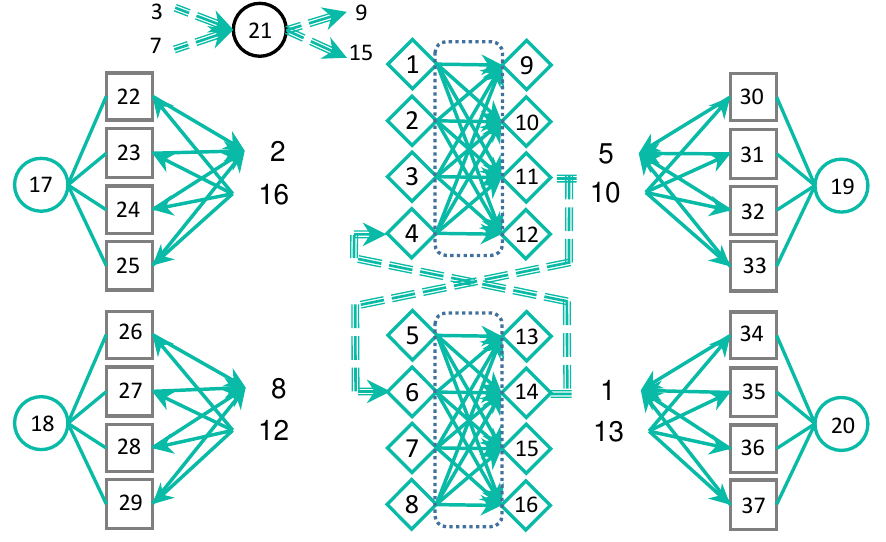}
\label{FIGURE5a}
}
\subfigure[]
{
\includegraphics[width=0.38\linewidth]{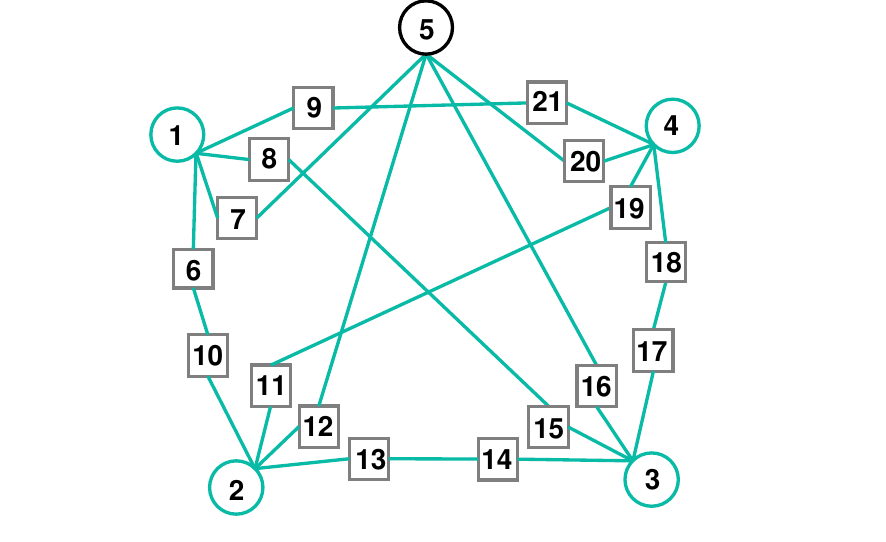}
\label{FIGURE5b}
}
\caption{PON-based DCNs graph representation. Squares and circles represent servers and switches, respectively. Cyan edges represent bidirectional links, while triple-dashed edges represent WDM links. Diamonds represent AWGR ports and rounded dashed rectangles represent AWGRs (a) PON3 and (b) PON5.}\label{FIGURE5} 
\end{figure*}

Six DCN topologies were considered. These are Fat-tree~\cite{REF34}, Spine-leaf~\cite{REF35}, BCube~\cite{REF36}, and DCell~\cite{REF37} which are electronic switching DCNs. In addition, the AWGR-centric and the server-centric  PON-based DCNs were considered. Each data centre is modelled as a graph with vertices in the set $ \mathbb{G}$ including the servers and the switches. The topology of each data centre is defined by a neighboring set ($\mathbb{G}_u$), where $u$ is a vertex in $\mathbb{G}$. Depending on these two sets, the edges of the graph are defined where each edge, denoted as $(u,v)$, represents a link between vertex $u$ and $v$, where $u,v \in \mathbb{G}$. The capacity of each edge per wavelength is set to 10 Gbps for all topologies. The power consumption and details of the electronic and optical equipment used in each topology are summarized in Table~\ref{TABLE2} and graph representations of the data centres are provided in Figures~\ref{FIGURE4} and~\ref{FIGURE5}.

The number of servers needed to accommodate map or reduce tasks is set to 16, to enable the comparison between the different data centre architectures.  To accommodate 16 servers, a $k$-ary Fat-tree network~\cite{REF34} with $k=4$ is sufficient, where $k$ is number of pods and also the number of ports in the switches. Such a Fat-tree requires a total of $3/4k^3=48$ links as modelled in Figure~\ref{FIGURE4a}. For BCube~\cite{REF36}, 16 servers can be connected in a $k=1, n=4$ configuration where $k$ and $n$ represents the layer index and the number of ports in the switches, respectively. In this configuration, a BCube$_0$ is composed of a 4-port switch and 4 servers and the BCube$_1$ is composed of 4 BCube$_0$ units and additional four 4-port switches as depicted in Figure~\ref{FIGURE4c}. A DCell$_{k}$ must be constructed recursively from $n-1$ DCell$_{k-1}$ units, where $k$ is the layer index and $n$ is the number of servers in each DCell$_0$~\cite{REF37}. The best configuration to connect 16 servers is a DCell$_1$ with five DCell$_0$ each with 4 servers~\cite{REF37} as in Figure~\ref{FIGURE4d}. To obtain results comparable to other topologies, the additional four servers are not assigned any additional tasks but can be used for the routing. The remaining topologies provide more flexibility with the number of servers in a rack and were configured as in Figures~\ref{FIGURE4b} and~\ref{FIGURE5}.

A 1 rack unit form-factor Cisco switch, model 3524X, was used as the switch in the spine-leaf DCN~\cite{REF38}. For the electronic Top-of-Rack (ToR) switches in the remaining data centres, SG500XG-8F8T was used~\cite{REF39}. In the servers of switch-centric topologies and in the ports of electronic switches, 10 Gbps Enhanced Small Form Factor Pluggable (SFP+) transceivers are considered~\cite{REF40}. In the servers of server-centric topologies (i.e. BCube, DCell, and PON5), PE10G2T-SR which is a commodity 10 Gbps NIC from Broadcom was considered~\cite{REF41}. PE10G2T-SR is based on short range Fiber (IEEE standard 802.3ae) connections and contains two 10 Gbps ports that can maximally provide a total of 18.7 Gbps capacity per port due to host and protocol overheads. A server consumes 14.29W to process offloaded 1 Gbps of traffic from this NIC~\cite{REF41}. 

Two rates ($\rho$) are considered for the transmission rate from each server. These rates are coupled with the data read speed from disks, memory or caches to the transceivers or the NIC of servers. The 2.8 Gbps rate matches the read speed of the MapReduce framework. It is used to read the results of map workers from Solid-State Drives (SSD) of the server before sending to the network. A rate of 8 Gbps can be considered if the framework uses memory, optimizes the use of Redundant Array of Independent Disks (RAID) or if it has caching capabilities in NICs. In this work, the completion time ($M$) calculation is done by only considering the transmission delay that results from the optimum routing and scheduling of the flows. Generally, there are four types of delay in communication networks which are the propagation delays, transmission delays, processing delays and queuing delays~\cite{REF42}. The first is mostly related to the speed of light in fiber links and can be ignored in DCN environments as the length of cables is typically small (i.e. less than 10 km) and the delay is estimated as 5$\mu$s per km~\cite{REF43}. The second is related to the capacity of links and the size of the transmitted packets and is considered dominant for elephant flows. The processing delay is related to the control overheads in the switches and NICs and depends on their CPU and RAM resources. The last delay (i.e. queuing delay) is determined by the limited buffer sizes in switches, the processing speed per packet, and the rate of packets arrival to each switch and can be estimated according to queuing theory which is complex in multi-path, and multi-hop connections in DCNs. The authors in~\cite{REF44} studied the performance of applications in Spine-leaf DCNs while assuming that leaf switches are ideal non-blocking switches and that the spine switch is a one large output-queued switch with infinite capacity. The authors in~\cite{REF44} assumed that the switches have unlimited buffer space and modelled them as shared-memory output queued switches that do not drop packets. Based on measurements, the processing latency in leaf switches was found to be 700 ns, while at spine switches 2 $\mu$s. The host networking stack  added a 10 $\mu$s delay resulting in a total RTT of 50 $\mu$s. Thus, in studies that optimize the routing and scheduling of large flows, the  processing, propagation, and queuing delays can be considered negligible compared to the transmission delay (i.e. the delay associated with transferring large files)~\cite{REF45}.

\begin{table*}[!ht]
\centering
\caption{Data centre-related parameters}
\resizebox{0.87\textwidth}{!}{%
\begin{tabular}{|c|c|c|c|c|c|c|c|}
\hline
\multirow{2}{*}{Topology} & \multirow{2}{*}{\begin{tabular}[c]{@{}c@{}}No of \\ servers\end{tabular}} & \multirow{2}{*}{\begin{tabular}[c]{@{}c@{}}No of\\ Switches\end{tabular}} & \multirow{2}{*}{\begin{tabular}[c]{@{}c@{}}No of \\ links\end{tabular}} & \multirow{2}{*}{\begin{tabular}[c]{@{}c@{}}Wavelengths \\ use $\mathbb{W}$\end{tabular}} & \multicolumn{3}{c|}{Networking Devices Characteristics} \\ \cline{6-8} 
 &  &  &  &  & Equipment & No & $O_{i(max)}$ Watts \\ \hline
Fat-tree~\cite{REF34} & 16 & 20 & 48 & Grey (colorless)& SG500XG-8F8T~\cite{REF39}  & 20 & 94.33   \\ \hline
Spine-leaf~\cite{REF35} & 16 & 6 & 24 & Grey (colorless) & Nexus 3524X~\cite{REF38} & 6 & 193  \\ \hline
\multirow{2}{*}{BCube~\cite{REF36}} & \multirow{2}{*}{16} & \multirow{2}{*}{8} & \multirow{2}{*}{32} & \multirow{2}{*}{Grey (colorless)} & SG500XG-8F8T~\cite{REF39}  &  8&  94.33  \\ \cline{6-8} 
 &  &  & & & PE10G2T-SR$^\dagger$~\cite{REF41} & 16 &  14  \\ \hline
\multirow{2}{*}{DCell~\cite{REF37}} & \multirow{2}{*}{20$^*$ } & \multirow{2}{*}{5} & \multirow{2}{*}{30} & \multirow{2}{*}{Grey (colorless)} & SG500XG-8F8T~\cite{REF39}  & 5  & 94.33   \\ \cline{6-8} 
 &  &  &  &  & PE10G2T-SR$^\dagger$~\cite{REF41} & 20 & 14   \\ \hline
\multirow{3}{*}{PON3~\cite{REF11,REF12}} & \multirow{3}{*}{16} & \multirow{3}{*}{7} & \multirow{3}{*}{$64^{\star\ddag}$} & \multirow{3}{*}{WDM} & 4$\times$4 Polymer back-plane & 4 & 12 \\ \cline{6-8} 
 &  &  &  &  & OLT with one card~\cite{REF46} & 1 & 217   \\ \cline{6-8} 
 &  &  &  &  &  $4 \times 4$ AWGR & 2 & 0   \\ \hline
\multirow{3}{*}{PON5~\cite{REF11,REF14}} & \multirow{3}{*}{16} & \multirow{3}{*}{5} & \multirow{3}{*}{23} & \multirow{3}{*}{Grey (colorless)} &  4$\times$4 Polymer back-plane & 4 & 12 \\ \cline{6-8} 
 &  &  &  & &   OLT with one card~\cite{REF46} & 1 & 217  \\ \cline{6-8} 
 &  &  &  &  &  PE10G2T-SR$^\dagger$~\cite{REF41} & 16  & 14   \\ \hline
\end{tabular}%
}
\begin{flushleft}
		\noindent \footnotesize{\quad $^*$The servers are 20 as it is a design scale requirement but workloads are allocated only in 16 servers, $^\star$ Excluding internal AWGRs links, $^\ddag$ directional.}\\
	\end{flushleft}
	\label{TABLE2}
\end{table*}

For the AWGR-based data centre (i.e. PON3), we considered the design presented in Figure~\ref{FIGURE1}. The corresponding system model is depicted in Figure~\ref{FIGURE5a}. PON group (i.e. rack) 1 contains servers 22, 23, 24, and 25, while the forth group contains servers 34, 35, 36, and 37. The OLT WDM port is in node 17, while nodes 1 to 16 are the ports of the two AWGRs. The power consumption of the OLT port is estimated by considering a single Ethernet card in the OLT in~\cite{REF46}. The power consumption required to operate the OLT was estimated to be 187 W by considering the maximum power consumption values for the power supply cards, common interface card, and the switching and control cards. The maximum power consumption of a single Ethernet interface card, which has 10G optical modules is 30 W.  A tuneable transceiver per server is required for the connections with the AWGRs and OLT port through the AWGRs. We considered SFP-10GDWZR-TC~\cite{REF47} which is a dual fiber 10 Gbps Tunable DWDM transceiver.  SFP-10GDWZR-TC consumes a maximum of 2 Watts and has span of 80 km which is more than sufficient in data centre environments. A Tuneable transceiver can only transmit at a single wavelength at a time, but can receive at multiple wavelengths if a wide band filter and appropriate receiver design and network interface are used. We considered a Field-Programmable Gate Array (FPGA)-based Network Interface card which has a power consumption between 11-12.3 W~\cite{REF48}. For the design presented in Figure~\ref{FIGURE1}, the servers can communicate with other servers in the rack only through an optical backplane.  For the optical backplane connections, we considered additional grey transceivers in the servers with total power consumption of 12 W per rack. The server-centric PON-based design, PON5, is assumed to have a WDM connection with the OLT, hence each of the 4 connected servers have a capacity of 10 Gbps. One cell of the server-centric PON-based DCN design was considered to have 4 servers per rack and a total of 4 racks in a single cell~\cite{REF1}. We considered PE10G2T-SR~\cite{REF41} for the NICs in the servers, in addition to the OLT and optical backplane equipment as in PON3.

\subsection{MapReduce Traffic Modeling}\label{SECIV:2}
In this evaluation, we considered a scenario where ten servers are dedicated for map tasks and six different servers are dedicated for reduce tasks. This configuration resembles a typical tasks ratio in the original Google's MapReduce~\cite{REF49}. The placement of map and reduce workers was randomly generated for each topology. To effectively examine network bottlenecks, sort workloads are considered. Sorting via MapReduce utilizes identity map functions to generate $<word,1>$ pairs from large text files. The entire intermediate data is to be shuffled according to words (i.e. keys) to reduce workers in order to be sorted and finally saved. Hence, input, intermediate and output data are all equal in size. The volume of total data to be sorted is varied from 1 Gbits to either 60 Gbits or\,120\,Gbits. A total of equivalent data is to be shuffled and transferred from map tasks to reduce tasks. We omit the details of assigning the data to individual tasks in each server and considered the traffic to be shuffled from a server containing several map tasks to one of the servers containing several reduce tasks as a single flow. This generates a total of 60 flows in the data centre network. Beside the shuffling traffic, Data File System (DFS) data transfers and control messages (i.e. heartbeats) are also required for the MapReduce framework. As input data placement is not deterministic in most of MapReduce-based frameworks and as the locality of map input data cannot be always ensured, a step before starting the map phase may include DFS input data transfers. Also, DFS final output data write can be assigned in servers different than the ones that are assigned to reduce tasks which will require additional transmission at the end of the MapReduce job. In this work we only considered the shuffling traffic and for simplicity, we assume that all map tasks finish at the same time and hence, all data is ready for transmission at the beginning of the shuffling phase. Such configuration can be realized by modifying the slow start option whose default configuration in Hadoop enables the shuffling to start when 3$\%$ of the map task output is ready~\cite{REF32}.

\begin{figure}[!ht]
	\centering
	\scalebox{.37}{\includegraphics{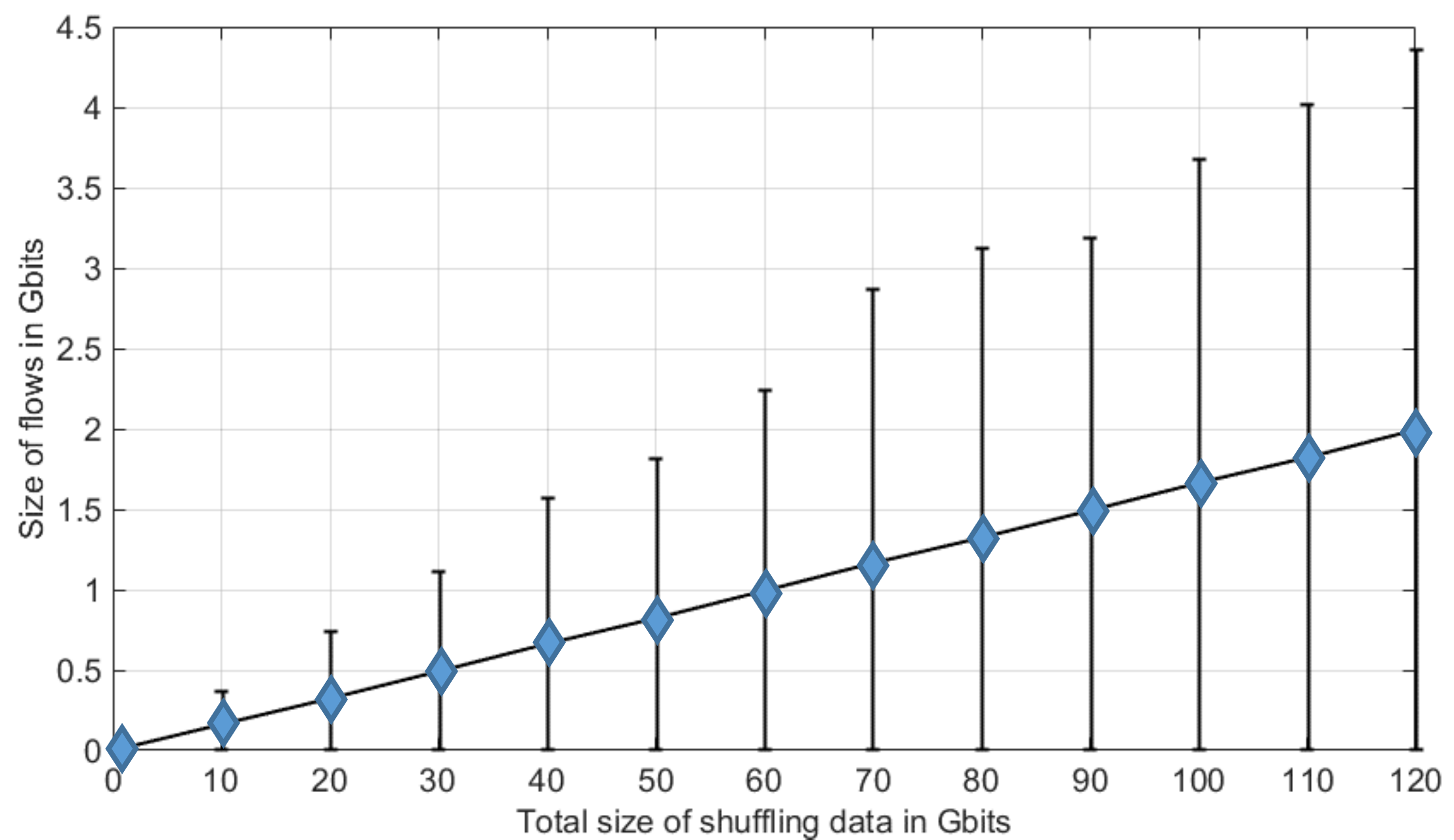}}
	\caption[Size of shuffling flows in Gbps.]{Size of shuffling flows in Gbps. Blue diamonds represent the uniform flow size with no data skew.}\label{FIGURE6}
\end{figure}

We considered two cases for the flows sizes distribution. In the first case, the results of all map tasks are of equal size, and hence, they generate equal size flows. Such workload is an equivalent to the Indy GraySort benchmark which has uniform intermediate key distributions due to balanced words count~\cite{REF50}. The second case considers uneven map task output sizes, which is equivalent to the Daytona GraySort benchmark~\cite{REF51}. The map output file sizes were generated randomly through a uniform distribution-based random generator with values that range between zero to the size of the total shuffling data volume. To ensure that the total sum of the randomly generated flow sizes (i.e. map output sizes) is maintained, proper scaling was performed. Figure~\ref{FIGURE6} shows the range of the skewed flow sizes as an error bar at each value of the total shuffling data volume.

\section{MILP Model for Optimizing the Scheduling and Routing of MapReduce Co-flows}\label{SECV}
Typically, in a Multi-Commodity Flow problem and based on the source, destination, and file size information of the flows in a co-flow, a MILP model can be used to optimize and determine the path and the data rate for each source. Then, the largest ratio of data to be sent divided by the data rate value determines the largest completion time required to transmit the co-flow. In the following MILP model, we alternatively considered a time-slotted approach, which allows the scheduling of the flows fully or partially in one of the time slots and in one of the routes between the servers containing the map and reduce tasks. In this case, the most congested link in the last  used time slot will determine the maximum transmission time of its flow, and will determine the completion time of transmitting the co-flow. By finding the most congested link, we have identified the link with the largest ratio of data to be sent over that link divided by the data rate of that link. Hence this is equivalent to replacing the multiple sources that send over this link by one source that has data to be sent equal to the sum of the data to be sent by each individual source, hence identifying the largest ratio. Based on the determined schedule and routing, the model also calculates the energy consumption based on the power consumption values of the networking equipment and the duration each element in the network is used while assuming that it will be on during the time slots it is utilized in. 

In what follows, the developed MILP model for this optimization problem is described. This model takes the topology of the data centre (i.e. connections and capacity of links), the power consumption for all equipment and the total shuffling traffic as input and provides the schedule (i.e. time slot and amount of assigned traffic for each link), the completion time and the total energy consumption. This is obtained under one of two objectives which are to reduce the energy consumption or to reduce the completion time and while considering the architectural and routing constraints. In our previous work, we utilized MILP in addition to different heuristics to reduce the energy consumption in different  network architectures and for different applications and systems. The work in~\cite{REF52,REF53} considered the energy efficiency of different optical networking architectures. In~\cite{REF54,REF55,REF56,REF57,REF58} the energy efficiency of IP over WDM core networks was addressed through optimizing the network design, data centres placement and renewable energy use. The study in~\cite{REF59} optimized virtual networks embedding in IP over WDM networks while the comprehensive work in~\cite{REF60} considered the energy efficiency of future IP over WDM network through design, network embedding and content placement joint optimization. In~\cite{REF61}, the bounds on the energy efficiency of future IP over WDM networks were evaluated, while in~\cite{REF62,REF63} the energy efficiency of core networks while accounting for their survivability was addressed. In~\cite{REF64}, the energy efficiency of IP over WDM networks was considered while accounting for recent Internet subscription models. The energy efficiency of content distribution systems such as cloud-based systems~\cite{REF65}, peer-to-peer systems~\cite{REF66} and video content caching in core network and fog data centres locations~\cite{REF67,REF68} was also considered. In~\cite{REF69}, the energy efficiency of virtual machine placement in cloud and fog environments was studied and in~\cite{REF70,REF71} optimizing big data transmission and progressive processing in core IP over WDM networks was addressed. The design of disaggregated data centres with optical interconnects  was optimized in~\cite{REF72}. Access networks design optimization with network function virtualization~\cite{REF73}, different Internet-of-Things (IoT) devices and systems~\cite{REF74,REF75,REF76} and with patient-centric systems and applications that are based on big data analytics~\cite{REF77,REF78} was addressed. Also, virtual machine placement optimization in the AWGR-centric PON-based data centre design was addressed in the study in~\cite{REF79} and the end-to-end delay when using the server-centric PON-based data centre design in fog environment was studied experimentally in~\cite{REF80}. For this work, the parameters and variables of the developed MILP model are provided below. All variables are set to be larger than or equal to zero. The sets are represented as double-lined letters while the small letters in the subscripts and postscripts indicate the indices of the parameters or variables.

\begin{table}[ht!]
\textbf{Sets and parameters:}\\
\resizebox{\columnwidth}{!}{%
\begin{tabular}{ll}
$\mathbb{G}$ &  Set of all vertices (servers and switches) in the data \\
& centre\\ 
$\mathbb{G}_u$ &  Set of neighbors of vertex; $u \in \mathbb{G}$\\ 
$\mathbb{R}$ & Set of servers in the data centre $(\mathbb{R} \subset \mathbb{G})$ \\
$\mathbb{S}$ & Set of switches in the data centre $(\mathbb{S} \subset \mathbb{G}, \mathbb{R} \cap \mathbb{S} =$\\
& $\varnothing)$ \\
$\mathbb{W}$ & Set of wavelengths \\
$\mathbb{T}$ & Set of time slots \\
$D$ & The duration of a time slot (in seconds) \\
$\Delta_{sd}$ & The total shuffling traffic to be transmitter from \\
&  server $s$ to server $d$; $s,d \in R$ (in Gbits) \\
$C_{uvw}$& Capacity of link ($u$,$v$); $u,v \in G$, at wavelength \\
&  $w \in W$ (in Gbps) \\
$P_{i(max)}$ & The maximum power consumption of a transceiver \\
&  in server $i$; $i \in \mathbb{R}$ or switch $i$; $i \in \mathbb{S}$ or an NIC in  \\
& server $i$; $i \in \mathbb{R}$ (in Watts) \\
$\epsilon$ & The performance per Watt in servers with an \\
& NIC (in Watts per Gbps) \\
$O_{i(max)}$ & The maximum power consumption of switch $i$; \\
& $i \in \mathbb{S}$  (in Watts)  \\
$\rho$ & The maximum data rate per server (in Gbps) \\
$\sigma$ & The maximum data rate per switch (in Gbps) \\
$L$ & A very large number \\
$Q$ & A weighting factor 
\end{tabular}
}
\end{table}

\begin{table}[ht!]
\textbf{Variables:}\\
\resizebox{\columnwidth}{!}{%
\begin{tabular}{ll}
$M$ & Completion time which is equal to  the time when the\\
&  last transmission ends \\
$E$ & The total energy consumption \\
$B_{iwt}$ & Binary variable which is equal to one if the transce- \\
& iver/NIC of server $i$ is used at wavelength $w$ and  \\
 & time slot $t$ and is equal to zero otherwise; $i \in \mathbb{R},$ \\
&  $w \in \mathbb{W},  t \in \mathbb{T}$   \\
$A_{iwt}$ & Binary variable which is equal to one if switch $i$   \\
& is used at wavelength $w$ and time slot $t$ and is equal  \\
 & to zero otherwise; $i \in \mathbb{S}, w \in \mathbb{W}, t \in \mathbb{T}$ \\
$\Gamma_{uvwt}$ & Binary variable which is equal to one if link $(u,v)$   \\
 &  is used at wavelength $w$ and time slot $t$ and is equal  \\
&  to zero otherwise; $u \in \mathbb{G}, v \in \mathbb{G}_u, w \in \mathbb{W}, t \in \mathbb{T}$ \\
$Z_{uvwt}$ & Binary variable which is equal to one at the link \\
 &$(u,v)$, wavelength $w$, and time slot $t$ where $M$  \\
& occurs (i.e. the last used link) and is equal to zero   \\
 &  otherwise; $u \in \mathbb{G}, v \in \mathbb{G}_u$, $w \in \mathbb{W}, t \in \mathbb{T}$ \\
$\chi_{uvwt}^{sd}$& Traffic in link $(u,v)$ that contributes to the shuffling  \\
 & data flow to be transmitted from server $s$ to server $d$  \\
& at wavelength $w$ and time slot $t$; $s, d \in  \mathbb{R}, s \neq d,$ \\
& $ u \in  \mathbb{G}, v \in  \mathbb{G}_u, w \in  \mathbb{W}, t \in  \mathbb{T}$ \\
$\psi_{uvwt}$& Total traffic in link $(u,v)$ at wavelength $w$ and time \\
& slot $t$; $u \in  \mathbb{G}, v \in  \mathbb{G}_u, w \in  \mathbb{W}, t \in  \mathbb{T}$
\end{tabular}
}
\end{table}

\begin{table}[ht!]
\resizebox{\columnwidth}{!}{%
\begin{tabular}{ll}
$\beta_{iwt}$ & The total ingress and egress traffic of the transceiver/ \\
& NIC of server $i$ at wavelength $w$ and time slot $t$;  \\
& $i \in \mathbb{R},$ $w \in \mathbb{W}, t \in \mathbb{T}$ \\
$\alpha_{iwt}$ & The total ingress and egress traffic of switch $i$ at wave-  \\
& length $w$ and time slot $t$; $i \in \mathbb{S}, w \in \mathbb{W}$, $t \in \mathbb{T}$ \\
$\delta_{sdt}$ & Traffic for shuffling data flows from server $s$ to server    \\
 & $d$ selected to be transmitted  at time slot $t$ \\
$\theta_{iwt}$ & Power consumption of the transciever of server $i$ at  \\
 & wavelength $w$ and time slot $t$; $i \in \mathbb{R}, w \in \mathbb{W}, t \in \mathbb{T}$ \\
$\phi_{iwt}$ & Power consumption of switch $i$ at wavelength $w$ and  \\
& time slot $t$; $i \in \mathbb{S}, w \in \mathbb{W}, t \in \mathbb{T}$ \\
$\Omega_{uvwt}$ & The earliest possible completion time of flow $\psi_{uvwt}$ \\
& that starts at time slot $t$ and is routed over link $(u,v)$ \\
&  at wavelength $w$; $u \in \mathbb{G}, v \in \mathbb{G}_u, w \in \mathbb{W}$, $t \in \mathbb{T}$\\
$\tau_{uvwt}$ & Variable that has the same definition as $\Omega_{uvwt}$ with  \\
& the exception that it takes a value of zero  if the link  \\
& $(u,v)$ is inactive; $u \in \mathbb{G}, v \in \mathbb{G}_u, w \in \mathbb{W}$, $t \in \mathbb{T}$
\end{tabular}
}
\end{table}

In the data centre networks considered, active devices are either transceivers or NICs in servers in addition to switches. The power consumption of the transceiver in server $i$ at wavelength $w$ and time slot $t$ with an ON/OFF power profile is equal to:
\begin{gather}
\theta_{iwt}= B_{iwt} \, P_{i(max)}, \nonumber \\ 
\forall i \in \mathbb{R}, \,  w \in \mathbb{W}, t \in \mathbb{T}.  \label{Eq19V}
\end{gather}

\noindent The power consumption of an NIC in server $i$ at wavelength $w$ and time slot $t$ is equal to:
\begin{gather}
\theta_{iwt}= B_{iwt} \, P_{i(max)} + \epsilon \, \beta_{iwt}, \nonumber \\ 
\forall i \in \mathbb{R}, \,  w \in \mathbb{W}, t \in \mathbb{T}.  \label{Eq20V}
\end{gather}

\noindent The power consumption of switch $i$ at wavelength $w$ and time slot $t$ is equal to:
\begin{gather} 
\phi_{iwt}= A_{iwt} \, O_{i(max)},  \nonumber \\ 
\forall i \in \mathbb{S}, \,  w \in \mathbb{W}, t \in \mathbb{T}.  \label{Eq21V}
\end{gather}

\noindent  Then, the total energy consumption ($E$) is equal to:
\begin{gather} 
E  =D\left[ \sum_{i \in \mathbb{R}, w \in \mathbb{W}, t \in \mathbb{T}} \theta_{iwt} + \sum_{i \in \mathbb{S}, w \in \mathbb{W}, t \in \mathbb{T}} \phi_{iwt} \right]. \label{Eq22V}
\end{gather}

\noindent The MILP model can have one of the following two objectives. The first objective is to minimize, $E$, the total energy consumption which can be expressed as:

\begin{gather} 
\min \left[  E + Q \, \sum_{s,d \in \mathbb{R}, t \in \mathbb{T}, s \neq d} \left( t \, \delta_{sdt}\right) \right], \label{Eq23V}
\end{gather}

\renewcommand{\arraystretch}{0.95}
\begin{table*}[]
\centering
\caption{Parameters related to the MILP model for optimizing the co-flows scheduling and routing of MapReduce traffic}
\resizebox{0.7\textwidth}{!}{%
\begin{tabular}{|c|c|c|}
\hline
\multicolumn{2}{|c|}{\textbf{Parameter}} & \textbf{Values} \\ \hline
\multicolumn{2}{|c|}{$\mathbb{G}, \mathbb{G}_u, \mathbb{R}, \mathbb{S}$} & Check Figures~\ref{FIGURE4}, and~\ref{FIGURE5} \\ \hline
\multirow{2}{*}{$\mathbb{T}, D$} & \begin{tabular}[c]{@{}c@{}}Fattree, Spineleaf, \\ BCube, DCell, PON5 \end{tabular} & Up to 6 slots, 1 second \\ \cline{2-3} 
 &PON3 &  Up to 6 slots, 0.25 seconds \\ \hline
\multicolumn{2}{|c|}{$\sum_{s,d \in \mathbb{R}, s \neq d} \Delta_{sd}$ } & \quad \quad \quad \quad 1-120 Gbits without skew and with skew \quad \quad \quad \quad \\ \hline
\multicolumn{2}{|c|}{$C$} & 10 Gbps \\ \hline
\multirow{2}{*}{$P_{i(max)}$} & Transceiver & 1 Watt \\ \cline{2-3} 
 & NIC & 14 Watts \\ \hline
\multicolumn{2}{|c|}{$\epsilon$} & 14.29 Watt/Gbps \\ \hline
\multicolumn{2}{|c|}{$O_{i(max)}$} & Check Table~\ref{TABLE2} \\ \hline
\multicolumn{2}{|c|}{$\rho$} & 8 Gbps, 2.8 Gbps \\ \hline
\multicolumn{2}{|c|}{$\sigma$} &The maximum switching capacity of the switch \\ \hline 
\multicolumn{2}{|c|}{$L$} & 5000 - 50000 \\ \hline
\multicolumn{2}{|c|}{$Q$} & 100 \\ \hline
\end{tabular}%
}\label{TABLE3}
\end{table*}

\noindent and the second objective is to minimize, $M$, the latest completion time of shuffling which can be expressed as:

\begin{gather}
\min \left[  M +  Q \, \sum_{s,d \in \mathbb{R}, t \in \mathbb{T}, s \neq d} \left( t \, \delta_{sdt}\right) \right]. \label{Eq24V}
\end{gather}

\noindent The second term in both objectives is to schedule the flows in the earliest time slots possible (i.e. encourage the use of first slots) and therefore the optimization is not skewed by large files which take the longest to be transmitted, hence causing the model to possibly schedule small files late (i.e. near completion time of the largest file). This term in effect improves the fairness for small files when large files are also present. Both objectives are to be calculated under the following capacity and architectural constraints in the data centre network represented by $\mathbb{G}$, $\mathbb{G}_u$, $\mathbb{R}$, $\mathbb{S}$ and $C_{uvw}$:

\begin{enumerate}

\item Flow conservation in the data centre: The allocation of the links to the flows follows the flow conservation law at each time slot $t$ and wavelength $w$:
\begin{gather} 
 \displaystyle\sum_{v \in \mathbb{G}_u} \chi_{uvwt}^{sd} -  \displaystyle\sum_{v \in \mathbb{G}_u} \chi_{vuwt}^{sd} = \left\{\begin{matrix}
\delta_{sdt} & u=s\\ 
-\delta_{sdt} & u=d \label{Cons25V}\\ 
 0 & otherwise
\end{matrix}\right. , \nonumber  \\ 
\forall s,d \in \mathbb{R}, s \neq d, u \in \mathbb{G}, w \in \mathbb{W}, t \in \mathbb{T}. 
\end{gather}

\item Constraint to ensure that the total egress traffic from a server does not exceed the maximum rate per server at each time slot $t$:
\begin{gather}
\displaystyle\sum_{v \in \mathbb{G}i, w \in \mathbb{W}} \psi_{ivwt} \leq \rho; \forall i \in \mathbb{R}, t \in \mathbb{T}.  \label{Cons26V}
\end{gather}

\item Constraint to ensure that the total ingress traffic of a switch does not exceed the maximum allowed rate per switch at each time slot $t$:
\begin{gather}
\displaystyle\sum_{u \in \mathbb{G}_i, w \in \mathbb{W}} \psi_{uiwt} \leq \sigma; \forall i \in \mathbb{S}, t \in \mathbb{T}.
 \label{Cons27V}
\end{gather}

\item Constraint to ensure that the total traffic for shuffling data flows in link $(u,v)$ at wavelength $w$ and time slot $t$ does not exceed its capacity:
\begin{gather}
\psi_{uvwt} \leq D \, C_{uvw}; \forall u \in \mathbb{G},  v \in \mathbb{G}_u, w \in \mathbb{W}, t \in \mathbb{T}.
 \label{Cons28V}
\end{gather}

\item Constraint to calculate $\psi_{uvwt}$ by summing the traffic for all shuffling data flows  between all servers that pass through link $(u,v)$ at wavelength $w$ and time slot $t$:
\begin{gather}
\psi_{uvwt} = \displaystyle\sum_{s,d \in \mathbb{R}, s \neq d} \chi_{uvwt}^{sd}, \nonumber \\
\forall u \in \mathbb{G}, v \in \mathbb{G}_u, w \in \mathbb{W}, t \in \mathbb{T}.
 \label{Cons29V}
\end{gather}

\item Constraint to ensure that the sum of shuffling data flow sizes to be sent from server $s$ to server $d$ in all time slots is equal to the total flow size:
\begin{gather}
\displaystyle\sum_{t \in \mathbb{T}} \delta_{sdt} = \Delta_{sd}; \forall s,d \in \mathbb{R}, s \neq d.
 \label{Cons30V}
\end{gather}

\item Constraints to find which transceivers or NICs are used (i.e. $B_{iwt}$ is equals to 1 only if $\beta_{iwt} >$ 0 and is equal to 0 otherwise):
\begin{gather}
\beta_{iwt}= \sum_{v \in \mathbb{G}_u} \psi_{ivwt}+ \sum_{u \in \mathbb{G}_u} \psi_{uiwt}, \quad  {\rm and} \label{Cons31V}  \\
L \, \beta_{iwt} \ge B_{iwt},\quad \, \, \, \quad \quad \quad \quad \quad \quad  \quad {\rm and} \label{Cons32V} \\
\beta_{iwt} \leq L \, B_{iwt}; \forall i \in \mathbb{R}, w \in \mathbb{W}, t \in \mathbb{T}. \label{Cons33V}
\end{gather}

\item Constraints to find which switches are used (i.e. $A_{iwt}$ is equal to 1 only if $\alpha_{iwt} >$ 0 and is equal to 0 otherwise):
\begin{gather}
\alpha_{iwt}= \sum_{v \in \mathbb{G}_u} \psi_{ivwt}+ \sum_{u \in \mathbb{G}_u} \psi_{uiwt}, \quad  {\rm and} \label{Cons34V}  \\
L \, \alpha_{iwt} \ge A_{iwt}, \quad \, \, \quad \quad \quad \quad \quad \quad \quad  {\rm and}  \label{Cons35V}  \\
\alpha_{iwt} \leq L \, A_{iwt}; \forall i \in \mathbb{S}, w \in \mathbb{W}, t \in \mathbb{T}. \label{Cons36V}
\end{gather} 

\item Constraints to find if link $(u,v)$ is active (i.e. $\Gamma_{uvwt}$ = 1 only if $\psi_{uvwt} >$ 0 and is equal to 0 otherwise):
\begin{gather}
L \, \psi_{uvwt} \ge \Gamma_{uvwt},\quad \quad \quad \quad \quad \, \,  \quad   {\rm and}  \label{Cons37V}  \\
\psi_{uvwt} \leq L \, \Gamma_{uvwt}; \forall u \in \mathbb{G}, v \in \mathbb{G}_u, w \in \mathbb{W}, t \in \mathbb{T}. \label{Cons38V}
\end{gather}

\item Constraints to find the transmission time in link $(u,v)$ at wavelength $w$ if it is used up to time slot $t$ and is active at it:
\begin{gather}
\Omega_{uvwt}  = D \,(t-1) + \frac{\psi_{uvwt}}{C_{uvw}}, \, \,  \quad \quad {\rm and} \label{Cons39V} \\
\tau_{uvwt} \leq L\, \Gamma_{uvwt}, \, \, \quad \quad \quad \quad \quad \quad \quad{\rm and}  \label{Cons40V} \\
\tau_{uvwt} \leq \Omega_{uvwt}, \quad \quad \quad \quad \quad \quad \quad \quad {\rm and}  \label{Cons41V} \\
\tau_{uvwt} \ge \Omega_{uvwt} - L \, \left(1-\Gamma_{uvwt} \right),  {\rm and} \quad   \label{Cons42V} \nonumber \\ 
\forall u \in \mathbb{G}, v \in \mathbb{G}_u, w \in \mathbb{W}, t \in \mathbb{T}. 
\end{gather} 

\item Constraints to calculate $M$, which is the completion time determined by the calculated transmission time at the last used link:
\begin{gather}
M \geq \tau_{uvwt}, \, \, \, \quad \quad \quad \quad \quad \quad \quad{\rm and}  \label{Cons43V} \\
M \leq \tau_{uvwt} + L \, \left[1-Z_{uvwt} \right], {\rm and}  \nonumber  \\
\forall u \in \mathbb{G}, v \in \mathbb{G}_u, w \in \mathbb{W}, t \in \mathbb{T}, \label{Cons44V} \\
\displaystyle\sum_{i \in \mathbb{G}, v \in \mathbb{G}_u, w \in \mathbb{W}, t \in \mathbb{T}} Z_{uvwt} = 1.  \label{Cons45V}
\end{gather} 

\noindent For PON3, the following additional set and constraints are required:
\begin{table}[h]
\resizebox{0.6\columnwidth}{!}{%
\begin{tabular}{ll}
$\mathbb{I}$ &  Set of input ports of the two AWGRs\\ 
\end{tabular}
}
\end{table}                  

\item  [12.] Constraint to ensure that servers do not forward the traffic of other servers:
\begin{gather}
\displaystyle\sum_{u \in \mathbb{R}, v \in \mathbb{G}_u, u \neq s} \chi_{uvwt}^{sd} \leq 0, \nonumber \\
\forall s \in \mathbb{R}, d \in \mathbb{R}, w \in \mathbb{W}, t \in \mathbb{T}, s \neq d.  \label{Cons46V}
\end{gather} 

\item [13.] Constraint to ensure that each server transmits only at one wavelength $w$ in a given time slot $t$:
\begin{gather}
\displaystyle\sum_{w \in \mathbb{W}} \Gamma_{uvwt} \leq 1, \nonumber \\
 \forall u \in \mathbb{R}, v \in \mathbb{G}_u \cap \mathbb{I}, t \in \mathbb{T}. \label{Cons47V}
\end{gather} 

\end{enumerate}

\begin{figure*}[!ht]
\centering
\subfigure[]
{
\includegraphics[width=0.42\linewidth]{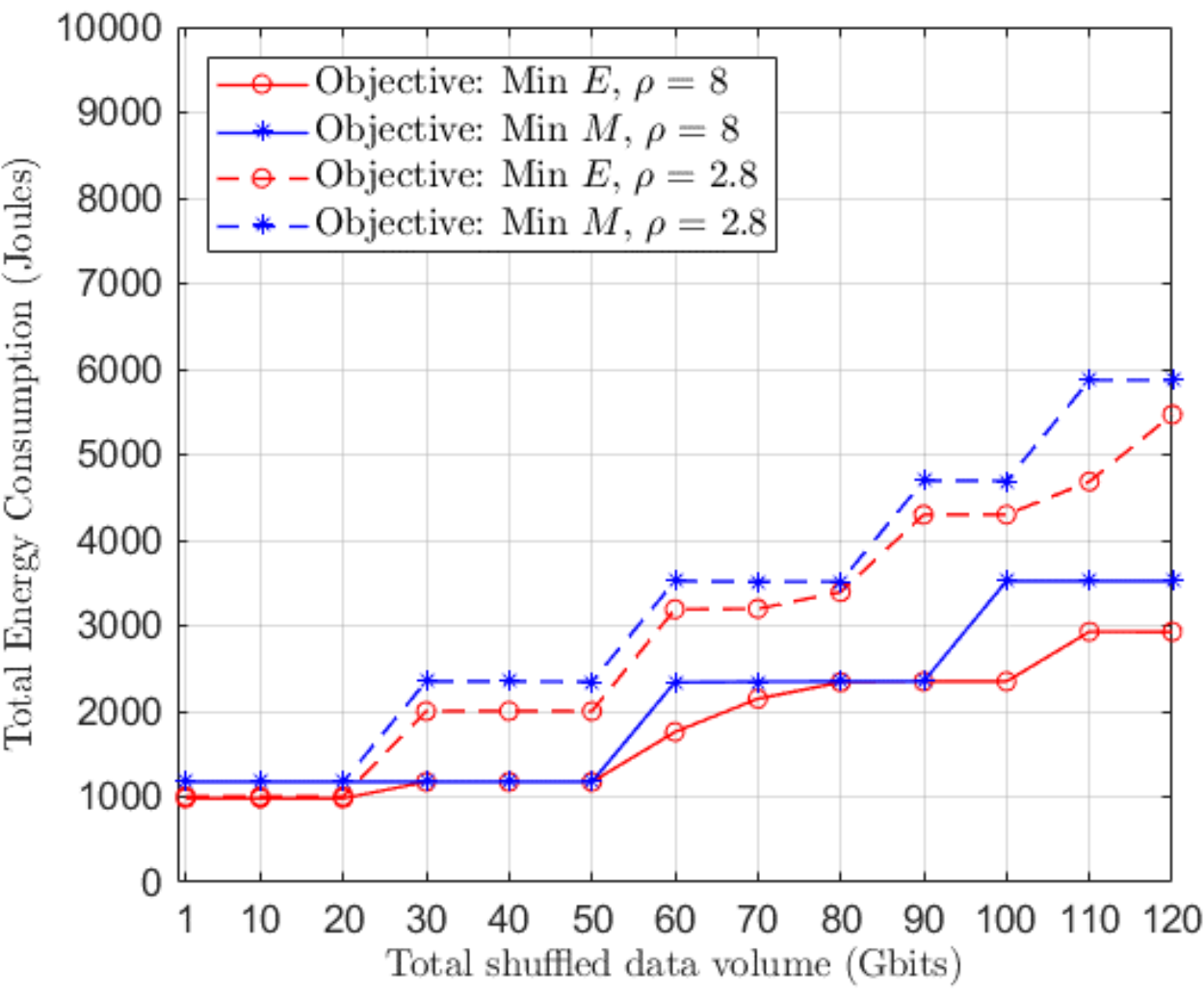}
\label{FIGURE7a}
}
\subfigure[]
{
\includegraphics[width=0.42\linewidth]{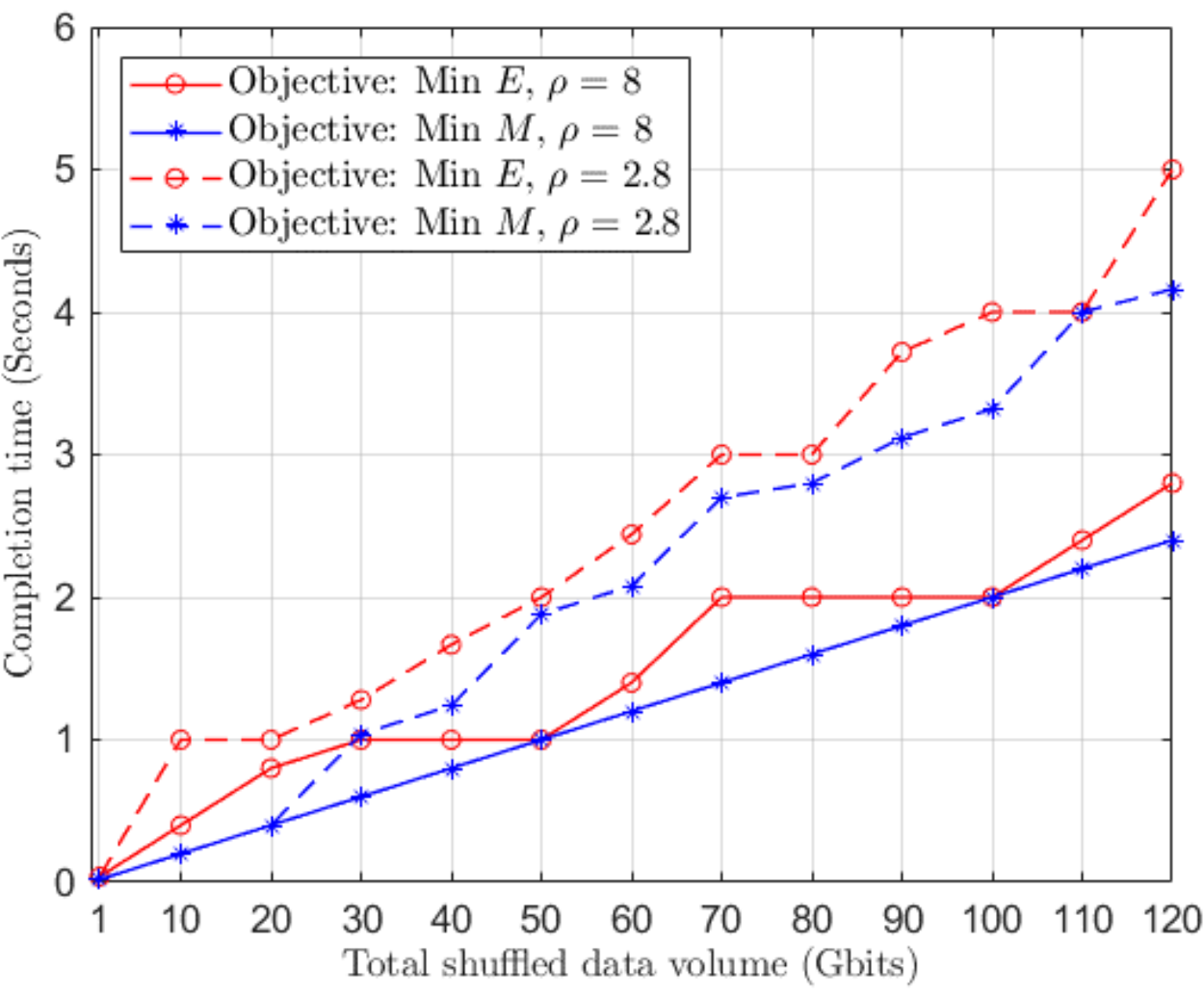}
\label{FIGURE7b}
}
\caption{Results for Spine-leaf DCN with no intermediate data skew: (a) Energy consumption, (b) Completion time.}\label{FIGURE7} 
\end{figure*}

\begin{figure*}[!ht]
\centering
\subfigure[]
{
\includegraphics[width=0.42\linewidth]{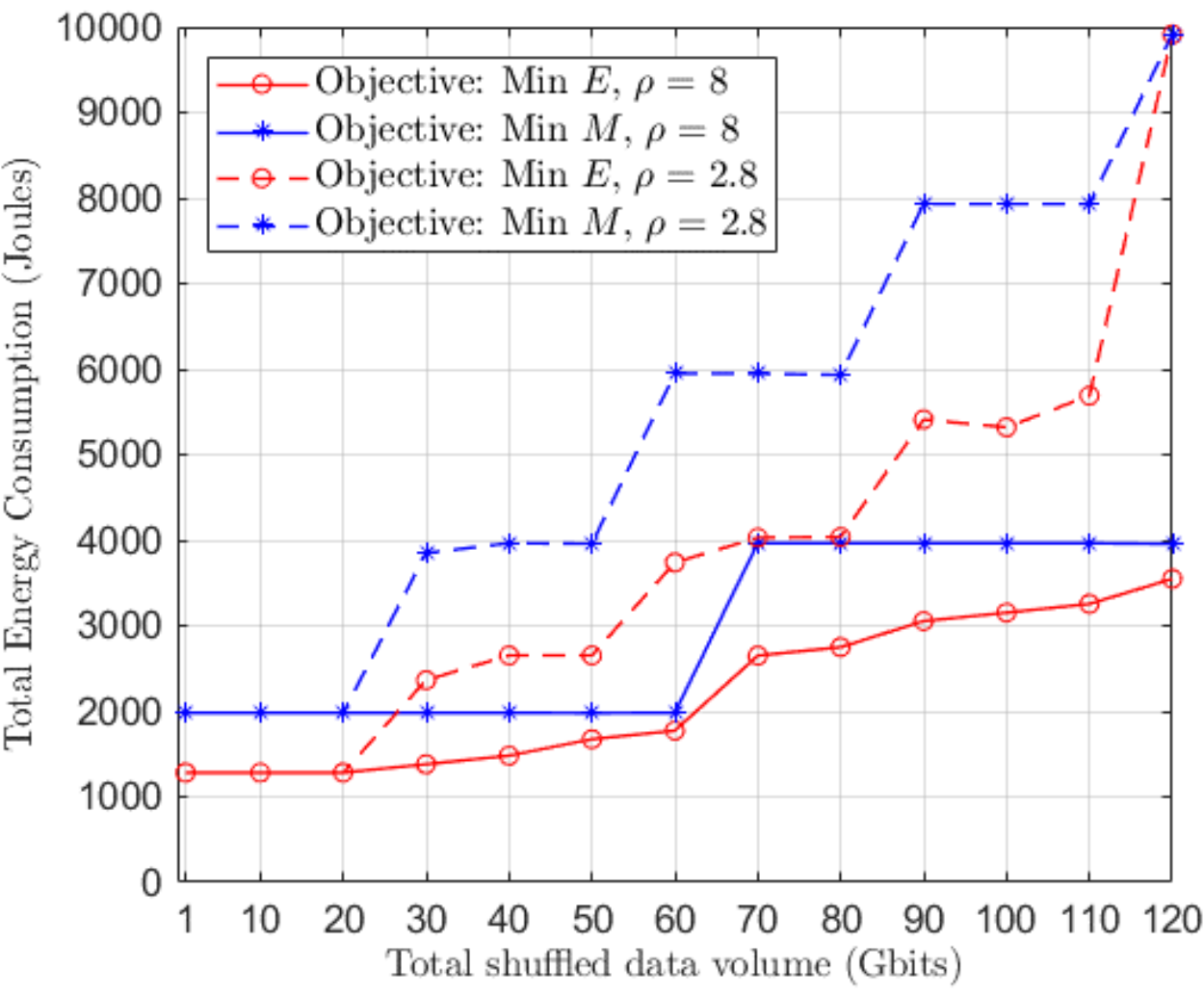}
\label{FIGURE8a}
}
\subfigure[]
{
\includegraphics[width=0.42\linewidth]{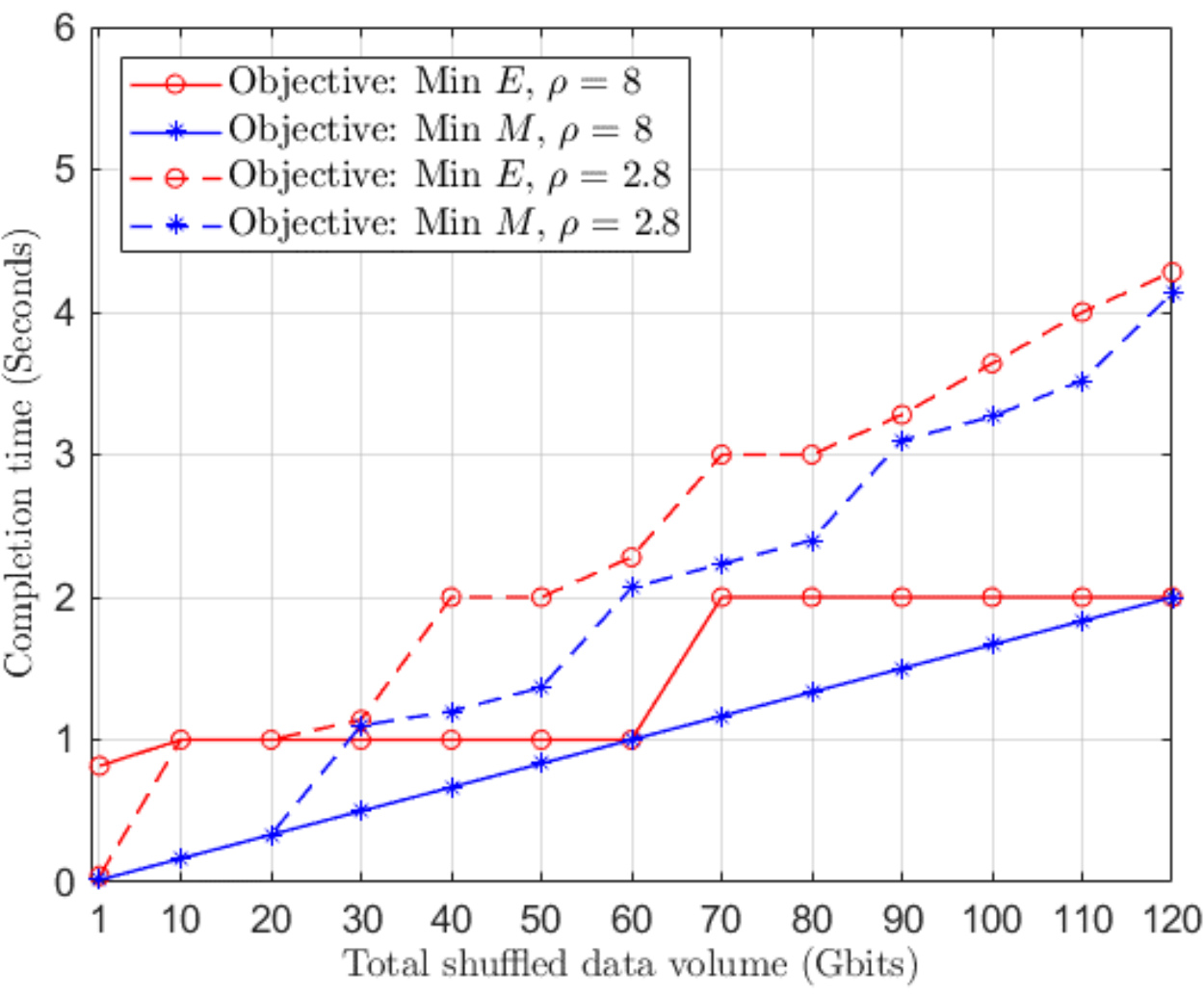}
\label{FIGURE8b}
}
\caption{Results for Fat-tree DCN with no intermediate data skew: (a) Energy consumption, (b) Completion time.}\label{FIGURE8} 
\end{figure*}

\section{Results and Discussion}\label{SECVI}

This Subsection provides the total energy consumption calculated by Equation~\ref{Eq22V} and the completion time estimated by Equations~\ref{Cons43V},~\ref{Cons44V}, and~\ref{Cons45V} when optimizing the routing and scheduling of shuffling traffic under the objective of minimizing the total energy consumption (i.e Equation~\ref{Eq23V}) or the completion time (i.e. Equation~\ref{Eq24V}). In all cases, the results are the outcome of the relevant MILP model and constraints. The results are generated for several scenarios while considering the parameters in Table~\ref{TABLE3} for the MILP model in Section~\ref{SECV}

\subsection{Electronic DCNs}

\subsubsection{Energy Consumption and Completion Time with ON/OFF power profile and no intermediate data skew}

\begin{figure*}[!h]
\centering
\subfigure[]
{
\includegraphics[width=0.42\linewidth]{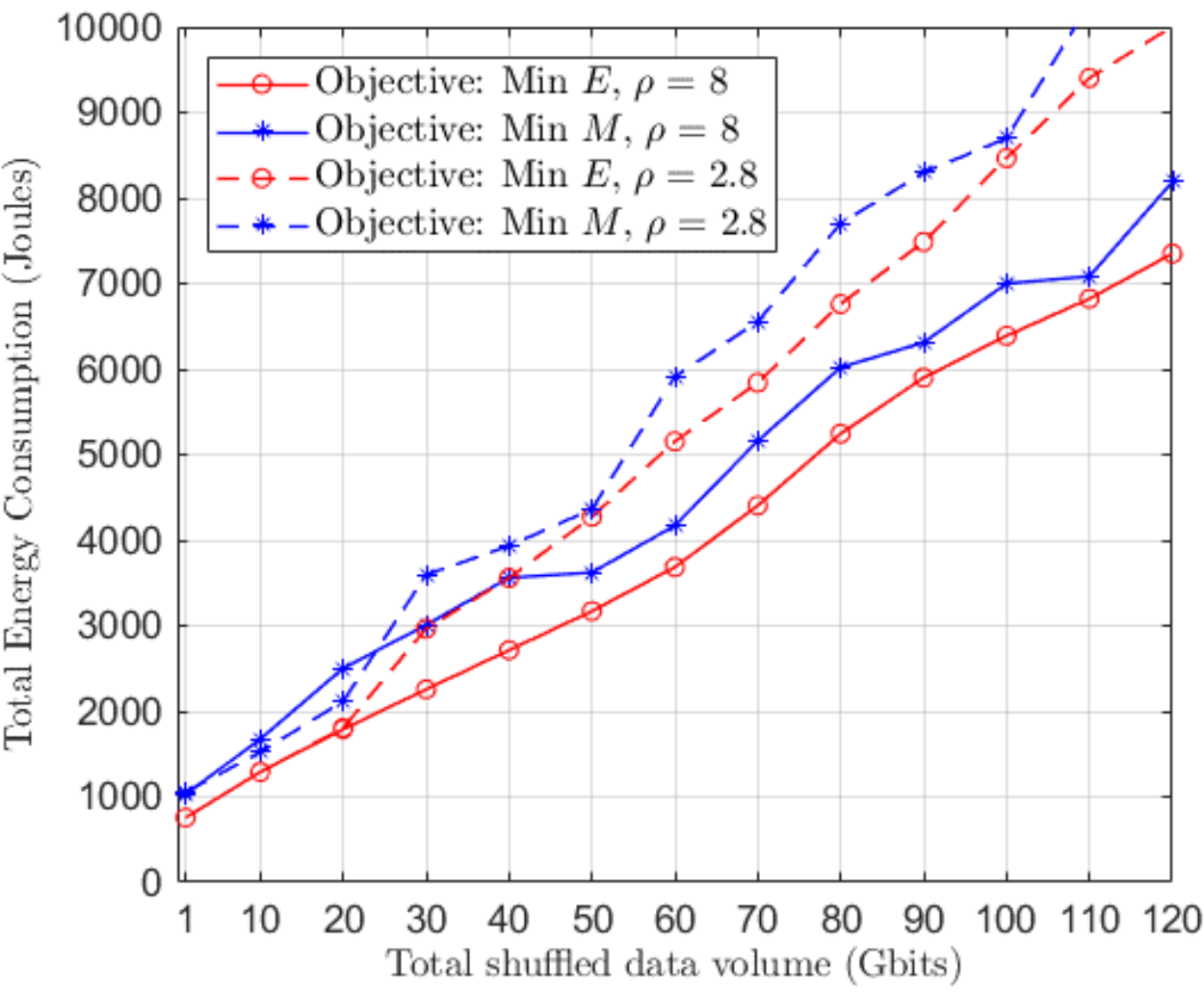}
\label{FIGURE9a}
}
\subfigure[]
{
\includegraphics[width=0.42\linewidth]{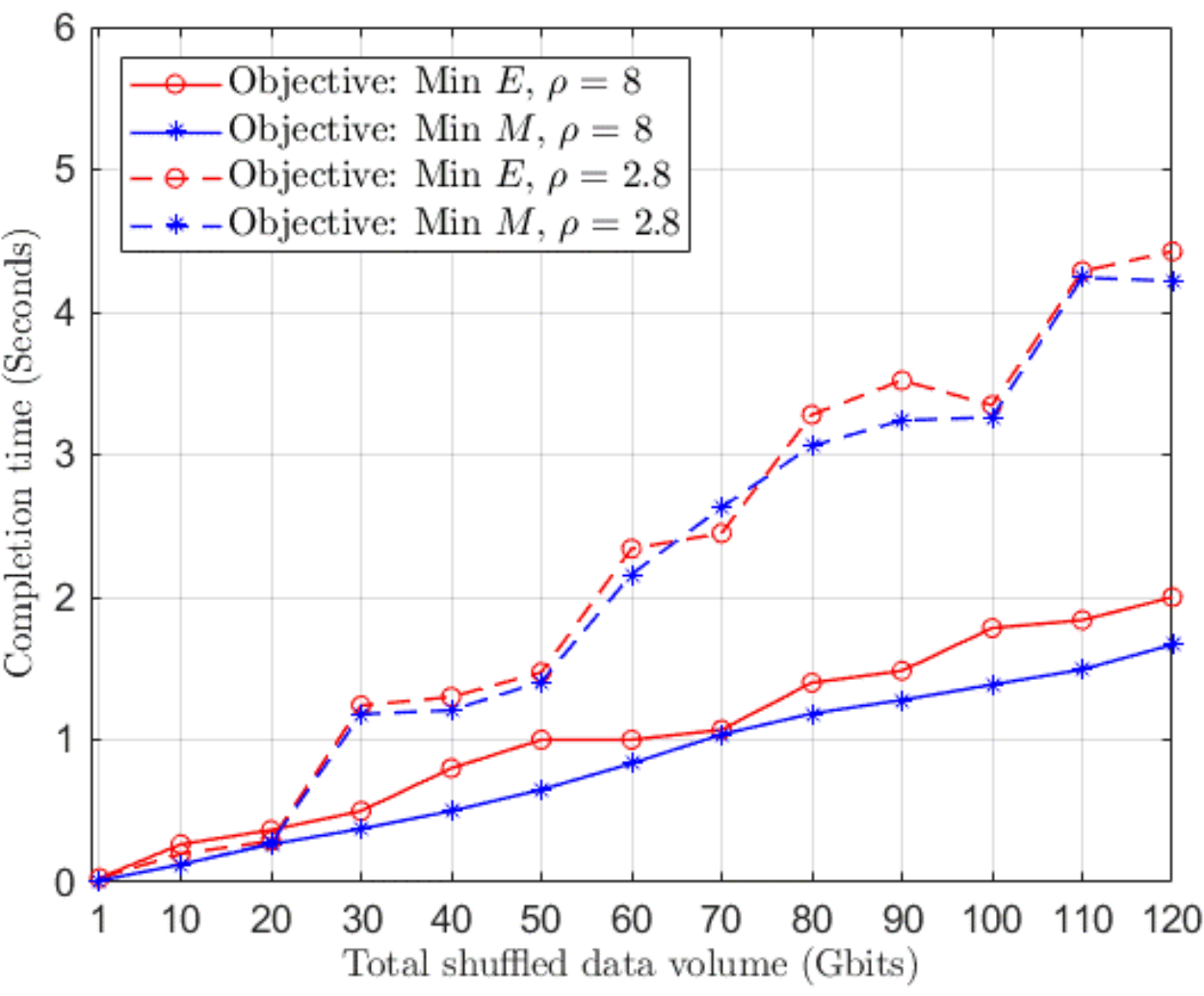}
\label{FIGURE9b}
}
\caption{Results for BCube DCN with no intermediate data skew: (a) Energy consumption, (b) Completion time.}\label{FIGURE9} 
\end{figure*}

\begin{figure*}[!h]
\centering
\subfigure[]
{
\includegraphics[width=0.42\linewidth]{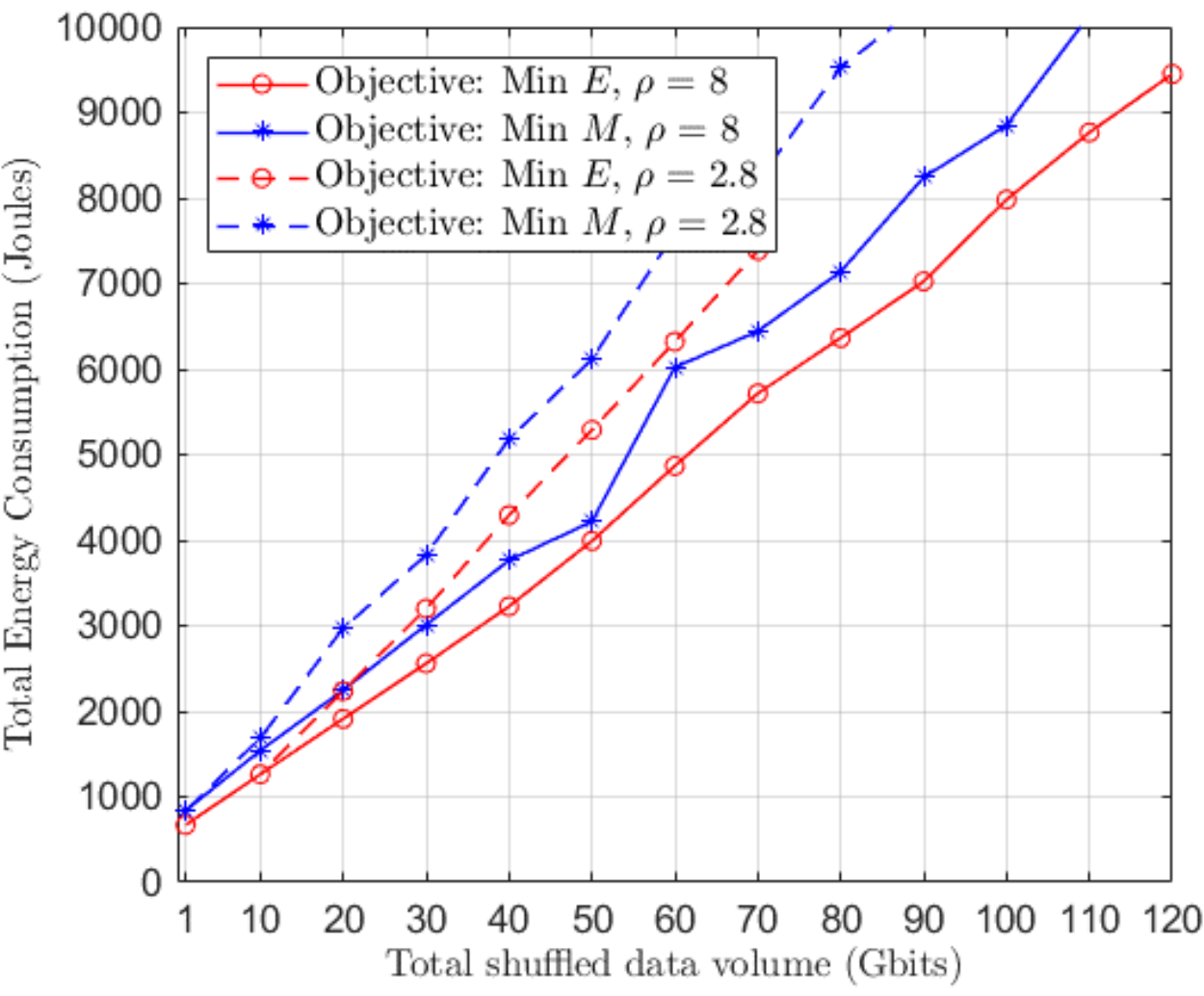}
\label{FIGURE10a}
}
\subfigure[]
{
\includegraphics[width=0.42\linewidth]{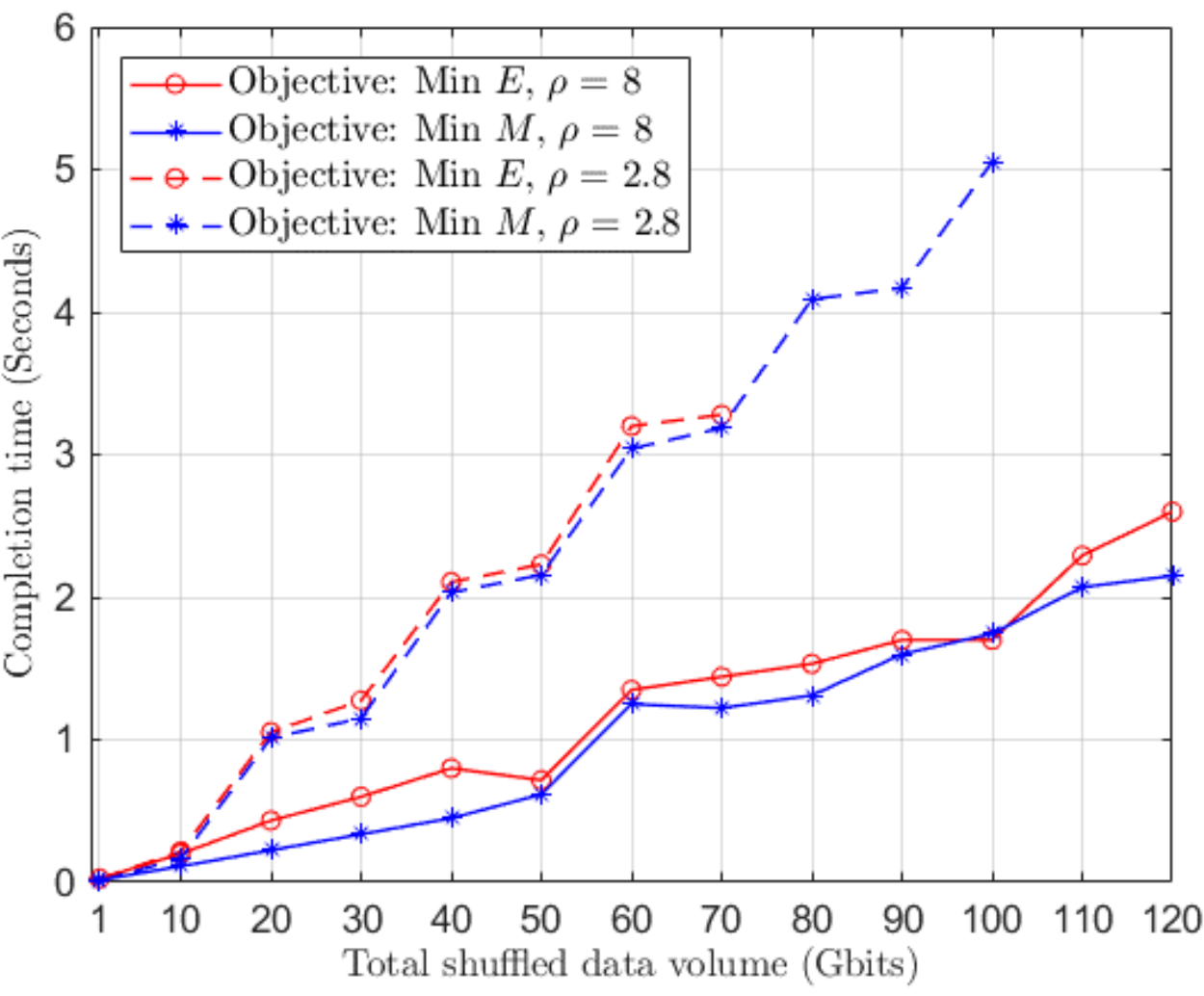}
\label{FIGURE10b}
}
\caption{Results for DCell DCN with no intermediate data skew: (a) Energy consumption, (b) Completion time.}\label{FIGURE10} 
\end{figure*}

Figures~\ref{FIGURE7},~\ref{FIGURE8},~\ref{FIGURE9} and~\ref{FIGURE10} show the results based on the MILP-obtained optimum routing and scheduling for shuffling traffic in Spine-leaf, Fat-tree, BCube, and DCell DCNs, respectively. The shuffling data is assumed to have no skew, ranging from 1 Gbits to 120 Gbits and the ON/OFF power profile was considered for the networking equipment with the specifications detailed in Table~\ref{TABLE2}. Two rates per server values ($\rho$) were considered which are 2.8 Gbps and 8 Gbps. As the power profile is ON/OFF for the switches, transceivers and NICs, they consume the same amount of power if their traffic is high or low. Higher server rates lead to lower energy use for the same amount of data to be sent due to higher utilization for shorter duration. Thus for the total energy consumption results in Figures~\ref{FIGURE7a},~\ref{FIGURE8a},~\ref{FIGURE9a} and~\ref{FIGURE10a}, the dashed curves for $\rho$ =2.8 Gbps show higher energy consumption compared to $\rho$ =8 Gbps results.

For the electronic DCNs, $D$, which is the time slot used for scheduling, was set to be 1 second. As the time is discrete and increases in integer multiples of $D$, a higher server rate does not necessarily provide an advantage if the file size is small. Namely if the file is small, the lower server rate may complete the transmission for example at 0.9 $D$ and a higher rate server may complete the transmission in 0.2 $D$. Both systems will need a full time slot, hence the advantage of a higher data rate in terms of the power efficiency is small at small file sizes. If the time is continuous (i.e. not discrete), then a higher data rate per server will mean that the data is transmitted in shorter time and hence the equipment can be switched off sooner leading to higher energy efficiency for higher data rates per server. The best strategy to minimize the energy consumption with an ON/OFF power profile is to transmit at the maximum available rate through fewer devices while switching off the remaining devices.

For all the data centres, the results show the trade-offs between the optimization results under the two objectives in Equation~\ref{Eq23V} (i.e. minimizing the energy consumption) and Equation~\ref{Eq24V} (i.e. minimizing the completion time). The results for the first and second objectives are represented in red and blue curves, respectively in Figures~\ref{FIGURE7}~-~\ref{FIGURE10}. With the objective of minimizing the energy consumption, the total energy consumption values are lower than with the objective of minimizing the completion time as can be seen in Figures~\ref{FIGURE7a}~-~\ref{FIGURE10a} and the completion time is higher than with the objective of minimizing the completion time as can be seen in Figures~\ref{FIGURE7b}~-~\ref{FIGURE10b}. For the configurations and parameters in Figure~\ref{FIGURE4}, Table~\ref{TABLE2} and Table~\ref{TABLE3} for the data centres, the completion time results under the same objective (i.e. minimizing $E$ or $M$) are comparable. The most energy efficient data centre is Spine-leaf and the least energy efficient is DCell due to excessive use of CPU in servers for processing offloaded traffic.

\begin{figure}[!h]
\centering
\subfigure[]
{
\includegraphics[width=0.64\linewidth]{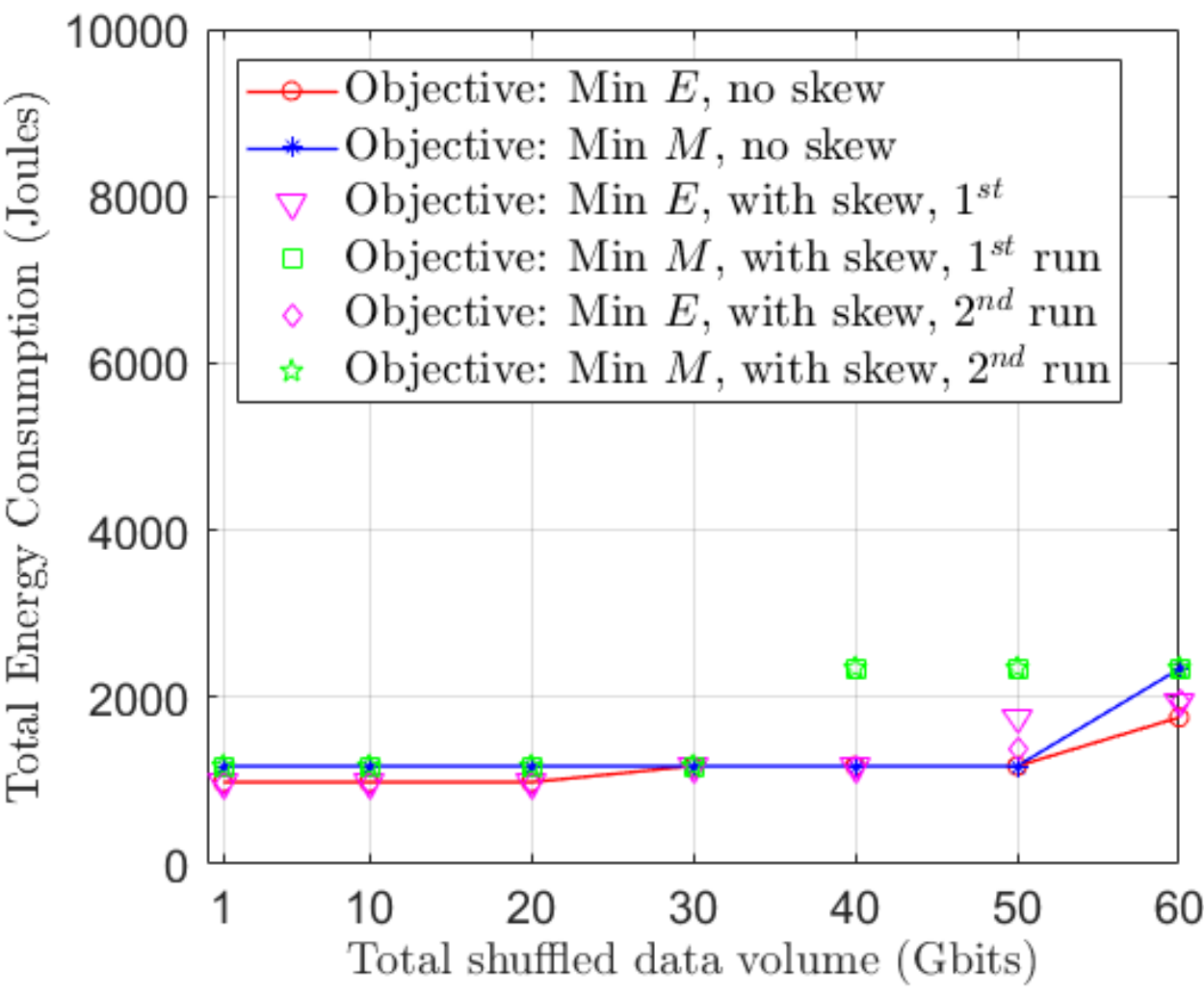}
\label{FIGURE11a}
}
\subfigure[]
{
\includegraphics[width=0.64\linewidth]{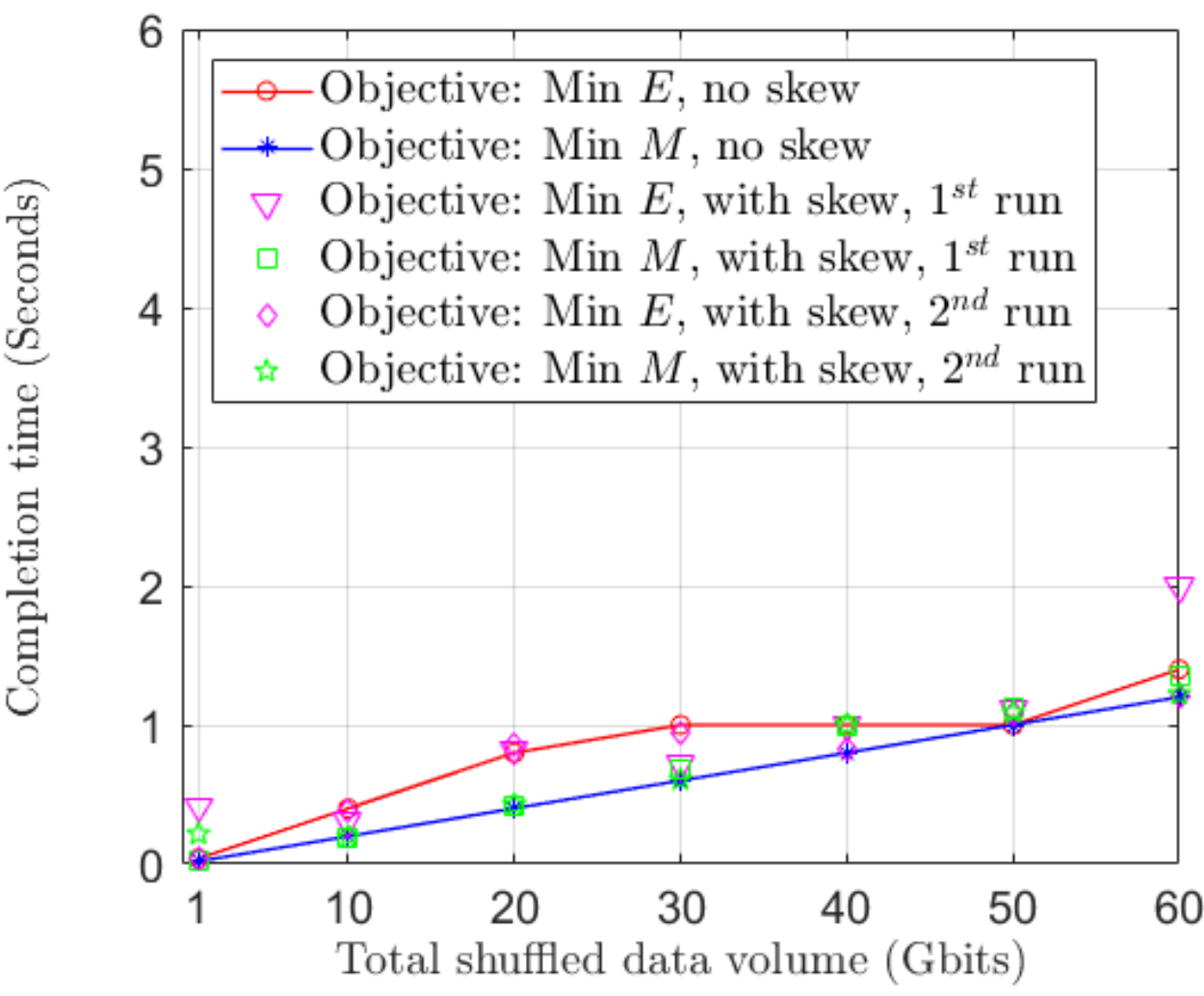}
\label{FIGURE11b}
}
\caption{Results for Spine-leaf DCN with intermediate data skew: (a) Energy consumption, (b) Completion time.}\label{FIGURE11} 
\end{figure}

\begin{figure}[!h]
\centering
\subfigure[]
{
\includegraphics[width=0.64\linewidth]{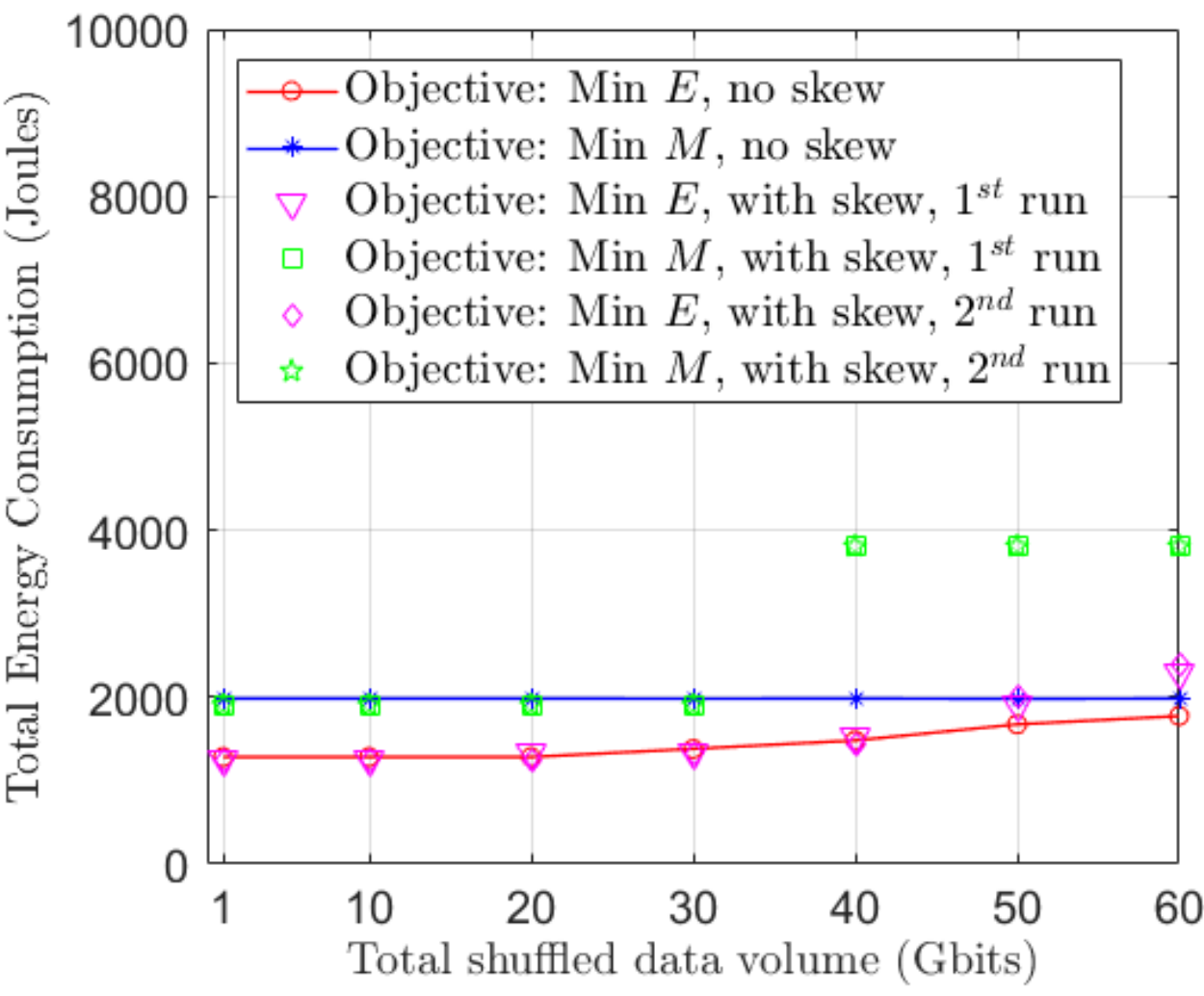}
\label{FIGURE12a}
}
\subfigure[]
{
\includegraphics[width=0.64\linewidth]{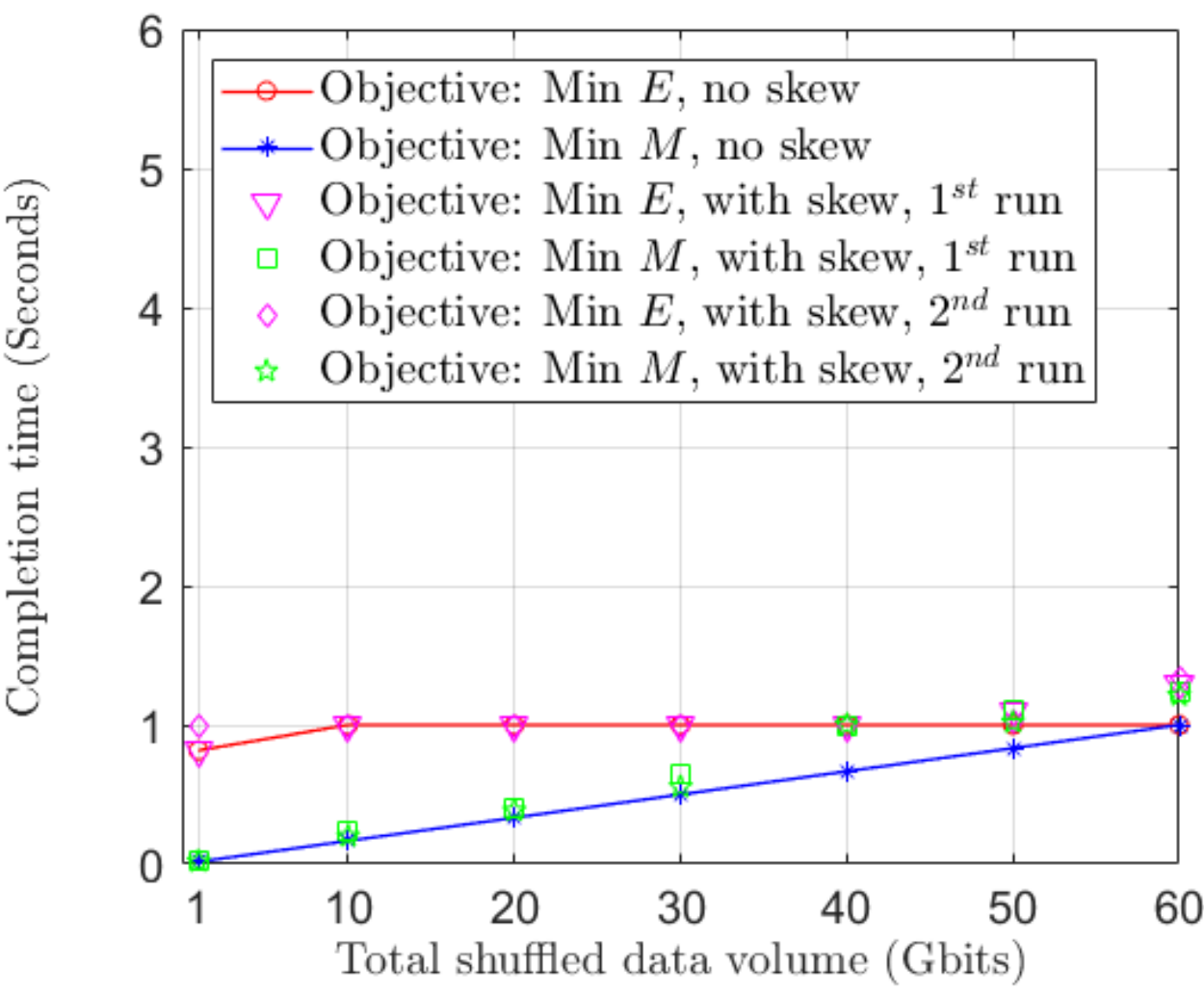}
\label{FIGURE12b}
}
\caption{Results for Fat-tree DCN with intermediate data skew: (a) Energy consumption, (b) Completion time.}\label{FIGURE12} 
\end{figure}

\begin{figure}[!h]
\centering
\subfigure[]
{
\includegraphics[width=0.64\linewidth]{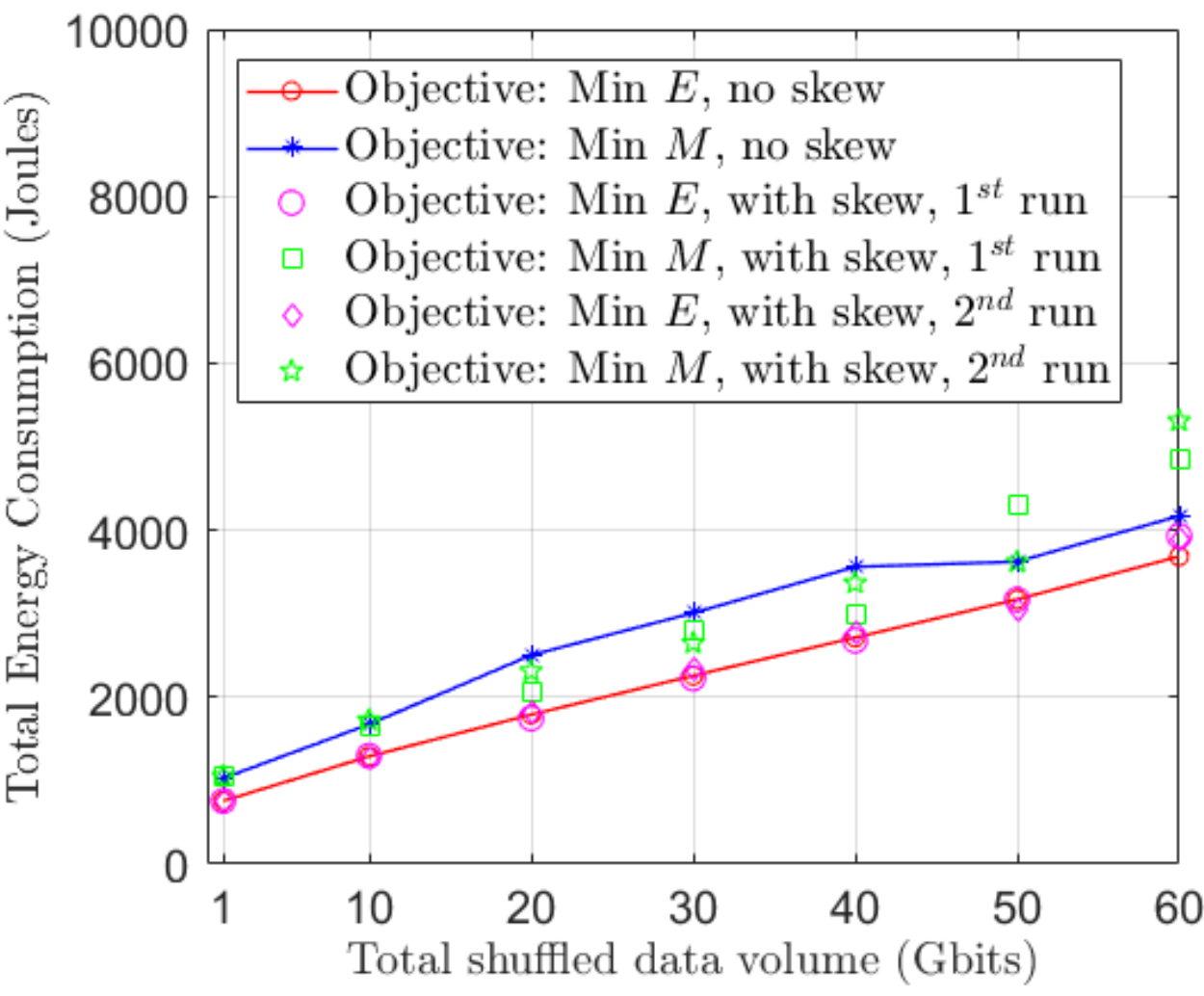}
\label{FIGURE13a}
}
\subfigure[]
{
\includegraphics[width=0.64\linewidth]{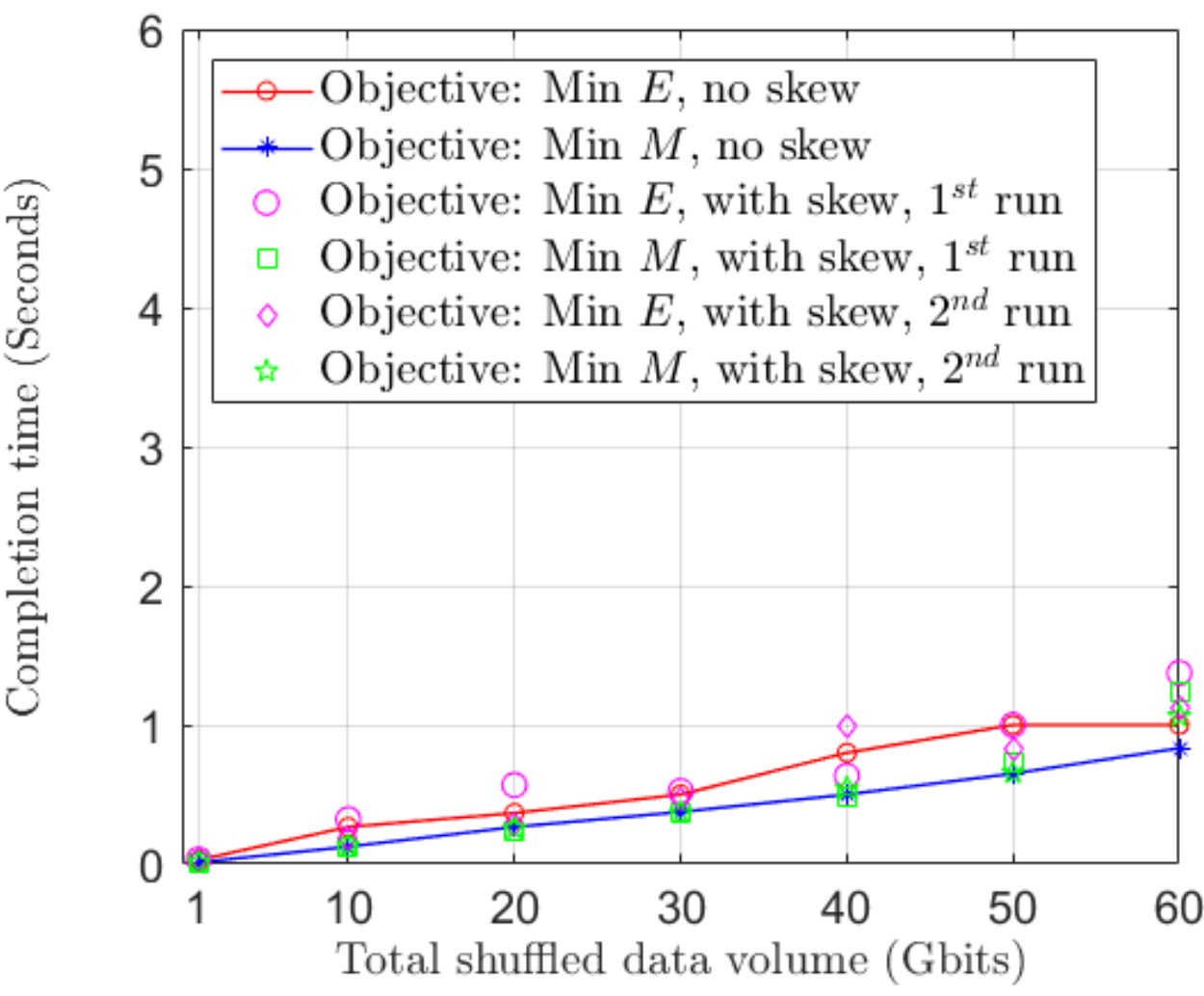}
\label{FIGURE13b}
}
\caption{Results for BCube DCN with intermediate data skew: (a) Energy consumption, (b) Completion time.}\label{FIGURE13} 
\end{figure}

\begin{figure}[!t]
\centering
\subfigure[]
{
\includegraphics[width=0.64\linewidth]{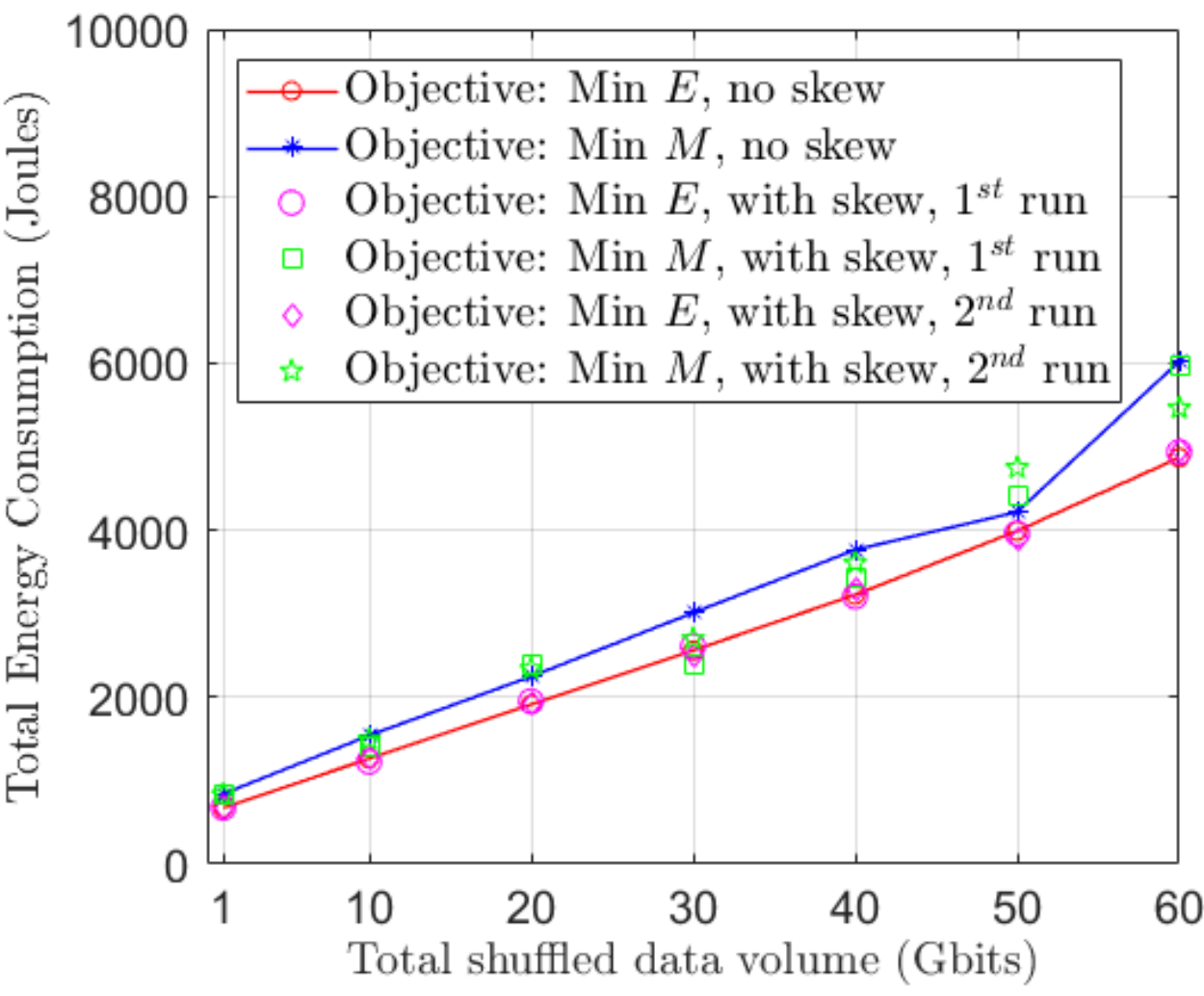}
\label{FIGURE14a}
}
\subfigure[]
{
\includegraphics[width=0.64\linewidth]{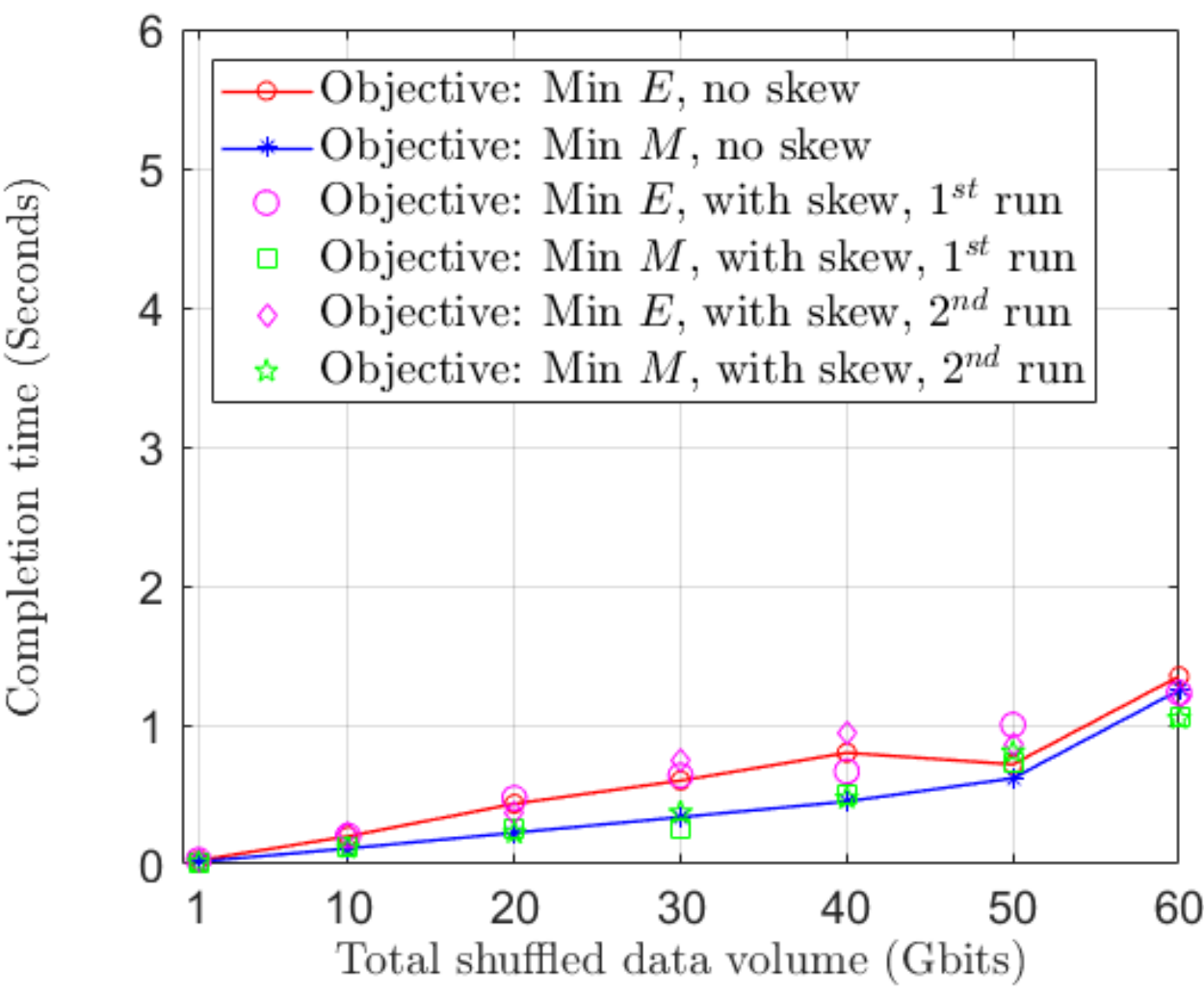}
\label{FIGURE14b}
}
\caption{Results for DCell DCN with intermediate data skew: (a) Energy consumption, (b) Completion time.}\label{FIGURE14} 
\end{figure}

\begin{figure*}[!t]
\centering
\subfigure[]
{
\includegraphics[width=0.42\linewidth]{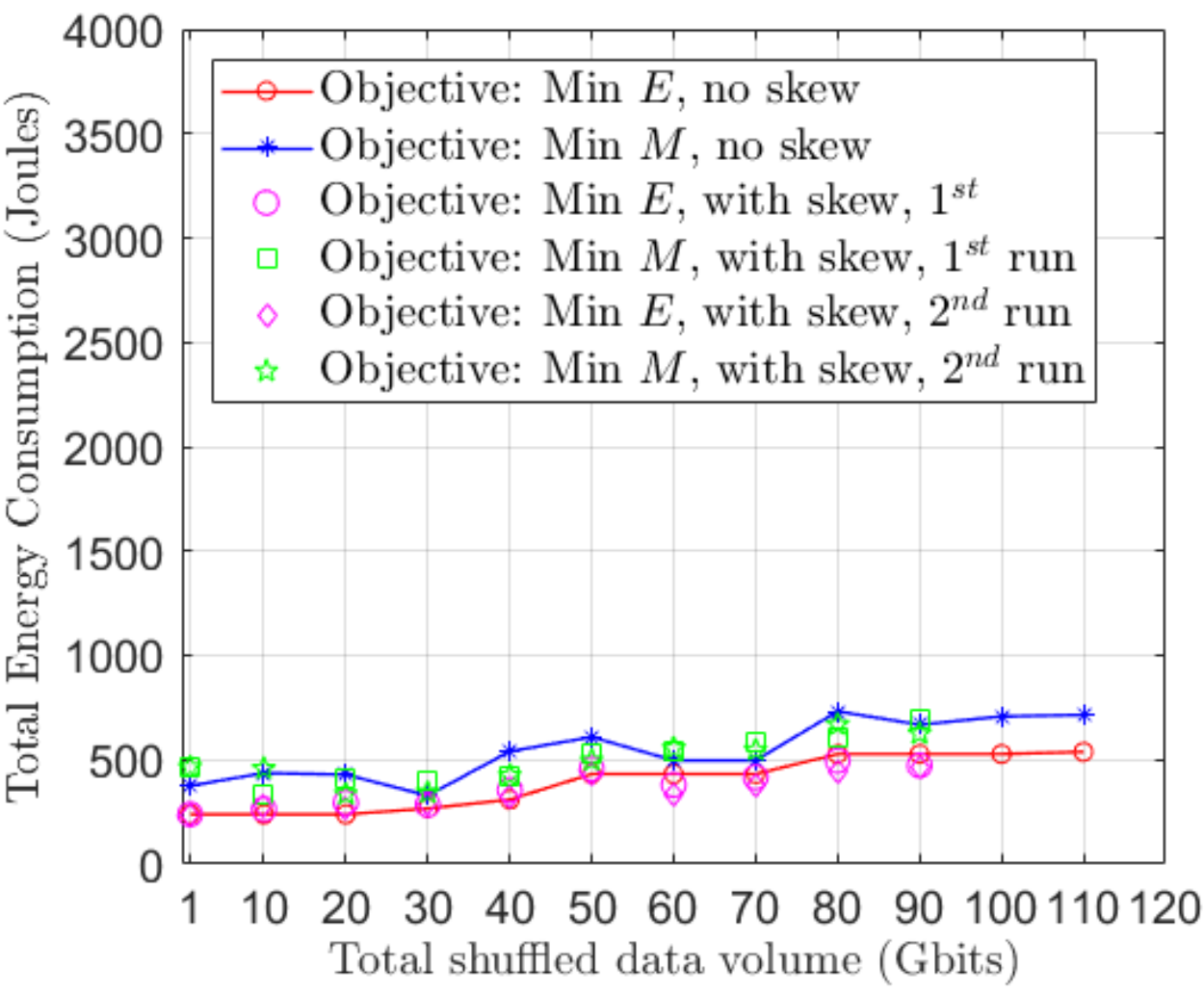}
\label{FIGURE15a}
}
\subfigure[]
{
\includegraphics[width=0.42\linewidth]{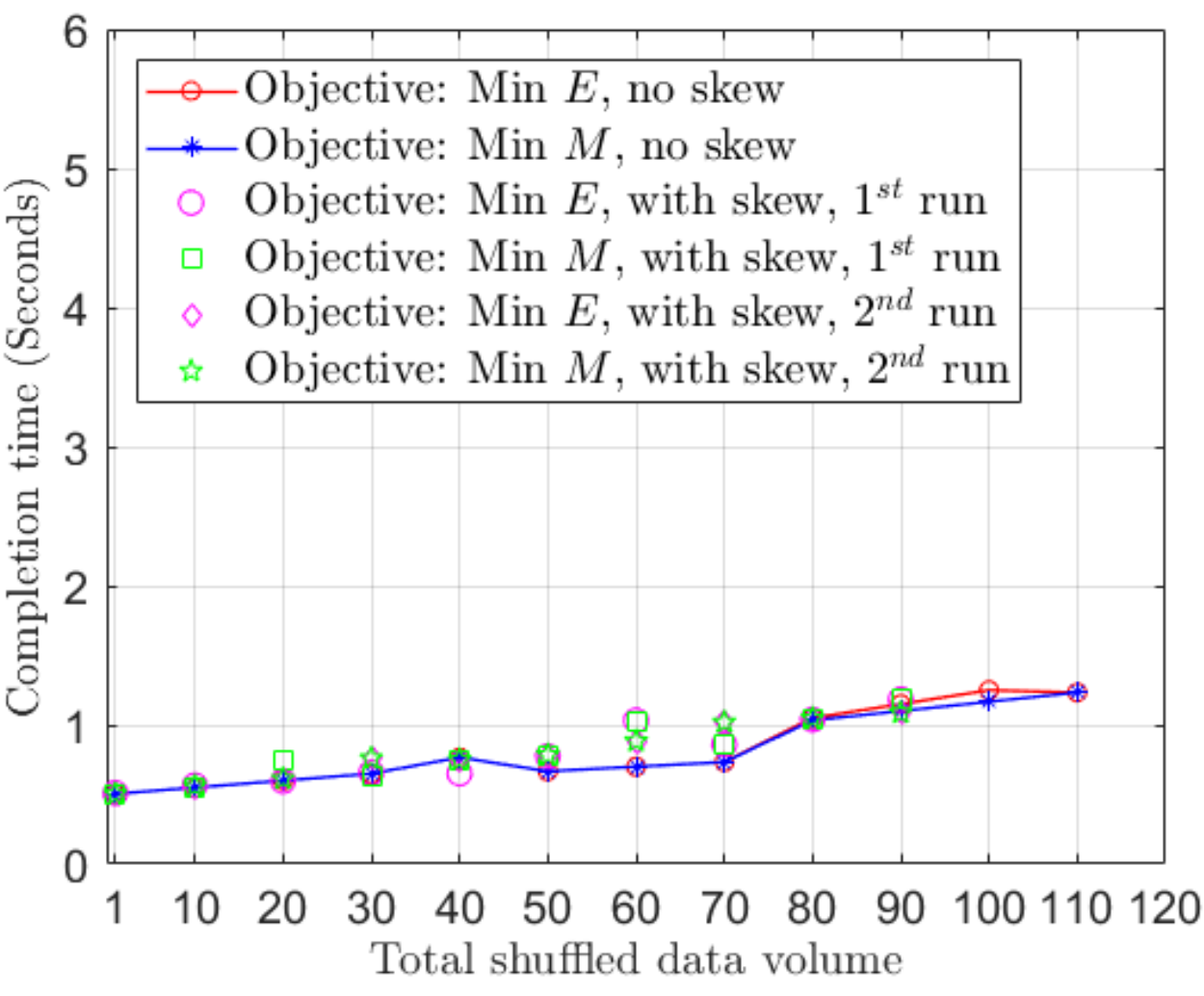}
\label{FIGURE15b}
}
\caption{Results for PON3 DCN without and with intermediate data skew: (a) Energy consumption, (b) Completion time.}\label{FIGURE15} 
\end{figure*}

\begin{figure*}[!t]
\centering
\subfigure[]
{
\includegraphics[width=0.42\linewidth]{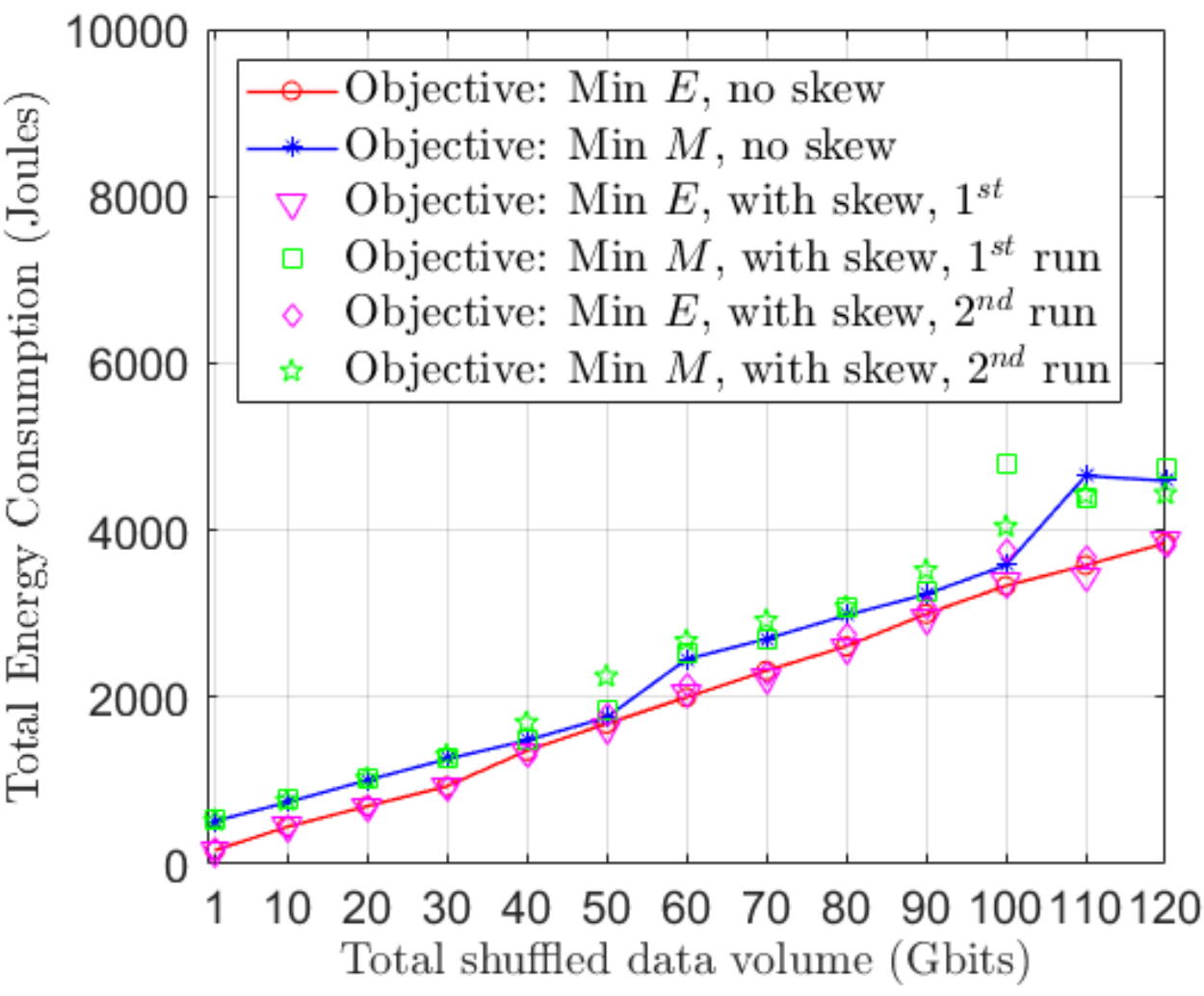}
\label{FIGURE16a}
}
\subfigure[]
{
\includegraphics[width=0.42\linewidth]{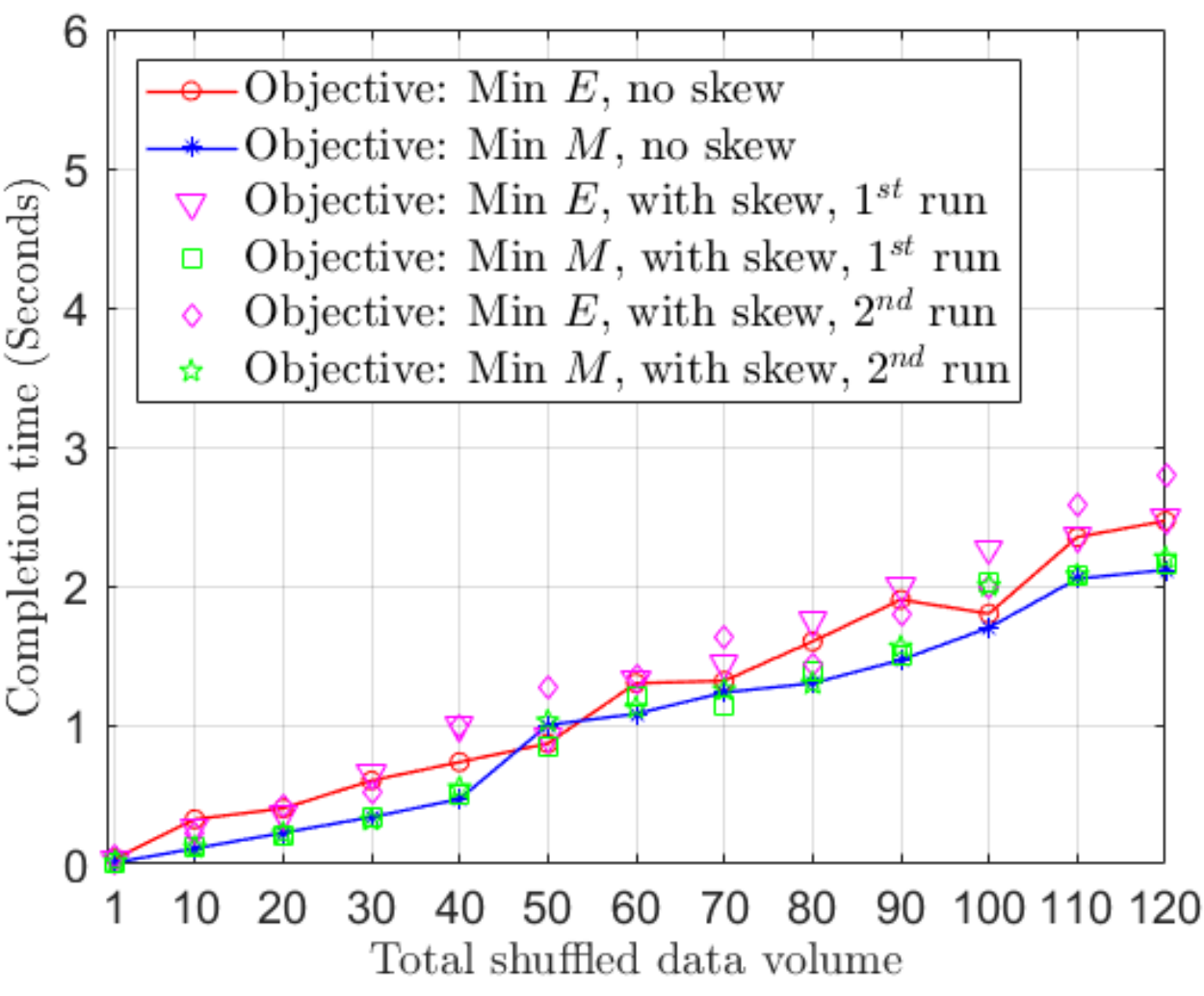}
\label{FIGURE16b}
}
\caption{Results for PON5 DCN without and with intermediate data skew: (a) Energy consumption, (b) Completion time.}\label{FIGURE16} 
\end{figure*}

\subsubsection{Energy Consumption and Completion Time with ON/OFF power profile and intermediate data skew}

The results for the energy consumption and completion time when the data is skewed were generated for optimizing the scheduling and routing for total shuffling data sizes of up to 60 Gbits with a data rate per server ($\rho$) of 8 Gbps. For each total shuffling data size and each data centre, two runs for the flow sizes, randomly generated, were utilized. Figures~\ref{FIGURE11},~\ref{FIGURE12},~\ref{FIGURE13}, and~\ref{FIGURE14} illustrate the results for Spine-leaf, Fat-tree, BCube, and DCell DCNs, respectively. 

The results  show that under the two different objectives (i.e. minimizing the energy consumption or the completion time), the abilities of different data centres to overcome the overheads of data skew are different. For Spine-leaf and Fat-tree data centres, the completion time results when minimizing the power consumption or the completion time for skewed data indicated almost negligible impact on either objective. However, to achieve this for the objective of minimizing the completion time for total data sizes larger than 30 Gbits, an increase by about 100$\%$ in the energy consumption is required. For BCube, a similar trend was observed but to achieve completion time similar to that when no skew is present for total data size larger than 40 Gbits, the increase in the power consumption is about 27$\%$. DCell results depicted in Figure~\ref{FIGURE14} indicated that optimizing the scheduling and routing with any of the two objectives resulted in better balancing for the impact of intermediate data skew compared to the other data centres.

\subsection{PON-based Optical DCNs}

We considered a data rate per server ($\rho$) of 8 Gbps for the evaluation of the completion time and energy consumption in the PON-based DCNs without and with intermediate data skew. For PON3, and as a tuneable transceiver allows each server to transmit at a single wavelength in a given time slot, $D$ was reduced to 0.25 seconds and an adequate number of slots was considered (i.e. up to 6 slots). This allows each map server to communicate with the six reduce servers in different time slots through the AWGR in case they are located in different racks. The results in Figure~\ref{FIGURE15} indicate that the completion time is reduced by about 50$\%$ compared to electronic data centres while reducing the energy consumption by up to about 88$\%$. This reduction is attributed to the many server to server routes at different wavelengths achieved by the PON design that allow better utilization of links. The results in Figure~\ref{FIGURE16} indicate that PON5 achieves similar completion time to that of electronic data centres while having lower energy consumption by about 55$\%$ and 40$\%$ compared to BCube and DCell server-centric data centres, respectively and similar or slightly higher power consumption compared to switch-centric data centres (i.e. Spine-leaf and Fat-tree). Intermediate data skew was also considered and the results show that PON3 and PON5 data centres are reasonably robust to data skew.

\section{Conclusions}\label{SECVII}
This paper introduced and studied in detail two PON-based data centre designs; an AWGR-centric design (PON3) and a server-centric design (PON5) that were proposed in~\cite{REF11,REF28}. Both designs utilize ports with OLT line cards for inter and possibly intra data centre networking in addition to passive interconnects for the intra data centre networking between different PON groups (i.e. racks) within a PON cell (i.e. number of PON groups connected to a single OLT port). The AWGR-centric design example presented in this paper allows up to 20 simultaneous connections between different racks and the OLT with bisection bandwidth of 200 Gbps when using 10 Gbps tuneable transceivers and 4$\times$4 AWGRs. The server-centric data centre provides a design with better resilience and cost effectiveness compared to the AWGR-centric design. To evaluate the performance and energy efficiency of these designs, a time-slotted MILP model was developed and used to optimize the scheduling and routing of the co-flows in the shuffling phase of MapReduce. Two objectives were considered, which are minimizing the total energy consumption or minimizing the completion time of the shuffling phase. For both objectives, an additional aim was to try to utilize earlier time slots which also helps in reducing the completion time. Thus, the completion time reduction objective aims to purely reduce completion time without any consideration for the energy consumption, while the energy minimization objective also targets reducing the completion time but as a lower priority. The smallest completion time was obtained for PON3 due to the use of WDM in the architecture which offers higher capacities per route. DCell was found to have the highest energy consumption when minimizing the completion time due to excessive use of CPU in servers to process offloaded traffic from NICs. At a data rate per server of 8 Gbps, the AWGR-centric DCN achieved completion time  reduction by about 50$\%$ compared to electronic data centres while reducing the energy consumption by up to about 88$\%$. The server-centric PON data centre design achieved similar completion time to that of electronic data centres while having lower energy consumption by about 55$\%$ and 40$\%$ compared to BCube and DCell server-centric data centres, respectively. The impact of data skew and the ability of data centres to reduce their overheads in terms of the energy consumption or the completion time was also examined. The least sensitive DCNs to intermediate data skew are the PON-based optical DCNs and the server-centric electronic data centres.

\section*{Acknowledgment}
Sanaa Hamid Mohamed would like to thank the UK Engineering and Physical Science Research Council (EPSRC) for funding her Ph.D. programme of study. This work was supported in part by the Engineering and Physical Sciences Research Council (EPSRC) through INTelligent Energy awaRe NETworks (INTERNET) project under Grant EP/H040536/1, SwiTching And tRansmission (STAR) project under Grant EP/K016873/1 and Terabit Bidirectional Multi-user Optical Wireless System (TOWS) for 6G LiFi project under Grant EP/S016570/1. All data are provided in full in the results section of this article. 

\bibliographystyle{IEEEtran}
\bibliography{REF_final}

\begin{thebibliography}{10}
\providecommand{\url}[1]{#1}
\csname url@samestyle\endcsname
\providecommand{\newblock}{\relax}
\providecommand{\bibinfo}[2]{#2}
\providecommand{\BIBentrySTDinterwordspacing}{\spaceskip=0pt\relax}
\providecommand{\BIBentryALTinterwordstretchfactor}{4}
\providecommand{\BIBentryALTinterwordspacing}{\spaceskip=\fontdimen2\font plus
\BIBentryALTinterwordstretchfactor\fontdimen3\font minus
  \fontdimen4\font\relax}
\providecommand{\BIBforeignlanguage}[2]{{%
\expandafter\ifx\csname l@#1\endcsname\relax
\typeout{** WARNING: IEEEtran.bst: No hyphenation pattern has been}%
\typeout{** loaded for the language `#1'. Using the pattern for}%
\typeout{** the default language instead.}%
\else
\language=\csname l@#1\endcsname
\fi
#2}}
\providecommand{\BIBdecl}{\relax}
\BIBdecl

\bibitem{REF1}
{S. H. Mohamed and T. E. H. El-Gorashi and J. M. H. Elmirghani}, ``{Energy
  Efficiency of Server-Centric PON Data Center Architecture for Fog
  Computing},'' in \emph{2018 20th International Conference on Transparent
  Optical Networks (ICTON)}, July 2018, pp. 1--4.

\bibitem{REF2}
Y.~{Luo}, F.~{Effenberger}, and M.~{Sui}, ``{Cloud computing provisioning over
  Passive Optical Networks},'' in \emph{2012 1st IEEE International Conference
  on Communications in China (ICCC)}, Aug 2012, pp. 255--259.

\bibitem{REF3}
M.~Taheri and N.~Ansari, ``{A feasible solution to provide cloud computing over
  optical networks},'' \emph{IEEE Network}, vol.~27, no.~6, pp. 31--35,
  November 2013.

\bibitem{REF4}
A.~H. {Helmy} and A.~{Nayak}, ``{Integrating Fog With Long-Reach PONs From a
  Dynamic Bandwidth Allocation Perspective},'' \emph{Journal of Lightwave
  Technology}, vol.~36, no.~22, pp. 5276--5284, Nov 2018.

\bibitem{REF5}
S.~H.~S. Newaz, W.~S. binti Haji~Suhaili, G.~M. Lee, M.~R. Uddin, A.~F.~Y.
  Mohammed, and J.~K. Choi, ``{Towards realizing the importance of placing fog
  computing facilities at the central office of a PON},'' in \emph{2017 19th
  International Conference on Advanced Communication Technology (ICACT)}, Feb
  2017, pp. 152--157.

\bibitem{REF6}
H.~Liu, F.~Lu, A.~Forencich, R.~Kapoor, M.~Tewari, G.~M. Voelker, G.~Papen,
  A.~C. Snoeren, and G.~Porter, ``{Circuit Switching Under the Radar with
  REACToR},'' in \emph{11th {USENIX} Symposium on Networked Systems Design and
  Implementation ({NSDI} 14)}.\hskip 1em plus 0.5em minus 0.4em\relax Seattle,
  WA: {USENIX} Association, 2014, pp. 1--15.

\bibitem{REF7}
C.~Kachris and I.~Tomkos, ``{Power consumption evaluation of hybrid WDM PON
  networks for data centers},'' in \emph{2011 16th European Conference on
  Networks and Optical Communications}, July 2011, pp. 118--121.

\bibitem{REF8}
P.~Ji, D.~Qian, K.~Kanonakis, C.~Kachris, and I.~Tomkos, ``{Design and
  Evaluation of a Flexible-Bandwidth OFDM-Based Intra-Data Center
  Interconnect},'' \emph{Selected Topics in Quantum Electronics, IEEE Journal
  of}, vol.~19, no.~2, pp. 3\,700\,310--3\,700\,310, March 2013.

\bibitem{REF9}
K.~Wang, L.~Zhao, H.~Gu, X.~Yu, G.~Wu, and J.~Cai, ``{ADON: a scalable
  AWG-based topology for datacenter optical network},'' \emph{Optical and
  Quantum Electronics}, vol.~47, no.~8, pp. 2541--2554, Aug 2015.

\bibitem{REF10}
P.~N. Ji, D.~Qian, K.~Kanonakis, C.~Kachris, and I.~Tomkos, ``{Design and
  Evaluation of a Flexible-Bandwidth OFDM-Based Intra-Data Center
  Interconnect},'' \emph{IEEE Journal of Selected Topics in Quantum
  Electronics}, vol.~19, no.~2, pp. 3\,700\,310--3\,700\,310, March 2013.

\bibitem{REF11}
\BIBentryALTinterwordspacing
J.~Elmirghani, T.~EL-GORASHI, and A.~HAMMADI, ``{Passive optical-based data
  center networks},'' 2016, wO Patent App. PCT/GB2015/053,604. [Online].
  Available: \url{http://google.com/patents/WO2016083812A1?cl=und}
\BIBentrySTDinterwordspacing

\bibitem{REF12}
A.~Hammadi, T.~El-Gorashi, and J.~Elmirghani, ``{High performance AWGR PONs in
  data centre networks},'' in \emph{Transparent Optical Networks (ICTON), 2015
  17th International Conference on}, July 2015, pp. 1--5.

\bibitem{REF13}
A.~Hammadi, T.~E.~H. El-Gorashi, and J.~M.~H. Elmirghani, ``{Energy-efficient
  software-defined AWGR-based PON data center network},'' in \emph{2016 18th
  International Conference on Transparent Optical Networks (ICTON)}, July 2016,
  pp. 1--5.

\bibitem{REF14}
A.~Hammadi, T.~E.~H. El-Gorashi, M.~O.~I. Musa, and J.~M.~H. Elmirghani,
  ``{Server-centric PON data center architecture},'' in \emph{2016 18th
  International Conference on Transparent Optical Networks (ICTON)}, July 2016,
  pp. 1--4.

\bibitem{REF15}
A.~E.~A. Eltraify, M.~O.~I. Musa, A.~Al-Quzweeni, and J.~M.~H. Elmirghani,
  ``{Experimental Evaluation of Passive Optical Network Based Data Centre
  Architecture},'' in \emph{2018 20th International Conference on Transparent
  Optical Networks (ICTON)}, July 2018, pp. 1--4.

\bibitem{REF16}
A.~E.~A. {Eltraify}, M.~O.~I. {Musa}, A.~{Al-Quzweeni}, and J.~M.~H.
  {Elmirghani}, ``Experimental evaluation of server centric passive optical
  network based data centre architecture,'' in \emph{2019 21st International
  Conference on Transparent Optical Networks (ICTON)}, 2019, pp. 1--5.

\bibitem{REF17}
{S. H. Mohamed, T. E. H. El-Gorashi, and J. M. H. Elmirghani}, ``{Impact of
  Link Failures on the Performance of MapReduce in Data Center Networks},'' in
  \emph{2018 20th International Conference on Transparent Optical Networks
  (ICTON)}, July 2018, pp. 1--4.

\bibitem{REF18}
J.~Han, M.~Ishii, and H.~Makino, ``{A Hadoop performance model for multi-rack
  clusters},'' in \emph{Computer Science and Information Technology (CSIT),
  2013 5th International Conference on}, March 2013, pp. 265--274.

\bibitem{REF19}
Z.~Kouba, O.~Tomanek, and L.~Kencl, ``{Evaluation of Datacenter Network
  Topology Influence on Hadoop MapReduce Performance},'' in \emph{2016 5th IEEE
  International Conference on Cloud Networking (Cloudnet)}, Oct 2016, pp.
  95--100.

\bibitem{REF20}
S.~H. Mohamed, T.~E.~H. El-Gorashi, and J.~M.~H. Elmirghani, ``{On the energy
  efficiency of MapReduce shuffling operations in data centers},'' in
  \emph{2017 19th International Conference on Transparent Optical Networks
  (ICTON)}, July 2017, pp. 1--5.

\bibitem{REF21}
L.~Wang, F.~Zhang, and Z.~Liu, ``{Improving the Network Energy Efficiency in
  MapReduce Systems},'' in \emph{Computer Communications and Networks (ICCCN),
  2013 22nd International Conference on}, July 2013, pp. 1--7.

\bibitem{REF22}
M.~Chowdhury, Y.~Zhong, and I.~Stoica, ``{Efficient Coflow Scheduling with
  Varys},'' \emph{SIGCOMM Comput. Commun. Rev.}, vol.~44, no.~4, pp. 443--454,
  Aug. 2014.

\bibitem{REF23}
S.~Luo, H.~Yu, Y.~Zhao, S.~Wang, S.~Yu, and L.~Li, ``{Towards Practical and
  Near-optimal Coflow Scheduling for Data Center Networks},'' \emph{IEEE
  Transactions on Parallel and Distributed Systems}, vol.~PP, no.~99, pp. 1--1,
  2016.

\bibitem{REF24}
Y.~Zhao, K.~Chen, W.~Bai, M.~Yu, C.~Tian, Y.~Geng, Y.~Zhang, D.~Li, and
  S.~Wang, ``{Rapier: Integrating routing and scheduling for coflow-aware data
  center networks},'' in \emph{2015 IEEE Conference on Computer Communications
  (INFOCOM)}, April 2015, pp. 424--432.

\bibitem{REF25}
R.~F. e~Silva and P.~M. Carpenter, ``{Energy Efficient Ethernet on MapReduce
  Clusters: Packet Coalescing To Improve 10GbE Links},'' \emph{IEEE/ACM
  Transactions on Networking}, vol.~25, no.~5, pp. 2731--2742, Oct 2017.

\bibitem{REF26}
K.~Christensen, P.~Reviriego, B.~Nordman, M.~Bennett, M.~Mostowfi, and
  J.~Maestro, ``{IEEE 802.3az: the road to energy efficient ethernet},''
  \emph{Communications Magazine, IEEE}, vol.~48, no.~11, pp. 50--56, November
  2010.

\bibitem{REF27}
M.~I. {Olmedo}, L.~{Suhr}, K.~{Prince}, R.~{Rodes}, C.~{Mikkelsen}, E.~{Hviid},
  C.~{Neumeyr}, G.~{Vollrath}, E.~{Goobar}, P.~{ \"{O}hl\'{e}n}, and I.~T.
  {Monroy}, ``{Gigabit Access Passive Optical Network Using Wavelength Division
  Multiplexing - GigaWaM},'' \emph{Journal of Lightwave Technology}, vol.~32,
  no.~22, pp. 4285--4293, Nov 2014.

\bibitem{REF28}
A.~A. Hammadi, ``{Future PON Data Centre Networks},'' Ph.D. dissertation,
  University of Leeds, School of Electronic and Electrical Engineering, Aug.
  2016.

\bibitem{REF29}
J.~Beals, N.~Bamiedakis, A.~Wonfor, R.~V. Penty, I.~H. White, J.~V. DeGroot,
  K.~Hueston, T.~V. Clapp, and M.~Glick, ``A terabit capacity passive polymer
  optical backplane based on a novel meshed waveguide architecture,''
  \emph{Applied Physics A}, vol.~95, no.~4, pp. 983--988, Jun 2009.

\bibitem{REF30}
A.~Hammadi, M.~Musa, T.~E.~H. El-Gorashi, and J.~H. Elmirghani, ``{Resource
  provisioning for cloud PON AWGR-based data center architecture},'' in
  \emph{2016 21st European Conference on Networks and Optical Communications
  (NOC)}, June 2016, pp. 178--182.

\bibitem{REF31}
R.~Ramaswami, K.~N. Sivarajan, and G.~H. Sasaki, \emph{{Optical Networks: A
  Practical Prespective}}, 3rd~ed.\hskip 1em plus 0.5em minus 0.4em\relax
  Morgan Kaufmann, 2010.

\bibitem{REF32}
T.~White, \emph{{Hadoop: The Definitive Guide}}, 1st~ed.\hskip 1em plus 0.5em
  minus 0.4em\relax O'Reilly Media, Inc., 2009.

\bibitem{REF33}
Y.~Shang, D.~Li, J.~Zhu, and M.~Xu, ``{On the Network Power Effectiveness of
  Data Center Architectures},'' \emph{Computers, IEEE Transactions on},
  vol.~64, no.~11, pp. 3237--3248, Nov 2015.

\bibitem{REF34}
M.~Al-Fares, A.~Loukissas, and A.~Vahdat, ``{A Scalable, Commodity Data Center
  Network Architecture},'' \emph{SIGCOMM Comput. Commun. Rev.}, vol.~38, no.~4,
  pp. 63--74, Aug. 2008.

\bibitem{REF35}
{Pall Beck, Peter Clemens, Santiago Freitas, Jeff Gatz, Michele Girola, Jason
  Gmitter, Holger Mueller, Ray O'Hanlon, Veerendra Para, Joe Robinson, Andy
  Sholomon, Jason Walker, and Jon Tate}, \emph{{IBM and Cisco: Together for a
  World Class Data Center}}.\hskip 1em plus 0.5em minus 0.4em\relax IBM
  Redbooks, 2013.

\bibitem{REF36}
C.~Guo, G.~Lu, D.~Li, H.~Wu, X.~Zhang, Y.~Shi, C.~Tian, Y.~Zhang, and S.~Lu,
  ``{BCube: A High Performance, Server-centric Network Architecture for Modular
  Data Centers},'' \emph{SIGCOMM Comput. Commun. Rev.}, vol.~39, no.~4, pp.
  63--74, Aug. 2009.

\bibitem{REF37}
C.~Guo, H.~Wu, K.~Tan, L.~Shi, Y.~Zhang, and S.~Lu, ``{DCell: A Scalable and
  Fault-Tolerant Network Structure for Data Centers},'' in
  \emph{SIGCOMM08}.\hskip 1em plus 0.5em minus 0.4em\relax Association for
  Computing Machinery, Inc., August 2008.

\bibitem{REF38}
\BIBentryALTinterwordspacing
{Cisco Nexus 3548-X, 3524-X, 3548-XL, and 3524-XL Switches Data Sheet}. (Cited
  on 2019, Sept). [Online]. Available:
  \url{https://www.cisco.com/c/en/us/products/collateral/switches/nexus-3548-switch/data_sheet_c78-707001.pdf}
\BIBentrySTDinterwordspacing

\bibitem{REF39}
\BIBentryALTinterwordspacing
{Cisco 500 Series Stackable Managed Switches Data Sheet}. (Cited on 2019,
  Sept). [Online]. Available:
  \url{http://www.cisco.com/c/en/us/products/collateral/switches/small-business-500-series-stackable-managed-switches/c78-695646_data_sheet.html}
\BIBentrySTDinterwordspacing

\bibitem{REF40}
\BIBentryALTinterwordspacing
{Cisco 10GBASE SFP+ Modules Data Sheet}. (Cited on 2019, Sept). [Online].
  Available:
  \url{https://www.cisco.com/c/en/us/products/collateral/interfaces-modules/transceiver-modules/data_sheet_c78-455693.htm}
\BIBentrySTDinterwordspacing

\bibitem{REF41}
R.~{Sohan}, A.~{Rice}, W.~M. {Andrew}, and K.~{Mansley}, ``{Characterizing 10
  Gbps network interface energy consumption},'' in \emph{IEEE Local Computer
  Network Conference}, Oct 2010, pp. 268--271.

\bibitem{REF42}
P.~X. Gao, A.~Narayan, S.~Karandikar, J.~Carreira, S.~Han, R.~Agarwal,
  S.~Ratnasamy, and S.~Shenker, ``{Network Requirements for Resource
  Disaggregation},'' in \emph{12th {USENIX} Symposium on Operating Systems
  Design and Implementation ({OSDI} 16)}.\hskip 1em plus 0.5em minus
  0.4em\relax Savannah, GA: {USENIX} Association, 2016, pp. 249--264.

\bibitem{REF43}
Y.~Cheng, M.~Fiorani, R.~Lin, L.~Wosinska, and J.~Chen, ``{POTORI: a passive
  optical top-of-rack interconnect architecture for data centers},''
  \emph{IEEE/OSA Journal of Optical Communications and Networking}, vol.~9,
  no.~5, pp. 401--411, May 2017.

\bibitem{REF44}
M.~Alizadeh and T.~Edsall, ``{On the Data Path Performance of Leaf-Spine
  Datacenter Fabrics},'' in \emph{High-Performance Interconnects (HOTI), 2013
  IEEE 21st Annual Symposium on}, Aug 2013, pp. 71--74.

\bibitem{REF45}
R.~Xie and X.~Jia, ``{Data Transfer Scheduling for Maximizing Throughput of
  Big-Data Computing in Cloud Systems},'' \emph{Cloud Computing, IEEE
  Transactions on}, vol.~PP, no.~99, pp. 1--1, 2015.

\bibitem{REF46}
\BIBentryALTinterwordspacing
{ZXA10 C300:The Industry's First Future-proof Optical Access platform}. (Cited
  on 2018, Apr). [Online]. Available:
  \url{http://www.zte.com.cn/global/products/access/xpon/PON-OLT/424194}
\BIBentrySTDinterwordspacing

\bibitem{REF47}
\BIBentryALTinterwordspacing
{MRV Pluggable Transceivers}. (Cited on 2019, Sept). [Online]. Available:
  \url{http://s1.dtsheet.com/store/data/001288813.pdf?key=b765b7adbd723bf63be5c121991bab76&r=1}
\BIBentrySTDinterwordspacing

\bibitem{REF48}
F.~Guo, O.~Ormond, L.~{Fialho de queiroz}, M.~Collier, and X.~Wang,
  ``\BIBforeignlanguage{English (US)}{Power consumption analysis of a netfpga
  based router},'' \emph{\BIBforeignlanguage{English (US)}{Journal of China
  Universities of Posts and Telecommunications}}, vol.~19, no. SUPPL. 1, pp.
  94--99, 6 2012.

\bibitem{REF49}
J.~Dean and S.~Ghemawat, ``{MapReduce: Simplified Data Processing on Large
  Clusters},'' \emph{Commun. ACM}, vol.~51, no.~1, pp. 107--113, Jan. 2008.

\bibitem{REF50}
A.~Rasmussen, G.~Porter, M.~Conley, H.~V. Madhyastha, R.~N. Mysore, A.~Pucher,
  and A.~Vahdat, ``{TritonSort: A Balanced and Energy-Efficient Large-Scale
  Sorting System},'' \emph{ACM Trans. Comput. Syst.}, vol.~31, no.~1, pp.
  3:1--3:28, Feb. 2013.

\bibitem{REF51}
\BIBentryALTinterwordspacing
{Sort Benchmark Home Page}. (Cited on 2019, Sept). [Online]. Available:
  \url{http://sortbenchmark.org/}
\BIBentrySTDinterwordspacing

\bibitem{REF52}
B.~G. Bathula, M.~Alresheedi, and J.~M.~H. Elmirghani, ``{Energy Efficient
  Architectures for Optical Networks},'' in \emph{Proc IEEE London
  Communications Symposium, London, Sept.}, 2009.

\bibitem{REF53}
B.~G. {Bathula} and J.~M.~H. {Elmirghani}, ``{Energy Efficient Optical Burst
  Switched (OBS) Networks},'' in \emph{2009 IEEE Globecom Workshops}, 2009, pp.
  1--6.

\bibitem{REF54}
{X. Dong, T. El-Gorashi, and J. Elmirghani}, ``{Green IP Over WDM Networks With
  Data Centers},'' \emph{Lightwave Technology, Journal of}, vol.~29, no.~12,
  pp. 1861--1880, June 2011.

\bibitem{REF55}
X.~Dong, T.~El-Gorashi, and J.~Elmirghani, ``{IP Over WDM Networks Employing
  Renewable Energy Sources},'' \emph{Lightwave Technology, Journal of},
  vol.~29, no.~1, pp. 3--14, Jan 2011.

\bibitem{REF56}
X.~{Dong}, A.~{Lawey}, T.~E.~H. {El-Gorashi}, and J.~M.~H. {Elmirghani},
  ``{Energy-efficient core networks},'' in \emph{2012 16th International
  Conference on Optical Network Design and Modelling (ONDM)}, 2012, pp. 1--9.

\bibitem{REF57}
X.~Dong, T.~El-Gorashi, and J.~Elmirghani, ``{On the Energy Efficiency of
  Physical Topology Design for IP Over WDM Networks},'' \emph{Lightwave
  Technology, Journal of}, vol.~30, no.~12, pp. 1931--1942, June 2012.

\bibitem{REF58}
T.~E. El-Gorashi, X.~Dong, and J.~M. Elmirghani,
  ``\BIBforeignlanguage{English}{{Green optical orthogonal frequency-division
  multiplexing networks}},'' \emph{\BIBforeignlanguage{English}{IET
  Optoelectronics}}, vol.~8, pp. 137--148, June 2014.

\bibitem{REF59}
L.~Nonde, T.~El-Gorashi, and J.~Elmirghani, ``{Energy Efficient Virtual Network
  Embedding for Cloud Networks},'' \emph{Lightwave Technology, Journal of},
  vol.~33, no.~9, pp. 1828--1849, May 2015.

\bibitem{REF60}
J.~M.~H. Elmirghani, T.~Klein, K.~Hinton, L.~Nonde, A.~Q. Lawey, T.~E.~H.
  El-Gorashi, M.~O.~I. Musa, and X.~Dong, ``{GreenTouch GreenMeter core network
  energy-efficiency improvement measures and optimization},'' \emph{IEEE/OSA
  Journal of Optical Communications and Networking}, vol.~10, no.~2, pp.
  A250--A269, Feb 2018.

\bibitem{REF61}
M.~O.~I. {Musa}, T.~E.~H. {El-Gorashi}, and J.~M.~H. {Elmirghani}, ``{Bounds on
  GreenTouch GreenMeter Network Energy Efficiency},'' \emph{Journal of
  Lightwave Technology}, vol.~36, no.~23, pp. 5395--5405, 2018.

\bibitem{REF62}
M.~{Musa}, T.~{Elgorashi}, and J.~{Elmirghani}, ``{Bounds for energy-efficient
  survivable IP over WDMnetworks with network coding},'' \emph{IEEE/OSA Journal
  of Optical Communications and Networking}, vol.~10, no.~5, pp. 471--481,
  2018.

\bibitem{REF63}
{M. {Musa} and T. {Elgorashi} and J. {Elmirghani}}, ``{Energy efficient
  survivable IP-over-WDM networks with network coding},'' \emph{IEEE/OSA
  Journal of Optical Communications and Networking}, vol.~9, no.~3, pp.
  207--217, 2017.

\bibitem{REF64}
{H. A. {Alharbi} and T. E. H. {Elgorashi} and J. M. H. {Elmirghani}}, ``{Impact
  of the Net Neutrality Repeal on Communication Networks},'' \emph{IEEE
  Access}, vol.~8, pp. 59\,787--59\,800, 2020.

\bibitem{REF65}
A.~Lawey, T.~El-Gorashi, and J.~Elmirghani, ``{Distributed Energy Efficient
  Clouds Over Core Networks},'' \emph{Lightwave Technology, Journal of},
  vol.~32, no.~7, pp. 1261--1281, April 2014.

\bibitem{REF66}
A.~Q. Lawey, T.~E.~H. El-Gorashi, and J.~M.~H. Elmirghani, ``{BitTorrent
  Content Distribution in Optical Networks},'' \emph{Journal of Lightwave
  Technology}, vol.~32, no.~21, pp. 4209--4225, Nov 2014.

\bibitem{REF67}
N.~I. Osman, T.~El-Gorashi, L.~Krug, and J.~M.~H. Elmirghani,
  ``{Energy-Efficient Future High-Definition TV},'' \emph{Journal of Lightwave
  Technology}, vol.~32, no.~13, pp. 2364--2381, July 2014.

\bibitem{REF68}
S.~H. {Mohamed}, M.~B.~A. {Halim}, T.~E.~H. {Elgorashi}, and J.~M.~H.
  {Elmirghani}, ``{Fog-Assisted Caching Employing Solar Renewable Energy and
  Energy Storage Devices for Video on Demand Services},'' \emph{IEEE Access},
  vol.~8, pp. 115\,754--115\,766, 2020.

\bibitem{REF69}
H.~A. {Alharbi}, T.~E.~H. {Elgorashi}, and J.~M.~H. {Elmirghani}, ``{Energy
  Efficient Virtual Machines Placement Over Cloud-Fog Network Architecture},''
  \emph{IEEE Access}, vol.~8, pp. 94\,697--94\,718, 2020.

\bibitem{REF70}
A.~M. Al-Salim, A.~Q. Lawey, T.~E.~H. El-Gorashi, and J.~M.~H. Elmirghani,
  ``{Energy Efficient Big Data Networks: Impact of Volume and Variety},''
  \emph{IEEE Transactions on Network and Service Management}, vol.~PP, no.~99,
  pp. 1--1, 2017.

\bibitem{REF71}
A.~M. {Al-Salim}, T.~E.~H. {El-Gorashi}, A.~Q. {Lawey}, and J.~M.~H.
  {Elmirghani}, ``{Greening big data networks: velocity impact},'' \emph{IET
  Optoelectronics}, vol.~12, no.~3, pp. 126--135, 2018.

\bibitem{REF72}
H.~M.~M. Ali, T.~E.~H. El-Gorashi, A.~Q. Lawey, and J.~M.~H. Elmirghani,
  ``Future energy efficient data centers with disaggregated servers,''
  \emph{Journal of Lightwave Technology}, vol.~35, no.~24, pp. 5361--5380,
  2017.

\bibitem{REF73}
A.~N. {Al-Quzweeni}, A.~Q. {Lawey}, T.~E.~H. {Elgorashi}, and J.~M.~H.
  {Elmirghani}, ``{Optimized Energy Aware 5G Network Function
  Virtualization},'' \emph{IEEE Access}, vol.~7, pp. 44\,939--44\,958, 2019.

\bibitem{REF74}
Z.~T. {Al-Azez}, A.~Q. {Lawey}, T.~E.~H. {El-Gorashi}, and J.~M.~H.
  {Elmirghani}, ``Energy efficient iot virtualization framework with peer to
  peer networking and processing,'' \emph{IEEE Access}, vol.~7, pp.
  50\,697--50\,709, 2019.

\bibitem{REF75}
H.~Q. {Al-Shammari}, A.~Q. {Lawey}, T.~E.~H. {El-Gorashi}, and J.~M.~H.
  {Elmirghani}, ``{Resilient Service Embedding in IoT Networks},'' \emph{IEEE
  Access}, vol.~8, pp. 123\,571--123\,584, 2020.

\bibitem{REF76}
{H. Q. {Al-Shammari} and A. Q. {Lawey} and T. E. H. {El-Gorashi} and J. M. H.
  {Elmirghani}}, ``{Service Embedding in IoT Networks},'' \emph{IEEE Access},
  vol.~8, pp. 2948--2962, 2020.

\bibitem{REF77}
M.~S. Hadi, A.~Q. Lawey, T.~E. El-Gorashi, and J.~M. Elmirghani, ``Big data
  analytics for wireless and wired network design: A survey,'' \emph{Computer
  Networks}, vol. 132, pp. 180 -- 199, 2018.

\bibitem{REF78}
M.~S. {Hadi}, A.~Q. {Lawey}, T.~E.~H. {El-Gorashi}, and J.~M.~H. {Elmirghani},
  ``{Patient-Centric HetNets Powered by Machine Learning and Big Data Analytics
  for 6G Networks},'' \emph{IEEE Access}, vol.~8, pp. 85\,639--85\,655, 2020.

\bibitem{REF79}
A.~E.~A. Eltraify, S.~H. Mohamed, and J.~M.~H. Elmirghani, ``{VM placement over
  WDM-TDM AWGR PON Based Data Centre Architecture},'' in \emph{IEEE 22nd
  International Conference on Transparent Optical Networks (ICTON)}, 2020.

\bibitem{REF80}
A.~E.~A. Eltraify, M.~O.~I. Musa, A.~Al-Quzweeni, and J.~M.~H. Elmirghani,
  ``{Evaluation of Applications Latency in Server Centric Passive Optical
  Network Based Data Centre Architectures},'' in \emph{IEEE 22nd International
  Conference on Transparent Optical Networks (ICTON)}, 2020.

\end{thebibliography}

\vfill
\begin{IEEEbiography}[{\includegraphics[width=1in,height=1.25in,clip,keepaspectratio]{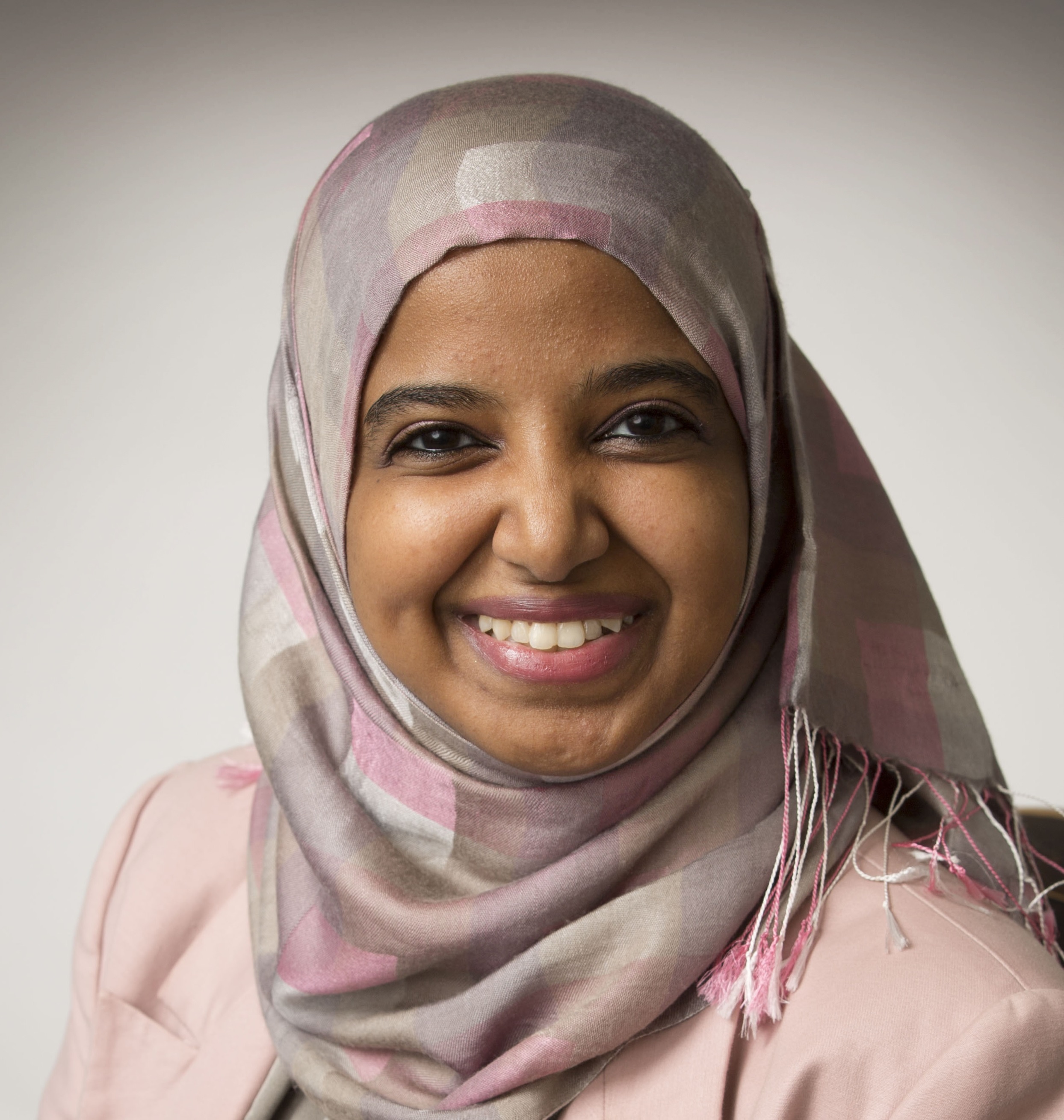}}]{Sanaa Hamid Mohamed}
received the B.Sc. degree (honors) in electrical and electronic engineering from the University of Khartoum, Sudan, in 2009, the M.Sc. degree in electrical engineering from the American University of Sharjah (AUS), United Arab Emirates, in 2013 and the PhD degree in electronic and electrical engineering from the University of Leeds, United Kingdom, in 2020. She received a full-time graduate teaching assistantship in the MSELE program, AUS in 2011. She received a Doctoral Training Award (DTA) from the UK Engineering and Physical Sciences Research Council (EPSRC) to fund her PhD studies at the University of Leeds in 2015. Currently, she is a research fellow at the Institute of Communication and Power Networks, University of Leeds, UK. She has held research and teaching positions at the University of Khartoum, AUS, Sudan University of Science and Technology, and Khalifa University between 2009 and 2013. She has been an IEEE member since 2008 and a member of the IEEE communication, photonics, computer, cloud computing, software defined networks, and sustainable ICT societies. Her research interests include wireless communications, optical communications, optical networking, software-defined networking, data centre networking, big data analytics, and cloud and fog computing.
\end{IEEEbiography}
\vfill
\begin{IEEEbiography}[{\includegraphics[width=1in,height=1.25in,clip,keepaspectratio]{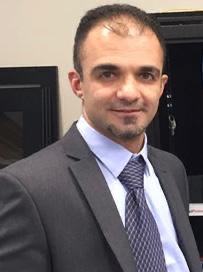}}]{Ali Hammadi} received the B.S. and M.S degrees in electrical engineering from the University of North Carolina at Charlotte, USA in 2001 and 2009 respectively and the Ph.D. degree from University of Leeds, Leeds, U.K., in 2016. He was a telecommunication engineer with Kuwait National Petroleum Company (KNPC) from 2001-2007. Currently, he is an assistant professor in the electronics engineering department at the Public Authority of Applied Education and Training, Kuwait. His research interests include energy efficiency in optical networks, passive optical data center architectures, cloud computing,  and Software Defined Networks in PON based architectures.
\end{IEEEbiography}
\vfill
\begin{IEEEbiography}[{\includegraphics[width=1in,height=1.25in,clip,keepaspectratio]{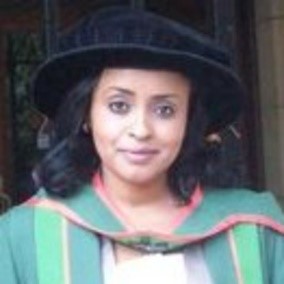}}]{Taisir E. H. El-Gorashi}
received the B.S. degree (first-class Hons.) in electrical and electronic engineering from the University of Khartoum, Khartoum, Sudan, in 2004, the M.Sc. degree (with distinction) in photonic and communication systems from the University of Wales, Swansea, UK, in 2005, and the PhD degree in optical networking from the University of Leeds, Leeds, UK, in 2010. She is currently a Lecturer in optical networks in the School of Electrical and Electronic Engineering, University of Leeds. Previously, she held a Postdoctoral Research post at the University of Leeds (2010- 2014), where she  focused on the energy efficiency of optical networks investigating the use of renewable energy in core networks, green IP over WDM networks with datacenters, energy efficient physical topology design, energy efficiency of content distribution networks, distributed cloud computing, network virtualization and big data. In 2012, she was a BT Research Fellow, where she developed energy efficient hybrid wireless-optical broadband access networks and explored the dynamics of TV viewing behavior and program popularity. The energy efficiency techniques developed during her postdoctoral research contributed 3 out of the 8 carefully chosen core network energy efficiency improvement measures recommended by the GreenTouch consortium for every operator network worldwide. Her work led to several invited talks at GreenTouch, Bell Labs, Optical Network Design and Modelling conference, Optical Fiber Communications conference, International Conference on Computer Communications, EU Future Internet Assembly, IEEE Sustainable ICT Summit and IEEE 5G World Forum and collaboration with Nokia and Huawei.
\end{IEEEbiography}
\vfill
\begin{IEEEbiography}[{\includegraphics[width=1in,height=1.25in,clip,keepaspectratio]{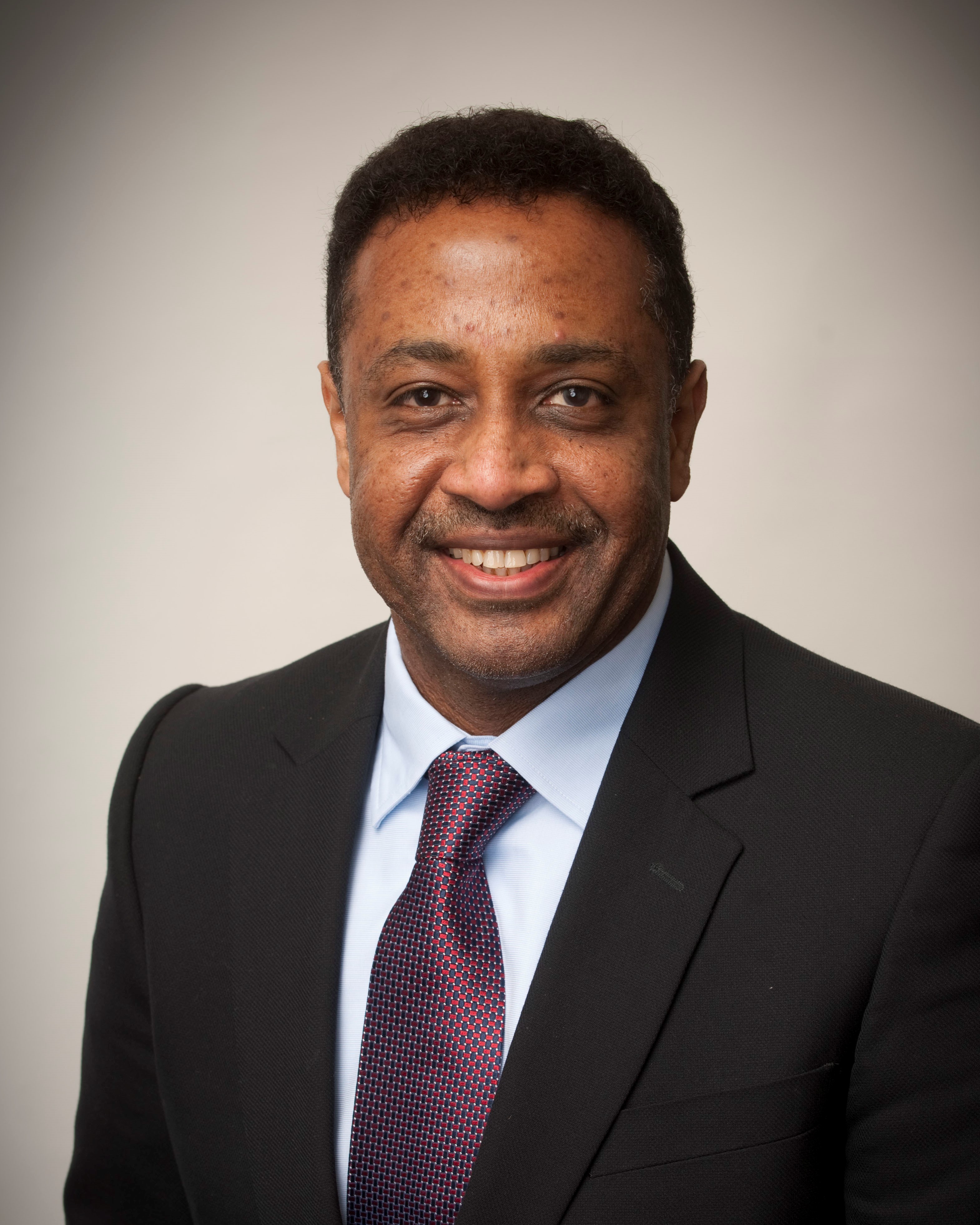}}]{Jaafar M. H. Elmirghani (M'92-SM'99)}
is the Director of the Institute of Communication and Power Networks within the School of Electronic and Electrical Engineering, University of Leeds, UK. He joined Leeds in 2007 and prior to that (2000-2007) as chair in optical communications at the University of Wales Swansea he founded, developed and directed the Institute of Advanced Telecommunications and the Technium Digital (TD), a technology incubator/spin-off hub. He has provided outstanding leadership in a number of large research projects at the IAT and TD. He received the Ph.D. in the synchronization of optical systems and optical receiver design from the University of Huddersfield UK in 1994 and the DSc in Communication Systems and Networks from University of Leeds, UK, in 2012. He has co-authored Photonic switching Technology: Systems and Networks, (Wiley) and has published over 550 papers. He has research interests in optical systems and networks. Prof. Elmirghani is Fellow of the IET, Fellow of the Institute of Physics and Senior Member of IEEE. He was Chairman of IEEE Comsoc Transmission Access and Optical Systems technical committee and was Chairman of IEEE Comsoc Signal Processing and Communications Electronics technical committee, and an editor of IEEE Communications Magazine. He was founding Chair of the Advanced Signal Processing for Communication Symposium which started at IEEE GLOBECOM'99 and has continued since at every ICC and GLOBECOM. Prof. Elmirghani was also founding Chair of the first IEEE ICC/GLOBECOM optical symposium at GLOBECOM'00, the Future Photonic Network Technologies, Architectures and Protocols Symposium. He chaired this Symposium, which continues to date under different names. He was the founding chair of the first Green Track at ICC/GLOBECOM at GLOBECOM 2011, and is Chair of the IEEE Sustainable ICT Initiative, a pan IEEE Societies Initiative responsible for Green and Sustainable ICT activities across IEEE, 2012-present. He is and has been on the technical program committee of 41 IEEE ICC/GLOBECOM conferences between 1995 and 2021 including 19 times as Symposium Chair. He received the IEEE Communications Society Hal Sobol award, the IEEE Comsoc Chapter Achievement award for excellence in chapter activities (both in 2005), the University of Wales Swansea Outstanding Research Achievement Award, 2006, the IEEE Communications Society Signal Processing and Communication Electronics outstanding service award, 2009, a best paper award at IEEE ICC'2013, the IEEE Comsoc Transmission Access and Optical Systems outstanding Service award 2015 in recognition of ``Leadership and Contributions to the Area of Green Communications'', received the GreenTouch 1000x award in 2015 for ``pioneering research contributions to the field of energy efficiency in telecommunications'', the 2016 IET Optoelectronics Premium Award, shared with 6 GreenTouch innovators the 2016 Edison Award in the ``Collective Disruption'' Category for their work on the GreenMeter, an international competition, and received the IEEE Comsoc Transmission Access and Optical Systems outstanding Technical Achievement award 2020 in recognition of ``Outstanding contributions to the energy efficiency of optical communication systems and networks'', clear evidence of his seminal contributions to Green Communications which have a lasting impact on the environment (green) and society. He is currently an editor/associate editor of: IEEE Journal of Lightwave Technology, IEEE Communications Magazine, IET Optoelectronics, Journal of Optical Communications, and is Area Editor for IEEE Journal on Selected Areas in Communications (JSAC) Series on Machine Learning in Communication Networks (Area Editor). He was an editor of IEEE Communications Surveys and Tutorials and IEEE Journal on Selected Areas in Communications series on Green Communications and Networking. He was Co-Chair of the GreenTouch Wired, Core and Access Networks Working Group, an adviser to the Commonwealth Scholarship Commission, member of the Royal Society International Joint Projects Panel and member of the Engineering and Physical Sciences Research Council (EPSRC) College. He was Principal Investigator (PI) of the \pounds6m EPSRC INTelligent Energy awaRe NETworks (INTERNET) Programme Grant, 2010-2016 and is currently PI of the \pounds6.6m EPSRC Terabit Bidirectional Multi-user Optical Wireless System (TOWS) for 6G LiFi Programme Grant, 2019-2024. He has been awarded in excess of \pounds30 million in grants to date from EPSRC, the EU and industry and has held prestigious fellowships funded by the Royal Society and by BT. He was an IEEE Comsoc Distinguished Lecturer 2013-2016.
\end{IEEEbiography}
\vfill

\end{document}